\documentclass[fleqn]{mnras}

\usepackage{newtxtext}
\usepackage[T1]{fontenc}

\DeclareRobustCommand{\VAN}[3]{#2}
\let\VANthebibliography\thebibliography
\def\thebibliography{\DeclareRobustCommand{\VAN}[3]{##3}\VANthebibliography}

\usepackage{graphicx}	% Including figure files
\usepackage{amsmath}	% Advanced maths commands
\usepackage{longtable}
\usepackage{amssymb}	% Extra maths symbols

\title[Colour evolution of Betelgeuse and Antares]{Colour evolution of Betelgeuse and Antares over two millennia,
derived from historical records, as a new constraint on mass and age}

\author[R. Neuh\"auser et al.]{
R. Neuh\"auser,$^{1}$\thanks{E-mail: ralph.neuhaeuser@uni-jena.de},
G. Torres$^{2}$,
M. Mugrauer$^{1}$,
D.L. Neuh\"auser$^{3}$,
J. Chapman$^{4}$,
D. Luge$^{1}$, 
M. Cosci$^{5}$
\\
$^{1}$ Astrophysikalisches Institut und Universit\"ats-Sternwarte Jena, Schillerg\"a\ss{}chen 2-3, 07745 Jena, Germany \\
$^{2}$ Center for Astrophysics $\vert$ Harvard \& Smithsonian, 60 Garden Street, Cambridge, MA 02138, USA \\
$^{3}$ Independent scholar, Meran/o, Alto Adige, Italy \\
$^{4}$ Department of East Asian Languages and Cultures, UC Berkeley, Berkeley, CA 94720, USA, now Independent scholar, Portland, OR 97202, USA \\
$^{5}$ Department of Philosophy and Cultural Heritage, Ca' Foscari University Venice, Malcanton Marcor\`a, Dorsoduro 3484/D, 30123 Venice, Italy
}

\date{Accepted 2022 July 7. Received 2022 July 7; in original form 2022 March 9}

\pubyear{2022}

\begin{document}
\label{firstpage}
\pagerange{\pageref{firstpage}--\pageref{lastpage}}

\maketitle

\begin{abstract}
After core hydrogen burning, massive stars evolve from blue-white dwarfs to red supergiants by expanding, 
brightening, and cooling within few millennia. 
We discuss a previously neglected constraint on mass, age, and evolutionary state of Betelgeuse and Antares,
namely their observed colour evolution over historical times:
We place all 236 stars bright enough for their colour to be discerned by the unaided eye (V$\le$3.3 mag)
on the colour-magnitude-diagram (CMD), and focus on those in the Hertzsprung gap.
We study pre-telescopic records on star colour with historically-critical methods
to find stars that have evolved noticeably in colour within the last millennia.
Our main result is that Betelgeuse was recorded with 
a colour significantly different (non-red) than today (red, B$-$V=$1.78 \pm 0.05$ mag).
Hyginus (Rome) and Sima Qian (China) independently report it 
two millennia ago as appearing like Saturn (B$-$V=$1.09 \pm 0.16$ mag)
in colour and `yellow' (quantifiable as B$-$V=$0.95 \pm 0.35$ mag), respectively
(together, 5.1$\sigma$ different from today). 
The colour change of Betelgeuse is a new, tight constraint 
for single-star theoretical evolutionary models (or merger models).
It is most likely located less than one millennium past the bottom of the red giant branch,
before which rapid colour evolution is expected.
Evolutionary tracks from MIST consistent with both its colour evolution
and its location on the CMD suggest a mass of $\sim$14M$_{\odot}$ at $\sim$14 Myr.
The (roughly) constant colour of Antares for the last three millennia
also constrains its mass and age.
Wezen was reported white historically, but is now yellow.
\end{abstract}

\begin{keywords}
stars: individual: Betelgeuse, Antares, Wezen -- stars: Hertzsprung-Russell and colour-magnitude-diagram --
stars: evolution -- stars: supergiants -- history and philosophy of astronomy
\end{keywords}

%%%%%%%%%%%%%%%%%%%%%%%%%%%%%%%%%%%%%%%%%%%%%%%%%%

%%%%%%%%%%%%%%%%% BODY OF PAPER %%%%%%%%%%%%%%%%%%

\section{Introduction}

Historical observations provide valuable input for many fields of astrophysics.
Examples include the reconstruction of past solar activity with sunspots and aurorae
(review in Vaquero \& Vazquez 2009; see also R. Neuh\"auser \& D.L. Neuh\"auser 2015),
the determination of cometary orbits (compilation in Kronk 1999; recent example in R. Neuh\"auser et al. 2021),
or the study of Galactic supernovae (review in Stephenson \& Green 2002; recent examples in, e.g.,
Rada \& R. Neuh\"auser 2015 and R. Neuh\"auser et al. 2016).
Here, we consider pre-telescopic colour records of stars with historical-critical methods
(R. Neuh\"auser, D.L. Neuh\"auser, \& Posch 2021) as a new test of theoretical evolutionary models,
and to better constrain masses, ages, and the evolutionary state of supergiants.
We aim to show that, two millennia ago, Betelgeuse ($\alpha$ Ori) was reported with a significant different (non-red)
colour than Antares ($\alpha$ Sco), which was always given as red, while today, both have almost the same red colour.
Hence, Betelgeuse has evolved rapidly in colour as it crossed the Hertzsprung gap,
which is consistent with its location in the colour-magnitude-diagram and constrains its parameters.
Only historical observations can provide such an empirical constraint.

The evolution of stars is often studied with the Hertzsprung-Russell diagram (HRD, luminosity versus temperature) 
or the colour-magnitude-diagram (CMD, brightness versus colour). During the stable, central hydrogen-burning phase, 
observable parameters hardly change for $10^{6-10}$ yr (the higher the stellar mass, the faster the evolution). 
Later, when H-to-He-fusion moves to an outer shell, the star expands, brightens, and reddens. 

Relatively few stars reside in the central part of the HRD or CMD, 
the so-called Hertzsprung gap: stars with $\sim$8--18~M$_{\odot}$ (solar mass) evolve from bluish-white main-sequence stars 
to red supergiants within only $\sim 10^{4}$ yr, so that the colour of some naked-eye (super-)giants may have changed 
noticeable from antiquity to today. However, the human eye can detect star colours only if brightness, hue (here B$-$V),
and saturation (admixture of white) are within certain ranges (Steffey 1992). Here we search for stars for which pre-telescopic 
records indicate a colour change compared to today that is consistent with their evolutionary status: 
theoretical calculations of stellar evolution provide tracks for different masses and predict how brightness, 
temperature, colour, mass, radius, etc. change with time due to fusion (e.g. Choi et al. 2016).

For thermal radiation (blackbody with absorption lines), star colour is quantified as an observed colour index, 
e.g. B$-$V, the difference in brightness between the blue (B centered around 440 nm) and the visible band V ($\sim$ 550 nm).
In Table 1, we define the colour terms `blue, white, yellow, orange', and `red' as applied to stars and planets.
Correcting B$-$V for interstellar extinction A$_{\rm V}$ leads to the intrinsic colour index (B$-$V)$_{0}$, 
related to the effective temperature.

Small, secular variations in brightness and/or colour over millennia are expected for evolving stars, 
but were rarely considered: Hearnshaw (1999) found that, statistically, there have been no 
significant brightness changes 
for giants and supergiants compared to Ptolemy's Almagest (2nd cent. AD).
One or a few individual stars might have evolved in brightness, but this would be hard
to prove because the magnitudes given in ancient star catalogs such as the Almagest scatter around
today's values by about $\pm 1$ mag (Hearnshaw 1999; Schaefer 2013).
There has been a debate on a possible colour change in Sirius, which is reported as `somewhat reddish' in
the Almagest, whereas it is given as blue or white in several other ancient works.
The colour in the Almagest
is likely due to either strong scintillation (some red rays) or as a late addition by a copying scribe 
(Ceragioli 1995 and references therein), rather than to an observation at low altitude.

\begin{table*}
\begin{tabular}{lrrccccl} \hline
\multicolumn{8}{l}{{\bf Table 1: The colour indices of stars:} colour, peak wavelength, effective temperature, colour index, and spectral type/} \\
\multicolumn{8}{l}{luminosity class (Allen 1973; Schmidt-Kaler 1982; Steffey 1992; Schaefer 1993; Drilling \& Landolt 2000)}  \\ \hline
colour  & wavelength & effective        & colour index  & \multicolumn{2}{l}{spectral type} & super- & examples \\
       & at peak [\AA] & temp.\,[K] & (B$-$V)$_{0}$ [mag] & dwarfs & giants & giants & and notes \\ \hline
red    & $\ge 5900$ & $\le 4000$       & $\ge 1.40$   & M0-9   & K4-M9  & K3-M9 & $\alpha$ Ori, $\alpha$ Sco \\ 
orange & 5800-5900  & 4000-5000        & 0.80 to 1.40 & K0-7   & G4-K3  & G1-K2 & $\alpha$ Boo, $\beta$ Gem \\
yellow & 5700-5800  & 5000-6000        & 0.60 to 0.80 & G1-9   & G0-3   & F8-G0 & Sun\,(1), $\alpha$ Aur\,(2) \\
green\,(3) & 4800-5700  & 6000-7300    & 0.30 to 0.60 & F      & F      & F4-7  & $\alpha$ CMi \\ 
white  &            & 7300-10000       & 0.00 to 0.30 & A      & A0-9   & A0-F3 & $\alpha$ Lyr (4) \\
blue   & 4400-4800  & $\ge 10000$      & $-0.33$ to 0.00 & OB  & OB     & OB    & $\alpha$ Vir, $\gamma$ Ori \\ \hline
\end{tabular}

\vspace{.1cm}

Notes: 
In astrophysical contexts, we use the colour terms as defined here,
e.g., a star is considered `red' if B$-$V$\ge$1.4 mag, or `orange', if B$-$V=0.8 to 1.4 mag. 
(1) Our Sun has B$-$V=0.65 mag (Livingston 2000). 
(2) $\alpha$ Aur has B$-$V=0.80 mag at the yellow-orange border. 
(3) Technically, we cannot perceive any stars as `green',
not even in the given temperature or spectral type range, 
but we see them instead as whitish. 
(4) The well-known photometric standard star Vega (spectral type A0) 
by definition has a colour index of 0.00 mag (white). 
NB: Due to absorption lines, the perceived colour of a star can be different
compared to the colour corresponding to the blackbody temperature, e.g., less red.
Within the general term `colour', we also include `white'.
\end{table*}

Cases of fast evolution through the H-R diagram are known from recent observations, such as Sakurai's object (V4334 Sgr):
it is considered to be a white dwarf that swelled up in its final shell helium flash to become briefly like a red giant;
currently, it has a quiescent brightness of V=14--15 mag (AAVSO) --
and it did move significantly through the H-R diagram within years from a temperature of
$10^{4}$ K in 1993 to $10^{5}$ K less than ten years later 
(Hajduk et al. 2005).
Other examples of rapid evolution include post-red-supergiant spectral evolution seen in IRC +10 420, 
or period changes in Cepheid variables (Smith 2014).

Since we deal with naked-eye detection of star colour, we give here a short primer on human colour perception:
During the bright day (the photopic vision regime), both cones and rods work to facilitate colour sensation.
During dark, moon-less nights (the scotopic regime), cones cannot be excited any more, and only rods are, so that
only light intensity variations can be noticed (loss of colour information).
During twilight and moon-lit nights (the mesopic regime), both cones and rods still work to
detect colour at least for sufficiently bright objects. The brightness limit for colour detection
lies at the border of the scotopic and mesopic regimes.
There are different cone types for red versus green and blue versus yellow;
colour perception originates from the balance between those differences
(humans are trichromates with the three main colours being red, blue, and green-yellow).
This differential operation has a lower S/N than the detection of light intensity variations,
and cones are also generally less sensitive to light than are rods (except above $\sim 6500$\AA);
this effect increases with age.
Colour vision operates on the parvocellular pathway, i.e. through cells smaller than on the
macrocellular pathway for intensity variations -- hence, the reduced sensitivity.
Colour sensation decreases with smaller brightness, because of insufficient stimulation of the cones.

Whether the colour of a star can be detected by the naked eye, depends on three properties:
brightness (also in contrast to fore- and background), hue (the colour's dominant wavelength),
and saturation (the hue's purity, or the amount of admixture of white).
In practice, we will consider the brightness and colour indices of stars.
When a star is fixated by the eye (direct vision), colour can be detected,
because the fovea, where the photons are focused, are populated by cones.
For more details, see Steffey (1992) and Schaefer (1993).

\bigskip

Here, we study systematically possible colour changes in historical times with two 
independent complementary approaches,
namely, astrophysically by selecting stars located in the Hertzsprung gap (Sect. 2),
and historically by considering pre-telescopic colour reports (Sect. 3).
We then use the historical colour constraints on Betelgeuse and Antares
to estimate their masses (Sect. 4). For Betelgeuse we consider here both
single-star evolutionary models (Sect. 4.1) as well as a merger scenario (Sect. 4.2).
We finish with a summary and conclusions in Sect. 5. 

\section{First approach: Bright stars in the Hertzsprung gap}

For the first, astrophysical approach in finding stars that may have evolved 
significantly in colour in historical times, 
we compile the colour indices 
of all bright stars other than the Sun down to V=3.3 mag.
We will assume here that yellow-orange-red colours of stars brighter than this can be discerned
by the naked eye.
This limit is represented by Edasich ($\iota$ Dra, V=3.3 mag, B$-$V=1.18 mag, i.e. orange from Table 1), 
which is the faintest orange-red star for which a colour has been reported in pre-telescopic times (see Sect. 3.1)\footnote{Ibn Qutayba
wrote on the Bedouine tradition: `al-\d{d}\textit{\={\i}}\d{h}: a red [Arabic: a\d{h}mar] star above
Bootes between Ursa Major and Vega', identified as $\iota$ Dra by Al-\d{S}\=uf\textit{\={\i}} (Kunitzsch 1961, p. 53).}.
Alternatively, if the identification of that red star as Edasich were considered to be uncertain,
a star of nearly equal brightness, $\gamma$ Hyi (V=3.26 mag, B$-$V=1.59 mag, spectral type of M1 III)
can be seen visually as red by naked eye by one of us (MM), although it is not found in pre-telescopic reports;
it is too far south for cultures which left written records.
Among these bright stars, we will focus here on those located in the Hertzsprung gap in more detail, 
which are the ones whose colours should evolve more quickly.

\subsection{Input data for the colour-magnitude-diagram of all bright stars}

In order to place all bright stars on the CMD, we use the most homogeneous sources available:
V-band magnitudes and B$-$V colour indices in the Johnson system and their measurement uncertainties are taken from the 
main part of the Tycho catalog (ESA 1997), if available, or otherwise from the Hipparcos main catalog (ESA 1997);
for HIP 61932 and HIP 36850, we take them from Kharchenko (2009) and Barrado (1998), respectively.
Extinction values A$_{\rm V}$ are adopted from Gontcharov \& Mosenkov (2017) for most stars that have them,
and for the remaining 11 stars we consulted Bobylev, Gontcharov, \& Bajkova (2006) and Melnik \& Dambis (2020).
Parallaxes are from the new Hipparcos reduction (van Leeuwen 2007).
For these bright stars for which naked-eye colour detection is possible, the parallaxes are always from
Hipparcos because the stars are too bright for Gaia (limit around 3.6 mag).

We calculate the absolute magnitude M$_{\rm V}$ from apparent V-band brightness, extinction, and parallax, and
the intrinsic colour index (B$-$V)$_{0}$ from B- and V-band brightness and reddening (adopting the usual
interstellar reddening law with R$_{\rm V}=3.1$).
The mean of the measurement uncertainties on the absolute magnitudes (mostly from the parallax uncertainties) is $0.09 \pm 0.16$ mag, 
and the mean of the intrinsic colour index error bars (from V and B uncertainties) is $0.006 \pm 0.012$ mag.
The accuracy of the extinction estimates is known to be better than 0.04 mag (Gontcharov \& Mosenkov 2017).
We also check and confirm that the dereddened colour indices are consistent with the spectral types as given by SIMBAD
(using the calibrations from Drilling \& Landolt 2000).
See the table in the appendix for the data.

\begin{center}
\begin{figure*}
\includegraphics[angle=270,width=1\textwidth]{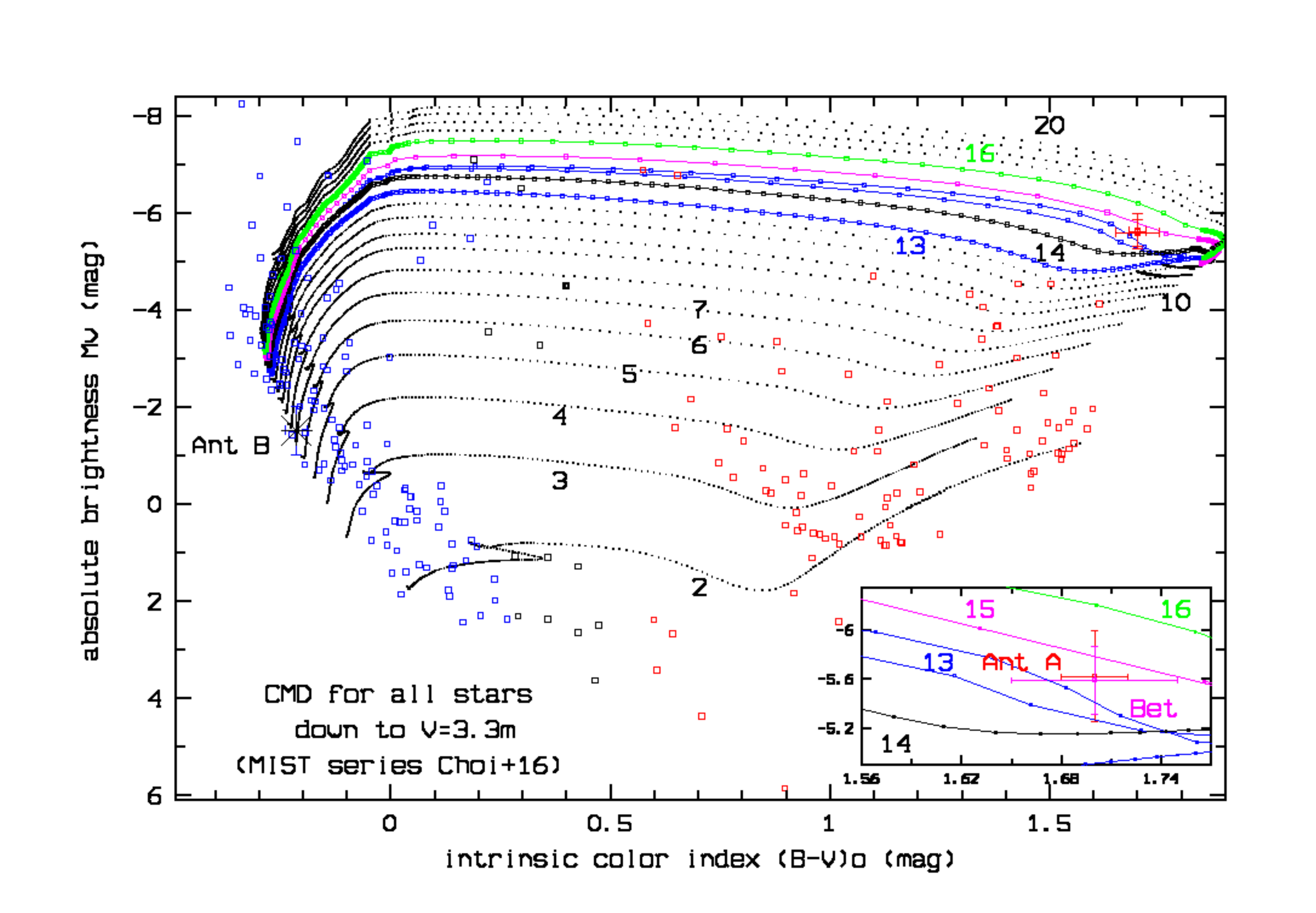}
\caption{{\bf Colour-magnitude-diagram for all bright stars down to V=3.3 mag.}
We plot absolute magnitude M$_{\rm V}$ versus intrinsic colour index (B$-$V)$_{0}$ with MIST tracks 
(Choi et al. 2016) for 2 to 20~M$_{\odot}$ (some labeled); dots indicate the (unequal) MIST time resolution. 
Tracks discussed for Betelgeuse
are enlarged in the inlay, such as the 13~M$_{\odot}$ blue loop (lower-mass blue loops are omitted 
for clarity, but see Fig. 2). Stars with B$-$V$\ge$0.6 mag are shown in red, 
those with B$-$V$\le$0.3 mag in blue. Antares A and Betelgeuse are seen in the
upper right with error bars (enlarged in the inlay, Betelgeuse plotted in pink).}
\end{figure*}
\end{center}

\begin{center}
\begin{figure*}
\includegraphics[angle=270,width=1\textwidth]{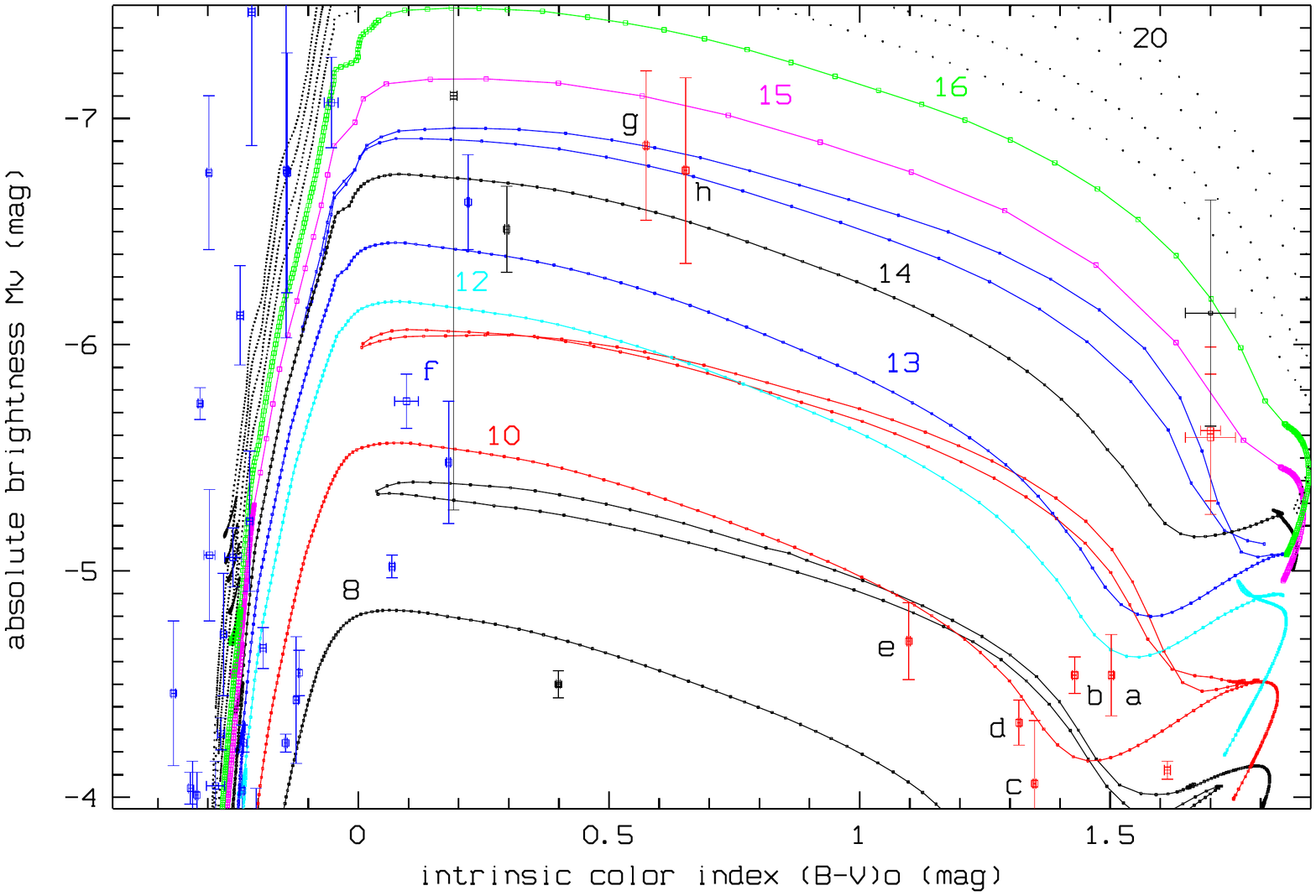}
\caption{{\bf Colour-magnitude-diagram (CMD) for the brighter stars} (M$_{\rm V} \le -3.9$ mag),
similar to Fig.\,1, with tracks for 8, 10, 12, 13, 14, ... 20~M$_{\odot}$ (most labeled) from the MIST series (Choi et al. 2016);
dots indicate the (unequal) MIST time resolution. 
While the MIST tracks for 8, 10, and 13~M$_{\odot}$ show pronounced blue loops, those for 12 and $\ge 14$~M$_{\odot}$ do not.
Betelgeuse and Antares A are seen on the upper right in red with overlapping error bars.
Betelgeuse is also plotted once in black for the larger Harper, Brown, \& Guinan (2008) distance.
Other stars in the Hertzsprung gap discussed in the Sects. 3.4 \& 3.5 are indicated as follows:
a: $\pi$ Pup (Ahadi), b: $\epsilon$ Peg (Enif), c: $\beta$ Ara, d: $\epsilon$ Gem (Mebsuta), 
e: $\epsilon$ Car (Avior), f: $\alpha$ Car (Canopus), g: $\gamma$ Cyg (Sadr), and h: $\delta$ CMa (Wezen);
the star just below the 8~M$_{\odot}$ track is Polaris.}
\end{figure*}
\end{center}

\subsection{The colour-magnitude-diagram for the brightest 236 stars}

The CMD for all 236 stars with V=3.3 mag or brighter is shown in Fig.\,1.
We include each single star or unresolved multiple stellar system seen as such by the naked eye.
Unresolved multiples appear slightly brighter than if they had been resolved (by up to 0.75 mag for equal brightness binaries),
but this should not matter much here, because the rapid evolution across the Hertzsprung gap is mostly parallel to the abscissa (colour axis).
Stars unresolved by the naked eye that appear to lie in the Hertzsprung gap, where fast colour evolution is possible, 
are then examined in detail to determine whether any of their components alone would experience current or recent rapid changes in colour.

In Fig.\,1, we compare the location of stars 
with theoretical stellar evolution tracks for single stars
from the MIST series (MESA Isochrones and Stellar Tracks; Choi et al. 2016; Dotter 2016), 
based on the MESA code (Modules for Experiments in Stellar Astrophysics).
The mass tracks are for 2 to 20~M$_{\odot}$, solar abundance, and a default rotation parameter of $v/v_{\rm crit}=0.4$.
These are among the more popular models used by the community, and are calibrated against a variety of
observational constraints. 
The track for 13~M$_{\odot}$ shows the most pronounced blue loop (Fig.\,2).

The two most promising (and prominent) stars located inside or towards the end of the Hertzsprung gap
are Betelgeuse and Antares (Figs. 1 \& 2).

\subsection{Betelgeuse: input data}

For Betelgeuse (HIP 27989, $\alpha$ Ori), the Tycho catalog gives V=0.57 mag and B$-$V=$1.736 \pm 0.018$ mag, 
while SIMBAD gives V=0.42 mag and B=2.27 mag (Johnson et al. 1966), i.e., B$-$V=1.85 mag 
(the difference may be due to its semiregular variability), i.e. clearly red (see Table 2 for details).
With an extinction of A$_{\rm V}$=0.24 mag (Gontcharov \& Mosenkov 2017), we obtain
(B$-$V)$_{0}=1.659 \pm 0.018$ mag from the Tycho data, and (B$-$V)$_{0}=1.770 \pm 0.018$ mag from data in Johnson et al. (1966).
These two values would correspond to a classification of K7-M0\,I and M2\,I, respectively (Drilling \& Landolt 2000).
The spectral type of Betelgeuse is M1-2\,Ia-Iab, corresponding 
to (B$-$V)$_{0}=1.70 \pm 0.05$ mag (Drilling \& Landolt 2000).
Since this is more consistent with the Tycho value, we use here V=0.57 mag (Tycho);
together with the intrinsic colour and the extinction mentioned above,
this leads to a colour range of B$-$V=1.73 to 1.83 mag (or $1.78 \pm 0.05$ mag).
This range encompasses the slight changes in the colour index before, during, and after the 
recent Great Dimming of Betelgeuse, and even possible colour differences 
stemming from an interstellar extinction law different than the usual one ($R_V=3.1$).

Betelgeuse is not known to have any confirmed bound companions. The Washington Double Star (WDS)
catalog (Mason et al. 2001) lists ten nearby stars, but many or all of them may be 
chance alignments rather than physical companions. They are all 10 to 13.5 mag fainter than Betelgeuse itself.
Early speckle imaging companion candidates with 3--5 mag brightness difference at 0.1--0.5$^{\prime \prime}$ separation
(Karovska et al. 1986) have not been confirmed as bound companions later (WDS).
The radial velocity variations seen in Betelgeuse are more consistent with pulsations
and other local fluctuations, rather than being caused by a low-mass companion (Goldberg 1984; Smith et al. 1989).
In any case, no such companions would significantly alter Betelgeuse's position in the CMD or HRD, 
i.e., they would not affect our mass, age, or colour estimates.
In particular, they would also not have affected pre-telescopic colour estimates.
The input data are listed in Table 2.

Note that the Hipparcos parallax from the re-reduction of the mission data ($6.55 \pm 0.83$ mas, 
corresponding to $151.5 \pm 19$ pc, van Leeuwen 2007) was revisited
using both Hipparcos and VLA data (Harper, Brown, \& Guinan 2008) resulting in 
a parallax of $5.07 \pm 1.10$ mas, implying a distance of $197 \pm 45$ pc.
However, we note that Harper, Brown, \& Guinan (2008) had to omit
the data from one of four VLA bands (X) in the final result, as it was discrepant. 
The distance determination for Betelgeuse is problematic,
because its apparent angular size is larger than its parallax
(convection cells on Betelgeuse's surface subtend a significant fraction of its disc, affecting the photometric centre)
-- this is a more severe problem in radio observations
(Harper, Brown, \& Guinan 2008) than in optical data from Hipparcos.
Hence, we favour the Hipparcos distance as revised in van Leeuwen (2007),
but also consider the value in Harper, Brown, \& Guinan (2008), see e.g. Fig. 2.
With the larger distance, a match to the observations becomes possible also with the 17 and 18~M$_{\odot}$ tracks (Fig.\,2).
And in any case, the two parallax measurements above are consistent within their $1\sigma$ uncertainties.
For a review of the open questions regarding Betelgeuse, see the summary of the 2012 Betelgeuse workshop (van Loon 2013).

\begin{table}
\begin{tabular}{llllll} \hline
\multicolumn{6}{l}{{\bf Table 2: Betelgeuse and Antares -- input data}} \\ \hline
Star           & V    & B            & (B$-$V)$_{0}$ [mag] & sources & comments \\ \hline
Betelgeuse     & 0.57 & 2.35         & $1.659 \pm 0.018$ & Tycho & HIP\,27989 \\
($\alpha$ Ori) & 0.42 & 2.27         & $1.770 \pm 0.018$ & SIMBAD & M1-2 I \\
           & \multicolumn{5}{l}{A$_{\rm V}$=0.24 mag, M1-2\,Ia-ab, i.e., (B$-$V)$_{0}$=$1.70 \pm 0.05$ mag} \\
           & \multicolumn{5}{l}{variability brightness range V=0.0 to 1.6 mag} \\ 
we use:    & 0.57    & \multicolumn{4}{l}{B$-$V=1.73-1.83 mag and (B$-$V)$_{0}$=1.65-1.75 mag} \\
           & \multicolumn{5}{l}{either M$_{\rm V}$=$-5.59 \pm 0.28$ mag at $151.5 \pm 19$ pc (a)} \\ 
           & \multicolumn{5}{l}{or M$_{\rm V}$=$-6.14 \pm 0.50$ mag at $197 \pm 45$ pc (b)} \\ \hline 
$\alpha$\,Sco\,A+B & 1.07           & 2.91          & $1.660 \pm 0.014$ & \multicolumn{2}{l}{A$_{\rm V}$=0.57 mag} \\
(HIP 80763) & \multicolumn{5}{l}{V=0.88-1.16 mag, M$_{\rm V}$=$-5.65 \pm 0.37$ mag, $\sim 170 \pm 30$ pc} \\
Antares B   & $\simeq 5.2$  & $\simeq 5.0$ & \multicolumn{3}{l}{B2-4\,V, (B$-$V)$_{0}$=$-0.24$ to $-0.19$\,mag}\\
Antares A   & $\simeq 1.1$  & $\simeq 3.1$ & \multicolumn{3}{l}{M1.5\,Iab, (B$-$V)$_{0}$=$1.70 \pm 0.02$\,mag} \\
            &               & \multicolumn{4}{l}{(B$-$V)$_{0}$ and A$_{\rm V}$ give B$-$V=$1.88 \pm 0.02$ mag} \\
           & \multicolumn{5}{l}{M$_{\rm V}=-5.62 \pm 0.37$ mag} \\ \hline
\end{tabular}

Notes: (a) van Leeuwen (2007), (b) Harper, Brown, \& Guinan (2008). 
\end{table}

\subsection{Antares: input data} 

For Antares (HIP 80763), the Hipparcos/Tycho catalog (ESA 1997) gives
V=1.07 mag and B$-$V=1.84 mag (the same colour index is reported in SIMBAD, Ducati 2002),
i.e. also clearly red (Table 2). With an extinction of A$_{\rm V}$=0.57 mag (Gontcharov \& Mosenkov 2017), 
we obtain (B$-$V)$_{0}=1.660 \pm 0.014$ mag. The revised Hipparcos distance for Antares is not disputed.
However, it is difficult to spatially resolve the close A+B pair well because of the large
dynamic range: the separation is small ($\sim 3^{\prime \prime}$) and the magnitude difference large ($\sim 4$ mag).
Spatially resolved spectra for Antares A and B have recently been obtained by Ma\'iz Apell\'aniz et al. (2021), 
who reported spectral classifications of M1.5\,Iab and B2\,Vn, respectively.
Previously reported spectral type determinations for the companion range from B2.5\,V (Garrison 1967) 
to B4\,V (Stone \& Struve 1954).
As an M1.5 supergiant, the intrinsic colour of Antares A should be (B$-$V)$_{0}=1.70$ mag,
whereas a B2--4 dwarf such as Antares B should have (B$-$V)$_{0}=-0.24$ to $-0.19$ mag, according to Drilling \& Landolt (2000).
For the resolved Antares B, we have V$\simeq 5.2$ mag and B$\simeq 5.0$ mag (Ma\'iz Apell\'aniz et al. 2021; Garrison 1967;
Kudritzki \& Reimers 1978),
which implies that Antares B lies blueward of the Hertzsprung gap (Fig.\,1).
Due to the uncertainties in the magnitudes and colour of the B companion, 
any calculated magnitudes and colour for the primary A would have relatively large uncertainties
(we obtain for Antares A alone V=$1.1 \pm 0.1$ mag, B=$3.1 \pm 0.1$ mag),
so it is advisable here to use the resolved spectral types for colour index determinations.
The resolved spectral type for Antares A (M1.5\,Iab, Ma\'iz Apell\'aniz et al. 2021) yields (B$-$V)$_{0}=1.70 \pm 0.02$ mag.
Since the spectral type of Antares A is determined with higher precision than for Betelgeuse,
its intrinsic colour index also has a smaller uncertainty.
See Table 2 for all our input data.
Antares A and B are the only components of a binary that is not resolvable by the naked eye
for which we plot the two resolved components in Fig.\,1.

Stars in the Hertzsprung gap are discussed further below.

\section{Second approach: Historical colour records}

In our second, historical approach to 
find stars that evolved in colour in historical times,
we search for pre-telescopic reports 
(i.e. the historical time until AD 1609) 
of star colours mainly from cultures that left written records.
We do this not only for Betelgeuse and Antares, but for any star
in our sample of the 236 brightest stars.
We restrict the study to observations from the pre-telescopic epoch, 
because observers could have been biased by knowing the colours from telescopic observations.

\subsection{Historical text corpus}

Records on star colours in Greek and Latin manuscripts from the Mediterranean antiquity as well as Assyrian and Babylonian sources
have been compiled by Boll \& Bezold (1918), who have consulted, among others, 
Aratos' Greek `Phaenomena' (last quarter of the 3rd century BC) as well as its translations and adaptations
to Latin by Cicero and Germanicus, 
and similar works by Eratosthenes, Manilius, Hyginus, Cleomedes, Avienus, and other authors
(see, e.g., Lippincott 2019 for a recent review on the Aratean text corpus). 
We consult them again in recent editions (some also in translations),
along with the pre-telescopic star catalogs by
Ptolemy (2nd century AD), Al-\d{S}\=uf\textit{\={\i}} (10th cent.), Ulug Beg (15th cent.), as well as Bayer and Brahe (around AD 1600).
For Assyrian texts including MUL.APIN (2nd millennium BC), we consult Kugler (1907), Reiner \& Pingree (1981),
Hunger \& Pingree (1989, 1999), and Hunger \& Steele (2019).
Classical Chinese sources (from the late 2nd millennium BC onward)
were studied in detail in Needham \& Wang (1959), Ho (1962, 1966) and Pankenier (2013),
who discuss explicitly also the colours of stars. We consult Sima Qian's work from the 
2nd/1st century BC (Pankenier 2013) and the 7th century {\it Jin shu} (Ho 1966).
Arabic manuscripts (since the 8th century AD) discussing stars and their colours
were studied in detail in Kunitzsch (1959, 1961, 1974).
Reports from the Bedouins were researched also by Adams (2018), and
Al-\d{S}\=uf\textit{\={\i}} also by Hafez (2010).
We consult Makemson (1939, 1941), Allen (1963), and Noyes (2019, 2021) 
as well for more reports on star colours, including from First Nations.\footnote{N.B.:
Wilk (1999) suggested that brightness (or even colour) changes of Betelgeuse may be hidden in the ancient Greek myth
of Pelops, who was killed and dismembered by his father, but was later restored to life and had
his lost shoulder replaced by a piece of ivory (which may have a somewhat yellowish colour). 
However, whether Pelops is really connected to Orion or its right shoulder (Betelgeuse), remains speculative --
Wilk (1999): `There is, in fact, little mythology associated with the figure (Orion), and the little
that exists is somewhat contradictory'.}

The relevant texts were analyzed with a historically-critical methodology 
(R. Neuh\"auser, D.L. Neuh\"auser, \& Posch, 2021; D.L. Neuh\"auser et al. 2021) --
also to transform them from a qualitative to a quantitative statement;
we also give a more literal translation to English.

Dealing with historically transmitted colour terms requires special care.
Colour terms translated to `red', for example, do not automatically correspond to
a colour index range we define today as red (see Table 1).
Furthermore, a term such as `red' may not have exactly the same meaning in different languages.
Nevertheless, proper understanding, calibration, and quantification of historical colour terms is possible 
when colours are given in a context in a systematic way, in direct relation to other colours
or in comparison to other stars or planets.

Here we will report mainly on Betelgeuse (Sect. 3.2), for which 
various credible pre-telescopic transmissions clearly
indicate a colour different from today.
In addition, for comparison, we consider Antares (Sect. 3.3), which is at a very similar location in the CMD 
as today's Betelgeuse.
In Sect. 3.4, we also discuss briefly Wezen ($\delta$ CMa), which now has B$-$V=0.7 mag and
which was reported to be `white' by the Bedouines in the 9th century. Since all
stars with B$-$V=0.0 to 0.6 mag are seen as white by the human eye, and since we 
have only one historical record, the case of Wezen is less compelling than Betelgeuse.
For other stars in the Hertzsprung gap as listed in Sect. 3.5 (see also Fig. 3),
we found no pre-telescopic transmissions indicating a colour change.
All other results on pre-telescopic reports on colours of stars will be published elsewhere.

\subsection{Betelgeuse}

We will first discuss reports by authors from the Mediterranean antiquity (Hyginus, Ptolemy, Germanicus,
Manilius, and Cleomedes), then from Classical China (Sima Qian), and finally other reports from the medieval period. 

\subsubsection{Hyginus}

Hyginus is the name of the author to which the Latin work 
{\it De Astronomia} is attributed, and which deals with constellations, stars,
and planets; see Vire (1992) for the most recent edition, Le Boeuffle (1983) for a Latin edition with
French translation of all four books, or Grant (1980) for an English translation of the first two books. 
Gaius Julius Hyginus (ca.\,BC 64 to AD 17) was originally from Iberia or Alexandria
and worked as superintendent of the Palatine library in Rome.\footnote{There 
is a scholarly discussion as to whether some of the works attributed to G.J. Hyginus were possibly
written in the 2nd century AD by some pseudo-Hyginus including {\it De Astronomia},
partly because the constellations are listed in the same order as in Ptolemy's Almagest from the mid 2nd cent.\,AD
(Huys 1996), but later authors date all work to around the BC/AD turn 
(Boriaud 1997, pp. VII--XIII; Schmidt \& Schneider 1998, column 778). 
In any case, even a dating to the 2nd century AD would be consistent with our results.}

In book IV, chapters 17--19 (lines 618--634),
Hyginus wrote in the context of the five naked-eye planets (Vire 1992, pp. 156--7):
\begin{quotation}
17. Iouis autem stella ... corpore est magnus, figura autem similis Lyrae ... \\
18. Solis stella ... corpore est magno, coloure autem igneo, similis eius stellae
quae est in humero dextro Orionis; ... Hanc stellam nonnulli Saturni esse dixerunt ... \\
19. Reliquum est nobis de Martis stella dicere, quae nomine Pyrois appellatur; hic autem non magno est corpore,
sed figura similis est flammae ...
\end{quotation}

Our literal translation to English is (our additions in brackets):  
\begin{quotation}
17. The star of Jupiter ... body is large (i.e. bright), and appearance (colour/colouration) similar to Lyra (i.e. Vega) ... \\
18. The Sun's star ... body is large (i.e. bright), and colour/colouration fiery/burning; 
similar to that star which is
in the right shoulder of Orion (i.e. Betelgeuse) ... Many have said that this star is (the star) of Saturn ... \\
19. It remains to speak about the star of Mars, which is also called by its name Pyrois (i.e. `the fiery');
its body of course is not large, yet its appearance (colour/colouration) is similar to a fire/flame ...
\end{quotation}

Hyginus gives first size (brightness) and then appearance/colour(ation), 
the latter called `figura' or `coloure'. 
He compares planets with stars in colour, which is quite rare,
namely Jupiter with Vega and Saturn with Betelgeuse. 
Indeed, Jupiter (B$-$V=$0.87 \pm 0.01$ mag; Tholen, Tejfel, \& Cox 2000; Mallama, Krobusek, \& Pavlov 2017) 
and Lyra = Vega (B$-$V=0.0 mag) are similar: 
while Vega's colour index is defined as white (Table 1), 
Jupiter was given by Plato empirically as `the whitest' ({\it Republic} X, 616E--617A), 
similarly (`white') by Ptolemy in {\it Tetrabiblos} II, 9.
Even though its B$-$V=0.87 mag pertains to orange (Table 1), 
Jupiter is seen as white by the unaided eye due to too much admixture of white (saturation).
(Note that the planet colours were constant over much longer than the historical time.)

Both Jupiter and Saturn were given as `large', i.e. bright, but they differ in colour.
Regarding Saturn (in antiquity called also the `star of the Sun', B$-$V$=1.09 \pm 0.16$ mag, orange according to Table 1), 
Hyginus compared its `fiery/burning' tint to the star on the right shoulder of Orion, 
i.e., Betelgeuse (right shoulder either in sky-view when facing us, as in the Hipparchus convention, 
or in back view on a globe). 
While the term `fiery/burning' used for Saturn is relative,
the comparison to Betelgeuse in colour is objective and quantifiable.
It seems striking that Hyginus compared Saturn (B$-$V=$1.09 \pm 0.16$ mag; Tholen, Tejfel, \& Cox 2000; Mallama, Krobusek, \& Pavlov 2017) 
with Betelgeuse and not with, 
e.g., Pollux (B$-$V=0.97 mag) or Arcturus (1.14 mag), 
because Betelgeuse today has B$-$V=1.73 to 1.83 mag (Table 2), which is very red and
much more comparable to Mars (B$-$V=$1.43 \pm 0.13$ mag; Tholen, Tejfel, \& Cox 2000; Mallama, Krobusek, \& Pavlov 2017) than Saturn.
Since the planets are constant in colour, Betelgeuse must have changed.
Hyginus is a credible scholar: the colour of Saturn was like that of Betelgeuse,
but the colour of Mars is reflected in its older name `Pyrois', i.e. `The Fiery' --
and also given as `similar to a fire/flame' (`similis est flammae').
Mars is uniformly reported as red in antiquity (Le Boeuffle 1981, p. 218, note 3 to chapt. 19);
no other planet was described that way.
We can conclude that Betelgeuse had a colour index like Saturn, i.e. B$-$V=$1.09 \pm 0.16$ mag, 
at the time of Hyginus. This is $4.1\sigma$ deviant from today's value for Betelgeuse ($1.78 \pm 0.05$ mag).

Hyginus relied on the colour theory of Plato in the chapters before the planets (Le Boeuffle 1981, pp. 215--216, note 1).
Plato (BC 428/427--348/347) described
Jupiter (B$-$V=$0.87 \pm 0.01$ mag) and Venus ($0.81 \pm 0.11$ mag) as white,
Saturn ($1.09 \pm 0.16$ mag) and Mercury ($0.97 \pm 0.03$ mag) as yellow,
and Mars ($1.43 \pm 0.13$ mag) as red, see {\it Republic} X, 616E--617A (Shorey 1969),
all empirically correct -- 
the bright Jupiter and Venus appear whitish due to too much admixture of white.
The unaided eye can clearly distinguish Saturn and Mars by colour;
their colour index ranges do not overlap. 
Plato has `red' for Mars and `yellow' for Saturn -- 
given the colour index ranges in Table 1, Mars is mostly in the red,
while Saturn is fully in the orange range.

\subsubsection{Ptolemy: Almagest and Tetrabiblos} 

Ptolemy worked in Alexandria (now Egypt) in the 2nd cent.\,AD.
His {\bf Almagest} is considered the most important work on astronomy of antiquity:
it was finished at the `beginning of the reign of Antonius', i.e. AD 138.
Ptolemy gave explicit colours for six stars as follows (translation in Toomer 1984): 
\begin{quotation}
Arcturus ($\alpha$ Boo): `The star between the thighs, called Arcturus, reddish' (V=0.16 mag, B$-$V=1.14 mag, Ptolemy: 1st mag), \\
Aldebaran ($\alpha$ Tau): `The bright star of the Hyades, the reddish one on the southern eye' (V=0.99 mag, B$-$V=1.48 mag, Ptolemy: 1st mag), \\
Pollux ($\beta$ Gem): `The reddish star on the head of the rear twin' (V=1.22 mag, B$-$V=0.97 mag, Ptolemy: 2nd mag), \\
Antares ($\alpha$ Sco): `The middle one of these, which is reddish and called Antares',
in Greek: `ho mesos aut\=on kai hypokirros kaloumenos Antar\=es' (V=1.07 mag, B$-$V=1.84 mag, Ptolemy: 2nd mag), \\
Betelgeuse ($\alpha$ Ori): `The bright, reddish star on the right shoulder',
in Greek: `ho epi tou dexiou \=omou lampros hypokirros' (V=0.57 mag, B$-$V=1.73 to 1.83 mag, Ptolemy: faint 1st mag), \\
Sirius ($\alpha$ CMa): `The star in the mouth, the brightest, which is called `the Dog' [Cyon] and is reddish'
(V=$-1.44$ mag, B$-$V=0.01 mag, Ptolemy: 1st mag). 
\end{quotation}

In the original Greek, the term used in the Almagest for the colours of these six stars 
is the compound adjective `hypokirros' (Heiberg 1903). 
Let us consider the meanings of the two components:
`hypo' means `below, pale, weak, somewhat';
`kirros' means something like `pale yellow' (German: `bla\ss gelb', Gemoll, Vretska, \& Kronasser 1954), 
`orange-tawny' (Liddell \& Scott 1940),
or `rose' (by comparison with wine colours give by Galen from the 2nd cent.\,AD, like Ptolemy).
For `hypokirros', we found `somewhat yellow' (Liddell \& Scott 1940);
in the Almagest, this word was translated to German as 
`rot' (red) or `r\"otlich' for reddish (Manitius 1898) and to 
`(orange-)rot' for orange-red (Kunitzsch 1974, pp.\,230 \& 267) --
as well as to `reddish' (Toomer 1984).
These translations have a purely philological quality.

We can now try to quantify the colour index and brightness ranges or limits
for the use of `hypokirros' in the Almagest
(and {\it Tetrabiblos}, see below).
That Sirius was included here could be due to the fact that it sometimes briefly does appear
with red rays due to scintillation at large brightness, 
or possibly as a late addition by a copying scribe (Ceragioli 1995).
The five other stars listed as `hypokirros' in the Almagest
are all indeed K- or M-type giants or supergiants, and their
B$-$V colour indices range today from 0.97 (Pollux) to 1.84 mag (Antares).
They are among the brightest stars in the sky.

The brightest star with a slightly smaller B$-$V colour index, but not 
qualified as `hypokirros', is Capella 
(V=0.08 mag, B$-$V=0.80 mag at the border between yellow and orange in Table 1),
but the five `hypokirros' stars in the Almagest are fainter, so that Capella's magnitude 
should not have been the limiting factor. Instead, practically, the combination of 
brightness and colour index (hue) defines the limit for colour sensation. 
The Almagest B$-$V colour index lower limit for the use
of `hypokirros' therefore lies between 0.80 and 0.97 mag.

Pollux with V=1.22 mag is the faintest star given as `hypokirros' by Ptolemy.
The faint magnitude limit for the use of `hypokirros' should then lie somewhere
between Pollux and the next faintest star that has at least B$-$V=0.97 mag -- that is
$\gamma$ Cru (Gacrux, M3.5 III, V=1.65 mag, B$-$V=1.52 mag).
However, it has a low culmination altitude at Ptolemy's epoch and place --
even if Ptolemy saw its colour,
he may have thought that it appears reddish merely because of strong atmospheric extinction.
Therefore, we should not choose $\gamma$ Cru for this limit.\footnote{Schlosser \& Bergmann (1985)
concluded from the fact that $\alpha$ Cen (V=$-0.01$ mag, B$-$V=0.71 mag) was not listed as
`hypokirros' in the Almagest, that the lower limit is B$-$V=0.97 mag (Pollux) --
but $\alpha$ Cen was far south for Ptolemy (culmination at $13^{\circ}$),
so that we exclude this and other stars too far south for the limit.
Capella is better suited.}
The next red-orange star slightly fainter than $\gamma$ Cru with B$-$V$\ge 0.97$ mag
is Dubhe ($\alpha$ UMa, V=1.82 mag, B$-$V=1.06 mag),
so that the fainter limit for using `hypokirros' in the Almagest
lies between Pollux (V=1.22 mag) and Dubhe (1.82 mag).
Given these limits (V$\le 1.22$ to 1.82 mag and B$-$V$> 0.80$ to 0.97 mag),
the Almagest list of those five stars with `hypokirros' is complete,
and no star that is both brighter and redder is missing.

Hence, we can derive a lower limit for the colour index of Betelgeuse to be B$-$V$> 0.80$ to 0.97 mag
at the epoch of the Almagest -- the other five stars listed as hypokirros were (at the very least almost) constant in
colour (and brightness) according to both historical reports and their location in the CMD (Figs. 1 \& 3).
Fainter red stars could have been omitted in the Almagest, either because they were too faint for colour detection by Ptolemy,
or because there was no general consensus about their colour -- it seems that Ptolemy wanted to list
the colour only for the brightest stars being `hypokirros'.

{\bf Tetrabiblos:} In Ptolemy's work on astrology, which includes some astronomical facts and
critique on astrology, explicit colours are given only for Antares, Arcturus, and Aldebaran 
(all `hypokirros'), as well as for Mars. We cite here {\it Tetrabiblos} I, 9
(here the translation from Robbins (1940), who translated `hypokirros' twice as `tawny' and once as `reddish'): 
\begin{quotation}
`... Taurus ... of the stars in the head, the one of the Hyades that is bright 
and somewhat reddish [hypokirros], called the Torch' (Aldebaran); \\
`Scorpius ... the three in the body, the middle one of which is tawny [hypokirros] and rather bright 
and is called Antares'
(Greek: `T\=on de en t\=o s\=omati tou Skorpiou ...  hoi de en t\=o s\=omati treis, h\=on ho mesos hypokirros kai lamproteros, 
kaleitai de Antar\=es'); and \\
`Bootes ... the bright, tawny [hypokirros] star, ... the star called Arcturus ...'.
\end{quotation}

In {\it Tetrabiblos}, 
neither Betelgeuse nor Pollux (or even Sirius) are given with colour; 
Pollux (V=1.22 mag, B$-$V=0.97 mag) is the faintest and (apart from Sirius) the 
star with the lowest B$-$V colour index among the `hypokirros' stars in the Almagest.
The very fact that Betelgeuse (now the 2nd reddest bright star, B$-$V=1.73--1.83 mag) 
was not listed as reddish (`hypokirros') in {\it Tetrabiblos} may well mean that it was not that red at that time.

Except the evolving Betelgeuse, the stars Antares, Arcturus, and Aldebaran (listed as hypokirros in {\it Tetrabiblos})
are the reddest stars among the Almagest `hypokirros' stars 
with a B$-$V colour index range of 1.14 to 1.84 mag (B$-$V=$1.49 \pm 0.35$ mag).
From the least red star listed as such in {\it Tetrabiblos} (Arcturus), 
we may derive B$-$V=1.14 mag as an upper limit for Betelgeuse (Table 3).

In sum, `hypokirros' as used by Ptolemy in the Almagest therefore means a colour index of B$-$V$> 0.80$ to 0.97 mag,
i.e. what we today consider as orange and red (Table 1).
Capella is bright enough (V=0.08 mag) for colour detection,
but with B$-$V=0.80 mag (yellow-orange border in Table 1), it did not satisfy Ptolemy's use of `hypokirros'
(Capella is often considered as yellow). 
In other words, in the Almagest, `hypokirros' would mean a colour range beyond yellow, towards and including red,
whilst in {\it Tetrabiblos}, only the three then reddest stars (B$-$V$\ge 1.14$ mag) plus Mars are given as `hypokirros'.
We will render `hypokirros' as `somewhat reddish'.

N.B.: The notion of what we today call `colour/tint' might have been different in antiquity:
``the attention of the (ancient) Greeks is not so much the qualitative difference
between colours as a quantitative difference between colours. Black and white are `colours',
and other colours are accounted for as shades between these extremes'' (Platnauer 1921, p.\,162).
And it was ``natural for the Greeks ... to think of the basic colour terms as being
arrangeable on a scale between melas [dark/black] to leukos [bright/white] at the end points''
(Lyons 1999, p.\,58), i.e., not only as sequence in wavelength, but also of brightness or shades of gray. 

Next, we will consider the other Mediterranean authors of antiquity who wrote about star colours,
first another Greek, then the Latin scholars. Like Ptolemy in {\it Tetrabiblos}, they do not include
Betelgeuse among the red stars, but all include Antares,
so that we can obtain the upper limits for Betelgeuse as listed in Table 3.

\subsubsection{Cleomedes} 

Cleomedes (ca. 1st cent. AD) from Greece wrote the following 
in {\it On the Circular Motions of the Celestial Bodies}  
about the colours of stars and planets (Cycl.\,theor.\,I, 8, line 46--49): 
\begin{quotation}
Dyo eisin asteres, kai t\=en chroan kai ta megeth\=e parapl\=esioi, diametrountes all\=elois:
ho men gar tou skorpiou, ho de tou taurou t\=en pentekaidekat\=en epechei moiran, meros \=on t\=on
Hyad\=on. Houtoi t\=o Arei t\=en chroan homoioi eisin hoi asteres ...
\end{quotation}
(Boll \& Bezold 1918, p.\,15 on this statement; Greek in Todd 1990; English in Bowen \& Todd 2004). \\
Our English translation: 
\begin{quotation}
There are two stars, almost identical in both colour and size [brightness], standing diametrically (opposite) to each other:
one is in Scorpion [Antares], the other in Taurus is at the 15th degree and part of the Hyades [Aldebaran].
These stars are similar in colour to Ares [Mars] ...
\end{quotation}

Cleomedes reported `almost identical' brightness and colour for Antares (V=1.09 mag, B$-$V=1.84 mag)
and Aldebaran (V=0.99 mag, B$-$V=1.48 mag), i.e. quite correct. And like Hyginus, he compared
stars with a planet in colour, here both Antares and Aldebaran with the red Mars (B$-$V=1.43 mag).
Since Cleomedes did not list Betelgeuse as red, the upper limit for the colour index of
Betelgeuse may be B$-$V$<1.48$ mag (Aldebaran).
(Aldebaran is indeed more exactly diametrically opposite to Antares than Betelgeuse.)
Cleomedes brings the ecliptic longitudes and colours of Antares and Aldebaran into the context of a 
discussion of the geometry of the universe and the size of the Earth (that we see always $180^{\circ}$ of the sky),
so that he may not have had the intention to list the reddest stars on sky; therefore, the limit we obtained
may be of lower significance.

Dating: Since Cleomedes cited the
astronomer Posidonius of Rhodes (ca.\,BC 133--51), he wrote not earlier than in the 1st century BC.
The latest suggested date is AD $371 \pm 50$ 
based on the assumption that the ecliptic longitudes $\lambda$ given by Cleomedes
(Tau $15^{\circ}$ for Aldebaran, i.e. $\lambda=45^{\circ}$, and -- opposite -- Sco $15^{\circ}$ for Antares, i.e. $\lambda=225^{\circ}$)
were obtained by a precession correction of the Almagest values (Neugebauer 1974, p.\,960).
However, Tau $15^{\circ}$ and Sco $15^{\circ}$ can also point to an epoch of ca.\,AD 220, if newly
and correctly observed. Recent work dates Cleomedes consistently to the 1st century AD (Ross 2000
and references therein), e.g. his understanding of
refraction is less elaborate compared to Ptolemy; 
he supported older Stoic ideas, and linguistic arguments also support this dating.
If Cleomedes wrote in the 1st century, his `15th degree' would either be $\sim 2^{\circ}$ off,
or it was just meant as a rough indication (Bowen \& Todd 2004, p.\,89).
In any case, a dating even to ca.\,AD 220 would not change our conclusions.

\subsubsection{Germanicus} 

Germanicus (BC 15 to AD 19), a Roman general, freely translated Aratos' {\it Phaenomena},
a poem on stars and constellations, in the 1st or 2nd decade AD (Possanza 2004); 
we use the edition with English translation by Gain (1976), which used 28 manuscripts,
and we consulted also the Latin-German version in Bischoff \& Klein (1989) based on the 
old 9th century MS Leiden Voss Q79 (which had no relevant variants).
Germanicus may have alluded to two star colours
($\beta$ And and $\alpha$ Sco).
We first discuss lines 201-204 with Mirach in Andromeda:
\begin{quotation} 
Nec procul Andromeda, totam quam cernere nondum \\
obscura sub nocte licet; sic emicat ore, \\
sic magnis umeris candet nitor ac mediam ambit \\
ignea substricta lucet qua zonula palla.
\end{quotation}
This was translated to English by Gain (1976), relatively close to the text:
\begin{quotation} 
Andromeda lies not far away; you can see all of her \\
when the night is not yet (allowed) dark; so great is the brightness \\ 
which shines in her face and large shoulders and surrounds her middle, \\
where her fiery belt gleams and her dress is tied.
\end{quotation}

Andromeda is a large, but clearly structured constellation with head, shoulder, arms, hands,
girdle/belt, knees, and feet (see Almagest, Toomer 1984). While $\alpha$ And is the head,
one of the three stars of the girdle is the red Mirach ($\beta$ And, V=2.08 mag, B$-$V=1.59 mag, M0\,III);
$\gamma$ and $\delta$ And are also reddish, but slightly fainter and clearly not in the belt.
Hence, the text on a `fiery' (`ignea') belt star can only mean Mirach (`where her dress is tied'
or `where her coat is girded up').

Also Antares is mentioned (line 660), again from the editon by Gain (1976):
\begin{quotation}
Scorpios ardenti cum pectore contigit ortus. 
\end{quotation}
which was translated to English by Gain (1976):
\begin{quotation}
when the Scorpion is touching the eastern horizon with its burning [lat. ardenti] breast.
\end{quotation}
The term `ardens/ardenti' obviously points 
to the {\it red} Antares, the middle of three stars in the breast (heart) and the reddest (B$-$V=1.84 mag).
We found no further colour information in the work of Germanicus;
there are no hints in the original work by Aratos on the colours of Mirach or Antares.

\subsubsection{Manilius} 

Manilius from Rome wrote the following in his work {\it Astronomica} 
(2nd/3rd decade AD), partly on stars (Latin in Fels 1990), first for Antares (I 268):
\begin{quotation}
ardenti fulgentem Scorpion astro
\end{quotation}
which we translate to English as follows: 
\begin{quotation}
Scorpius with its fiery burning star. 
\end{quotation}

Then, he also has Canopus (I 216--218): 
\begin{quotation}
nusquam invenies fulgere Canopon ... ille ... ignis
\end{quotation}
which we translate as: 
\begin{quotation}
You will never find the burning Canopus ... that [Canopus'] fire.
\end{quotation}
Both stars are specified as red, Antares with `ardenti' (`fiery') and Canopus with `ignis' (`fire').
Canopus appears red only due to its low altitude (few degrees above horizon) 
in the Mediterranean region (B$-$V=0.17 mag, 
i.e. white, but with a declination $-52.5^{\circ}$ at the epoch of Manilius). 
It seems that Manilius mentioned the reddest star (Antares) and the brightest (presumably) red star (Canopus, V=$-0.63$ mag).

\subsubsection{Summary of Greek and Latin Mediterranean authors}

In sum, Ptolemy's {\it Tetrabiblos}, Cleomedes, Germanicus, and Manilius all appear to
list the brightest and reddest stars down to their respective magnitude limit,
but they do not mention Betelgeuse, even though it is now one of the brightest and reddest stars.
The least red star listed by the above is Arcturus (B$-$V=1.14 mag), which sets the upper limit 
in colour index for Betelgeuse for this time range (1st/2nd century AD). 
We list the limits on Betelgeuse in Table 3.
Except Betelgeuse, all stars listed by these authors as red/reddish were indeed constant in colour since 
the last few millennia according to their location in the CMD and their MIST tracks (Figs. 1 \& 3).

The largest B$-$V colour index range for yellow-to-red stars is found in the Almagest
(namely B$-$V=0.80 to 1.84 mag), so that both Pollux and Betelgeuse were included
(plus Sirius, which is in a different category). 
Thus, we obtained B$-$V$>0.80$ to 0.97 mag as a lower limit for Betelgeuse from Ptolemy's Almagest,
from Capella and Pollux.
From the additional upper limit obtained for Betelgeuse from {\it Tetrabiblos}, we may constrain
the colour index of Betelgeuse to be B$-$V=0.80 to 1.14 for Ptolemy's time (mid 2nd century AD);
this value (B$-$V=$0.97 \pm 0.17$ mag) is 4.6$\sigma$ deviant from Betelgeuse's current value of $1.78 \pm 0.05$ mag --
but since it is derived indirectly, we consider it to be less significant than the two constraints from the
direct statements from Hyginus (Sect. 3.2.1) and the Simas (Sect. 3.2.7).
This is consistent with the colour index obtained from Hyginus (like Saturn),
namely B$-$V$=1.09 \pm 0.16$ mag (1st century AD).

We conclude that naked-eye observers could distinguish colours equivalent to B$-$V colour index variations down 
to $\pm 0.1-0.2$ mag;\footnote{E.g., Tycho Brahe mentioned the colour difference between Aldebaran (now B$-$V=1.48 mag) and Betelgeuse
(although B$-$V=1.73 to 1.83 mag, but it would have had
B$-$V$\simeq 1.7$ mag from the lower, best-fit curve in Fig. 5 at Brahe's time):
`it [SN 1572] was like Aldebaran, or the one, which is red in the right shoulder of Orion [Betelgeuse].
But it was not as red like the one in the shoulder, but more like the color of Aldebaran'
(Latin in Dreyer 1913, pp. 28-29).
Similarly, seven observations of the colour (index) of SN 1572 by Tycho Brahe and others are all consistent
with the colour indices expected at those phases for a normal supernova type Ia (as SN 1572), within the
$1\sigma$ error bars and within $\pm 0.1$--0.2 mag with respect to the expected values (Ruiz-Lapuente 2017).} 
with this, the B$-$V limits we have established, 
can constrain or revise the tracks and/or the exact mass of Betelgeuse.
The only slight discrepancy between a B$-$V colour index derived here from historical observations (Table 3) 
and the lower best fit curve for Betelgeuse (Fig. 5) is found in {\it Tetrabiblos} for Arcturus (Sect. 3.2.2).
However, Arcturus itself is slightly variable, V=$-0.03$ to 0.16 mag (HIP, Kukarkin et al. 1981).
In SIMBAD, we find B$-$V=1.23 mag (Ducati 2002), although B$-$V=1.14 mag (appendix table) is more
consistent with its spectral type of K1.5\,III.
Since Antares (B$-$V=1.84 mag) was named after Ares=Mars (B$-$V=$1.43 \pm 0.13$ mag) because of similar colour, 
they should have (had) a similar colour index, but the indices now differ by slightly more than the 
above $\pm 0.1$ to 0.2 mag; however, their colour indices may have been within $\pm 0.2$ mag 
up until the mid 2nd millennium BC (Fig. 5 lower pink line for Antares).

\subsubsection{Sima Qian and Sima Tan from China}

Sima Qian (BC 145--87), `Senior Archivist' (or `Prefect Grand Scribe Astrologer') 
during the Western Chinese Han dynasty (BC 206--AD 9), compiled his
work `Tianguan shu' (`Treatise on the celestial offices', i.e., asterisms and stars),
a treatise in the `Shiji' (`Historical Records' or `Records of the Senior Archivist'), around BC 100. 
The `Tianguan shu' is perhaps partly based on work started by his father, Sima Tan (ca.\,BC 165--110); 
see Pankenier (2013, pp.\,444--511) for introduction and translation.
The work summarizes astronomical knowledge, in particular lists and explanations of asterisms.
It also includes numerous astrological considerations.
For the five naked-eye planets, omina are formulated in combination with one of the five {\it wuxing} colours 
(white, yellow, red, blue, black/dark), e.g. `if Venus' rays are red, there will be war; if white there will be
obsequies', etc (Pankenier 2013, p. 484). Yet, the colours themselves are not defined by those omina.
Outside of the main omina parts, each of the five planets is given with one such colour,
e.g. Saturn as `yellow' (Pankenier 2013, p.\,479), Mars as `fire' (p.\,477), and Venus as `Surpreme White' (p.\,481).
In other Classical Chinese texts, we can also find Mercury as dark and Jupiter as blue (e.g. Jin shu);
hence, some planet colours are empirically correct, others are chosen by the construction (the need to include 
five distinct colours).
In the context of Venus, which is discussed most extensively, there is a unique sentence
about star colours and colour definition, namely `Shiji' 26.1325 in Chinese (edition Sima Qian 1959):
\begin{quotation}
Taibo bai, bi Lang; chi, bi Xin; huang, bi Shen zuo jian; 
cang, bi Shen you jian; hei, bi Kua da xing
\end{quotation}
This is translated to English as
\begin{quotation}
For [Venus] white, compare Lang (Sirius); \\
for red, compare Xin ($\alpha$ Sco); \\
for yellow, compare the Left Shoulder of Shen ($\alpha$ Ori); \\
for blue, compare the Right Shoulder of Shen ($\gamma$ Ori); \\
and black[/dark], compare the large star of Kui ($\beta$ And).
\end{quotation}
This translation in Jiang (1993) underlines the idea of standardization;
we added `Venus', which is clearly given in the Chinese text 
(`Taibo', lit. `Supreme White', the name for Venus);
a similar translation is found in Pankenier (2013, p.\,484).

This text can be understood as a definition of colours for stars as seen on sky.
Bo Shuren, Jianmin, \& Jinyi (1978), commenting on this text, wrote: The Chinese `generally adopted
the five stars as standards for comparison with the colours' (our translation
to English from their Chinese).

The Simas specified colour terms with example stars, e.g. Antares as `red' and Betelgeuse as `yellow', 
i.e. definitely different. For those five stars, the Simas purposely referenced each of the five colours 
(for dark/black, they picked the red, but fainter Mirach). Except for Betelgeuse, 
the colour indices of the other stars have not changed for millennia (Fig. 5). 
Indeed, Sirius is mainly seen as white (V=$-1.44$ mag, B$-$V=0.01 mag), 
Antares as red (V=1.07 mag, B$-$V=1.84 mag), and 
Bellatrix as blue (V=1.66 mag, B$-$V=$-0.14$ mag) --
see Table 1 for the colour definition.
The Chinese word `cang' (conflation of blue and green) is usually translated as blue or azure,
because it describes the colour of the day-time sky or sea.
While Mirach of course cannot be seen as `black',
yet as a relatively faint and red star (V=2.08 mag, B$-$V=1.59 mag), 
it takes the role of dark (`hei') as a prototype of such stars.
The Chinese term `hei' is usually translated as `black' (e.g., `hei qi' for `black spot/s' on the Sun), 
but for celestial objects on the night sky, a rendering as `dark' is better (Bo Shuren 1981).
Bo Shuren, Jianmin, \& Jinyi (1978) point out: `Even today in our country (China), there are some places where dark red candy is called black candy'.

In reporting Betelgeuse as `yellow', the Simas actually constrained its colour index to be larger than the 
blue-white Bellatrix and Sirius (B$-$V=0.01 mag) and lower than the red Antares and Mirach (1.59 mag). 
Since pre-telescopic naked eye observers could discriminate colour indices to $\pm 0.1$ to 0.2 mag (Sect. 3.2.6), 
Betelgeuse's colour was even $\le 1.5$ mag. Because human eyes see stars with B$-$V=0.3 to 0.6 mag also as white (Steffey 1992), 
see Table 1, this would imply that Betelgeuse had B$-$V$\ge 0.60$ mag. 
Putting this all together, we can reasonably argue that Betelgeuse had 
B$-$V=0.60 to 1.5 mag ($1.05 \pm 0.45$ mag), 
i.e. 1.6$\sigma$ different from today (1.73 to 1.83 mag or $1.78 \pm 0.05$ mag). In the same `Tianguan shu', 
the Simas also wrote that `Saturn is yellow' and Mars `fiery' red (Pankenier 2013), so that their `yellow' is 
limited to B$-$V$\le 1.3$ mag (Mars has B$-$V=$1.43 \pm 0.13$ mag). Then, the `yellow' Betelgeuse had B$-$V=0.6 to 
1.3 mag ($0.95 \pm 0.35$ mag), 
also considering the precision of naked-eye colours of $\pm 0.1$ to 0.2 mag
(and that Mars and Saturn were clearly distinguishable).
This is 2.3$\sigma$ different from today. 

For the `yellow' colour, the Simas gave Betelgeuse as representative, which is now red (B$-$V=1.73 to 1.83 mag) like Antares.
Is the identification certain?
According to Western (Babylonian-Greek) tradition, Orion is facing us in sky-view;
from Hipparchus: `Orion ... the end of its setting with the right shoulder' (Manitius 1894, p.\,71).
Thus, the star on Orion's `right shoulder' is clearly Betelgeuse -- and the star on his left shoulder is Bellatrix.
In Chinese star mapping it was seen differently; for example, in the astronomical chapters of the {\it Jin shu}, 
we can find: `Shen ... The NE is the left shoulder ..., while the NW forms the right shoulder' 
(Ho 1966, p.\,102)\footnote{The {\it Jin shu} is the history of the Jin dynasty (AD 266--420) and includes an astronomical treatise compiled 
under emperor Taizong (AD 627--649) written by Li Chunfeng (AD 602--670),
director of the astrological/astronomical service of the emperors ca.\,AD 648--664; see Ho (1966, p.\,13).}.
The asterism `Shen' is quite similar to our Orion and also considered a hunter/warrior (Pankenier 2013, p.\,468). 
Except for Betelgeuse, the stars mentioned by the Simas were constant in colour index 
given the location in the CMD and their MIST tracks (Figs.\,1 \& 5).

Sima Qian and Sima Tan were not working ex nihilo, but rather, compiled their treatise from materials 
in the imperial archives. As early as BC 168, the Mawangdui `Prognostics of the Five Planets' shows correspondences 
between the five visible planets and a set of five materials (wood, fire, soil, metal, water) 
and five directions: east, south, center, west, north (Cullen 2011).  
The Simas further developed these correspondences by introducing 
a set of five colours for stars (white, yellow, red, blue/green, dark/black).

In sum, the Simas give Betelgeuse as `yellow', as well as Saturn 
(B$-$V=$1.09 \pm 0.16$ mag, i.e. orange in Table 1).

The above text from the Simas was reproduced in a few other Classical Chinese texts
(Bo Shuren, Jianmin, \& Jinyi 1978; Jiang 2013), but it was always a more or less verbatim quotation
from the Shiji treatise and not based on actual new observations,
and they suffer from corruptions due to mistakes by copying scribes.
For example, in the {\it Han shu} 26.1284 
(History of the Western/Former Han, BC 206 to AD 9, compiled ca. AD 100, edition Ban Gu et al. 1962),
the characters `zuo' and `you' for `left' and `right' (shoulders of Shen/Orion)
are exchanged -- they are quite similar and were exchanged by mistake more often in historical texts;
furthermore, instead of `cang', it used `qing', also meaning blue-green. 

In the {\it Jingzhou zhan} as cited in the `Kaiyuan zhanjing' (45.9a) compiled by Gautama 
Siddh\=artha (aka Qutan Xida, originally from India, flourished around AD 729, text edition in Siddh\=artha Gautama 2007) 
and included in the `Siku quanshu', 
black/dark with Mirach and blue/green with Bellatrix are missing, and Shen for Orion is not
explicitly mentioned; also, Vega is added to Sirius for white;
it has foot (jiao) instead of shoulder (jian) as a scribal error; the two characters do share some graphic similarities
that might contribute to transcription errors if copied manuscripts are in poor condition
(the right foot of Shen is Rigel, also blue). 

In {\it Jin shu} 12.320 ({\it 109}) and {\it Sui shu} 20.558 (editions Fang Xuanling 1974; Wei Zheng 1973), 
which are the histories of the Jin (AD 265--420) and the Sui (AD 581--618) dynasties 
(astronomical chapters of both were compiled by Li Chunfeng in the 7th century, see footnote 6), 
the characters `zuo' and `you' for `left' and `right' are exchanged (shoulders of Shen/Orion)
and have the same text (just with slightly different punctuation in modern Zhonghua editions).
The translation of the Jin shu in Ho (1966) is missing the end of one line and the beginning of the next,
namely (missing portion in italics): 
`red (means the same as) {\it Xin da Xing (Antares); yellow (means the same as)}'.
We were not able to find any version of the Jin shu that omits this portion of the text. 
The later {\it Song shi} 52.1074 (edition Tuotuo 1977), the history of the Song dynasty (AD 960-1279), 
is somewhat modified -- not only Venus is given for the five different colours, but also other planets, 
e.g. `Mars colour red, compare Xin Great Star ($\alpha$ Sco)'.

To our knowledge, the publication by Bo Shuren, Jianmin, \& Jinyi (1978) 
in Chinese was the first to qualitatively notice the relevance of Sima's passage on
the colour of Betelgeuse in {\it Tianguan shu}: `The well-known red giant Cen (Shen) Lodge 4 (Betelgeuse) was 
yellow 2000 years ago ... or at least, that it was not so red as Heart Lodge 2 (Antares)';
and by consulting evolutionary models of their time (yielding 15 to 20~M$_{\odot}$, see their table), 
they noticed that, according to Chiosi \& Summa (1970),
`the time ... from yellow to red, requires only a few thousand years'
(our translation to English from their Chinese). 
Bo Shuren, Jianmin, \& Jinyi (1978) did not consider any of the other
independent records from the Mediterranean or the Near East, nor other quantitative details.

\subsubsection{Medieval and pre-modern reports on Betelgeuse (always red)}

The {\bf Bedouin} tradition is transmitted by Ibn Qutayba (AD 828--889), 45, 13 on Orion:
\begin{quotation}
Arabic: kawkab\=an azhar\=an f\textit{\={\i}} a\d{h}adihim\=a \d{h}umra
\end{quotation}
\begin{quotation}
two bright stars, one of them reddish
\end{quotation} 
(Kunitzsch 1961, p.\,116).
These two stars (Arabic dual `kawkab\=an') of the sixth lunar station (in Orion) 
were identified as Betelgeuse and Bellatrix by Al-\d{S}\=uf\textit{\={\i}} (AD 903--986).
We also find Canopus (only due to always appearing at very low altitude),
Capella (ca. AD 1000, Lis\={a}n, Jacob 1887), Arcturus, Aldebaran, Antares, Schedar, and Edasich as `red/reddish' 
(Ibn Qutayba 1956; Kunitzsch 1959, 1961; Adams 2018).
The star listed as red with the lowest B$-$V colour index is Capella, with B$-$V=0.80 mag.
We can therefore conclude that the limit for what they call `red/reddish', i.e. the B$-$V limit for Betelgeuse,
is B$-$V$\ge 0.80$ mag (Table 3).

{\bf Al-\d{S}\=uf\textit{\={\i}}} described Betelgeuse as follows (epoch AD 964): 
\begin{quotation}
in Orion ... the 2nd is the great bright red (Arabic: al-\d{h}umra) star located on the right shoulder
\end{quotation}
(Hafez 2010, pp.\,275--280), 
`right shoulder' again in sky-view looking at us, i.e. Betelgeuse.
Al-\d{S}\=uf\textit{\={\i}} revised the Almagest and listed as reddish stars those of the Almagest, but
without Sirius. In other words, he included Antares, Betelgeuse, Aldebaran, Arcturus, and Pollux, and he also  
correctly added Alphard ($\alpha$ Hya, V=K3\,III, V=1.99, B$-$V=1.45 mag), and somewhat surprisingly
Algol ($\beta$ Per, B8\,V, V=2.11--3.39 mag, B$-$V$=-0.01$ mag), which is now blue-white.
Algol is definitely not red -- it is an eclipsing binary, but the colour index changes by only 
$\sim 0.1$ mag during the primary eclipse, and even less during secondary eclipse (Kim 1989).
According to van Gent (1989), ``a similar and independent statement
[that Algol would be red] is found in the `Liber Hermetis de XV Stellis', a medieval Latin astrological treatise
[13th cent. AD] which contains Hellenistic astrological material probably dating from the first few centuries AD''.
Van Gent (1989) also pointed out that Angelo Secchi (1818--1878) had classified Algol first as red, 
but later as blue-white, and that Algol
``experienced a major period-jump around 1854'', possibly related to a mass-transfer event and, hence, a colour change.
Indeed, strong star-stream interactions in the Algol system were noticed (Richards 1992),
and tentative signals of at least five companion candidates have been detected with the light-time travel effect
in 236 years of Algol data, with periods from 1.863 to 219 yr (Jetsu 2021) -- possibilities for interaction.
Also Julius Schmidt at Athens, Greece, reported Algol to be `reddish-yellow' in 1841 (Ceragioli 1995).
Earlier, in Tycho Brahe's star catalog, Algol was described as 'Fulgens in dextro latere' (Verbunt \& van Gent 2010),
i.e. `burning in the right part' (of Perseus), which might also indicate a reddish colour.
For Al-\d{S}\=uf\textit{\={\i}}, we can derive the same lower limit for Betelgeuse as for the Almagest,
B$-$V$>0.80$ to 0.97 mag, because Pollux is listed and Capella is omitted.

{\bf Ulug Beg} (died AD 1449) 
lists the five reddish stars from the Almagest 
(Antares, Aldebaran, Arcturus, Betelgeuse, and Pollux, but not Sirius) with the
Persian `sur\d{h}\textit{\={\i}}' for `red/redness' (Kunitzsch 1974),
so that we again derive the same lower limit for B$-$V for Betelgeuse 
as for the Almagest and Al-\d{S}\=uf\textit{\={\i}}.

{\bf Tycho Brahe} mentioned implicitly that Betelgeuse was redder than Aldebaran:
\begin{quotation}
it [SN 1572] was like Aldebaran, or the one, which is red in the right shoulder of Orion [Betelgeuse].
But it was not as red like the one in the shoulder, but more like the color of Aldebaran
\end{quotation}
(Latin in Dreyer 1913, pp. 28-29).
From this, we obtain a lower limit of B$-$V$> 1.48$ mag (colour index of Aldebaran) for Betelgeuse.
This is the most stringent constraint since the data and limits from antiquity: 
The B$-$V colour index of Betelgeuse increased from $1.09 \pm 0.16$ mag (Saturn) 
to the above 1.48 mag (Aldebaran) between antiquity and Brahe.

All limits are given in Table 3.

\subsubsection{Final remarks on Betelgeuse}

In Figs.\,4 \& 5, we plot all these data and limits for the full possible (conservative) time ranges, 
usually the author's life time or dating range (see Table 3).
While some oral tradition from First Nations may well originate from several centuries ago, 
they were recorded recently by anthropologists; 
their datings, star identifications, and colour term meanings are more uncertain.
In Hawaii, Betelgeuse, Antares, Aldebaran, and Mars were all called `Hoku-ula' for `red star' 
(Makemson 1941; Noyes 2019; Brosch 2008),
so that they considered these stars to be roughly as red as Mars.
A somewhat reddish Betelgeuse may be included also in traditions from Tonga, the Central American Lacandon people,
as well as the Japanese Taira clan (Makemson 1941; Milbrath 1999; Renshaw \& Ihara 1999).

To add the significance from Hyginus (4.1$\sigma$) to the significance from the Simas (2.3$\sigma$),
being independent,
we first convert these significances to probabilities (values from 0 to 1),
then multiply the two probabilities, and then convert the result back to a significance (in $\sigma$).
Adding in this way 4.1$\sigma$ and 2.3$\sigma$ properly, yields a total of 5.1$\sigma$.
The lower and upper limits from Ptolemy's Almagest (B$-$V$> 0.80$ mag) and {\it Tetrabiblos} ($< 1.14$ mag), respectively,
see Sect. 3.2.2 and Table 3,
yield a B$-$V range of 0.80 to 1.14 mag, or $0.97 \pm 0.17$ mag, i.e. 4.6$\sigma$, different from today's Betelgeuse;
adding this additional significance properly, we obtain a total significance of 7.2$\sigma$.
That the scatter of the nominal values (Hyginus 1.09 mag, Simas 0.95 mag, Ptolemy 0.97 mag) 
around their mean (1.00 mag) is smaller than the individual error bars (in particular $\pm 0.35$ mag from the Simas),  
may indicate that we have overestimated (conservatively) one or more error bars.

It may be possible to find some additional colour observations of Betelgeuse for the time
after Ptolemy, which could point to a colour index of B$-$V$\simeq$1.4 to 1.6 mag
in that time, namely a somewhat quantitative and objective comparison of its colour
to a star or planet (e.g. in colour like Aldebaran).

\begin{table*}
\begin{tabular}{llllll|ll} \hline
\multicolumn{8}{l}{{\bf Table 3: Betelgeuse and Antarss colour indices from historical reports:}} \\ \hline
        &              &                   & \multicolumn{2}{c}{Betelgeuse}                              &  & Antares & Antares \\
Author  & dating       & range             & Sect. & B$-$V [mag]~~~~~~~~~sign.\,(1) & pred.\,(2) & B$-$V [mag] & evidence \\ \hline
Oracle bones & BC\,1300 & BC\,1400-1200    & 3.3   & -                                    & -    & $\ge 1.43$  & like fire/Mars \\
Book Odes    & BC\,1000 & BC\,1100-900     & 3.3   & -                                    & -    & $\ge 1.43$  & like fire/Mars \\
Zuozhuan     & BC\,600  & BC\,722-468      & 3.3   & -                                    & -    & $\ge 1.45$  & Alphard \\
Greek   & BC\,200      & BC\,1000-AD\,150  & 3.3   & -                                    & -    & $\ge 1.43$  & `like Mars' \\
Simas   & BC\,90       & BC\,150$-$87      & 3.2.7 & $0.95 \pm 0.35$~(3)~~~~~2.3$\sigma$  & 1.16 & $\ge 1.43$  & like Mars \\ 
Hyginus & AD\,15       & BC\,50$-$AD\,17   & 3.2.1 & $1.09 \pm 0.16$~(4)~~~~~4.1$\sigma$  & 1.23 & -           & -  \\ 
Germanicus & AD\,7     & BC\,5$-$AD\,19    & 3.2.4 & $< 1.59$ ($\beta$ And)      & 1.22   & $\ge 1.59$  & Mirach \\ 
Manilius   & AD\,20    & AD\,10$-$30       & 3.2.5 & $< 1.84$ ($\alpha$ Sco)     & 1.23   & -           & - \\
Cleomedes  & AD\,60    & BC\,100$-$AD\,220 & 3.2.3 & $< 1.48$ ($\alpha$ Tau)     & 1.25   & $\ge 1.48$  & Aldebaran \\
Almagest    & AD\,138  & AD\,100$-$170     & 3.2.2 & $> 0.80$ to 0.97 ($\alpha$ Aur - $\beta$ Gem) (5) & 1.30 & $\ge 0.89$ & as Betelgeuse \\
Tetrabiblos & AD\,150  & AD\,100$-$170     & 3.2.2 & $< 1.14$ ($\alpha$ Boo) (5)  & 1.31   & $\ge 1.14$  & Arcturus \\
Hephaistos & AD\,415   & AD\,415           & 3.3   & -                           & -      & $\ge 1.43$ & like Mars \\
Bedouins   & AD\,889        & AD\,700$-$1000      & 3.2.8   & $\ge 0.80$ ($\alpha$ Aur) & 1.60 & $\ge 0.80$ & Capella \\
Al-\d{S}\=uf\textit{\={\i}} & AD 964 & AD\,920$-$964 & 3.2.8 & $> 0.80$ to 0.97 ($\alpha$ Aur - $\beta$ Gem) & 1.61 & $\ge 0.89$ & as Betelgeuse \\ 
Al-Biruni  & AD\,1000       & AD\,1000               & 3.3   & -                           & -      & $\ge 1.43$ & like Mars \\
Al-Tusi    & AD\,1200       & AD\,1200               & 3.3   & -                           & -      & $\ge 1.43$ & like Mars \\
Ulug Beg   & AD\,1440       & AD\,1400$-$1449      & 3.2.8   & $> 0.80$ to 0.97 ($\alpha$ Aur - $\beta$ Gem) & 1.66 & $\ge 0.89$ & as Betelgeuse \\ 
T. Brahe   & AD\,1572/3     & AD\,1572/3            & 3.2.8   & $> 1.48$ ($\alpha$ Tau) & 1.67 & - & - \\  \hline
\end{tabular}

Notes: (1) Significance for being different (smaller) than the current value of B$-$V=$1.78 \pm 0.05$ mag. \\
(2) Predicted B$-$V (mag) colour index for Betelgeuse from the lower, best-fit line for Betelgeuse in Fig. 5. \\
(3) `yellow', (4) like Saturn. \\ 
(5) Upper and lower limit together yield a range of 0.80 to 1.14 mag, or $0.97 \pm 0.17$ mag (4.6$\sigma$ significance).
\end{table*}

\subsection{Antares}

Since Betelgeuse has changed its colour in historical times,
we should also check Antares in detail, as it is located at a simliar position in the CMD (Figs. 1 \& 2).

Above, we also quoted the colour specifications for Antares by Sima Qian (2nd cent.\,BC),
Germanicus (1st/2nd decade AD), Manilius (early 1st cent.\,AD), 
Cleomedes (ca.\,1st cent.\,AD), and Ptolemy (mid 2nd cent.\,AD).
Here, we list all additional pre-telescopic colour records for Antares we have found.

{\bf China:} In one of the oldest extant `guest star' records, found on oracle bones (around BC 1300), it reads 
\begin{quotation}
On the 7th day of the month, a chi-ssu day, a great new star
appeared in company with Antares (=Huo=Fire-star)
\end{quotation}
(Needham \& Wang 1959, p.\,424).
The `Fire-star' (Antares) is displayed with an ancient character resembling a fire --
indicating its red colour.
The Classical Chinese name for Antares is `Dahuo' and means `Great Fire'.
Antares is also the focus of ode 154 in the `Odes' classic (Shijing) from the late 2nd or early 1st millennium BC:
`In the 7th month the Fire [star] {\it declines}' (Pankenier 2013, p.\,461, note 61).

Then, in the work `Zuozhuan', a tradition associated with the ancient Chinese `Annals' (Chunqiu, BC 722--468), we can read
(round brackets from Pankenier 2013, square brackets from us):
\begin{quotation}
The ancient Regulator of Fire was renumerated either (at the season of) asterism {\it Heart} [Xin with $\alpha$ Sco]
or asterism {\it Beak} [Qixing with $\alpha$ Hya] in order (that he should) take out and bring in the fire.
For this reason {\it Beak} is (called) {\it Quail Fire} and {\it Heart} is (called) {\it Great Fire}. Tao Tang's
(i.e., legendary Emperor Yao's) Regulator of Fire, Yan Bo, dwelt at Shangqiu and sacrificed to {\it Great Fire} ($\alpha$ Sco, Antares),
using fire to mark the seasons there
\end{quotation}
(translation by Pankenier 2013, p.\,274).
Obviously, the names of {\it Quail Fire} and {\it Heart} reflect the red colours of
Alphard (B$-$V=1.45 mag) and Antares (B$-$V=1.84 mag), respectively (Pankenier 2013, pp.\,274--275).

{\bf Egypt:}
In ancient Egypt (BC), Antares was called `the red one of the prow' (Lull \& Belmonte 2009). 

{\bf Greece:}
The name `Antares' is derived from the red(dest) planet: the Greek preposition `anti' can mean `like' 
or `in place of', i.e., {\it `ant Ares' = `like Mars'} (Liddell \& Scott 1940; Kunitzsch \& Smart 2006).
The naming of Antares after the red Mars (Greek) is most certainly from the 1st millennium BC --
the first known appearance is in the Almagest. 

{\bf Rome:}
The Roman poet Virgil (BC 70-19) mentioned the colour of Antares in the beginning of his work Georgics I, lines 33-35 
(published probably BC 29), which we cite here in the edition by Thilo \& Hagen (1881): 
\begin{quotation}
qua locus Erigonem inter Chelasque sequentis \\
panditur (ipse tibi iam bracchia contrahit ardens \\
Scorpius et caeli iusta plus parte reliquit).
\end{quotation}

This has been translated freely to English by Fairclough (2000):
\begin{quotation}
between the Virgin and the grasping Claws (of Sco), a space is opening -- lo! for you (Caesar)
even now the blazing [lat. ardens, lit. also fiery] Scorpion draws in his arms.
\end{quotation}
The `blazing' (fiery) star in Scorpio towards Virgo is quite obviously Antares
(today with Libra in between, which was then often considered as claws of the Scorpion).
However, since we do not know the colour terminology or scaling of Virgil,
and since he did not compare Antares with any other star or planet in colour,
we cannot convert this statement into a colour index.

{\bf Hephaestion} of Thebes, born AD 380, a Hellenized Egyptian astrologer/astronomer, 
wrote around AD 415 in {\it Apotelesmatica} 1, 3:
\begin{quotation}
(Scorpion) ... in his body are three (stars), of which the middle one is reddish [hypokirros] --
it is called Antares
\end{quotation}
(our English translation from the Greek in Engelbrecht 1887, p.\,69). 

In {\bf Hawaii}, it is called `Hoku-ula' meaning `red star' (Makemson 1941; Noyes 2019)
or `Mehakua-koko', where `koko' means `red' (Noyes 2021);
note that `Hoku-ula' just means `red star' and was applied to both Betelgeuse and Antares.
`Te Mata-heko' for `red star' is the Tuamotuan (an archipelago of what is now the French Polynesian Isles)      
name for Antares (Makemson 1941, p.\,230),
and `Rerehu', for `burning', was used by the Maori (Makemsom 1941, pp.\,251, 270). 

The {\bf Bedouin} tradition is transmitted by Ibn Qutayba (edition in Ibn Qutayba 1956) from the 9th century (AD 828--889):
\begin{quotation}
{\it al-Qalb} ... is the red star behind {\it al-ikl\textit{\={\i}}l} ($\beta$,$\delta$,$\pi$ Sco) between two stars 
named {\it al-niy\=at} ($\sigma$,$\tau$ Sco)
\end{quotation}
(Kunitzsch 1961, p. 89; Adams 2018, pp.\,278, 298, 300).

{\bf Al-\d{S}\=uf\textit{\={\i}}} (AD 903--986) wrote about Antares: 
\begin{quotation}
in Scorpius ... the middle one of these which is reddish and called Qalb ...
The 8th is the bright red that is close to the 7th. It is of 2nd mag ... The bright red 8th star on the body is called al-Qalb [heart]
\end{quotation}
(Kunitzsch 1974; Hafez 2010). Ptolemy's Almagest also had Antares with 2nd mag. 

{\bf Al-Batt\=an\textit{\={\i}}} (died AD 929) mentioned the red
colour of Antares: `al-a\d{h}mar' for `the red' (Kunitzsch 1974). 

{\bf Al-B\textit{\={\i}}r\=un\textit{\={\i}}} (AD 973--1048) and {\bf al-\d{T}\=us\textit{\={\i}}} (12th cent.\,AD) 
both wrote that Antares has the same `nature' as Mars, referring to colour (Kunitzsch 1974). 

{\bf Ulug Beg} and {\bf Tycho Brahe:} In the late pre-telescopic star catalogs of Ulug Beg (died AD 1449, Kunitzsch 1974) 
and Tycho Brahe (ca.\,AD 1600), 
Antares is given as red. For example, from Brahe: `Inmed.\,rutlinas, Antares seu Cor dicta',
i.e., `in the middle the red, called Antares or heart' (Verbunt \& van Gent 2010).

Whenever Antares is compared in colour to Mars (B$-$V$\ge 1.43 \pm 0.13$ mag), 
we can consider this colour index as a lower limit for Antares,
because there is no planet redder than Mars.
(And from comparisions of historical supernovae in brightness or colour with planets,
it is known that the {\it typical} brightness or colour is meant (e.g. Stephenson \& Green 2002),
so that we can use 1.43 mag as limit instead of the lower bound being $\sim 1.3$ mag). 
Antares was never compared to any other planet (in colour).

And since Antares was always included among the red(dest) stars, it most likely was always the reddest (naked-eye) star.
For those scholars (Ptolemy, Germanicus, Cleomedes, Ibn Qutayba, Al-\d{S}\=uf\textit{\={\i}}, and Ulug Beg),
who listed Antares together with other stars as red, we use the limits to Betelgeuse also as lower limits for Antares.
In Figs.\,4 \& 5, we shift them by 50 yr for clarity.
Colour indices for Antares are also listed in Table 3.

In sum, while several independent historical records show that Betelgeuse was significantly less red than today, e.g. 
like Saturn (Hyginus) or `yellow' (Simas), Antares was always reported as being red.

\subsection{Wezen}

Wezen ($\delta$ CMa, HIP 34444, V=1.84 mag, B$-$V=0.70 mag, F8\,I, its WDS companion is 7 to 8 mag fainter)
is plotted in Figs.\,1 \& 2 at (B$-$V)$_{0}\simeq 0.6$ mag and M$_{\rm V}\simeq-7$ mag (13 to 15~M$_{\odot}$).
As seen in Fig. 3, it is expected to have evolved from whitish to yellow by 
increasing its B$-$V index by $\sim 0.7$ mag
in the past $\sim 3000$ yr, if on the 1st crossing with 14 or 15~M$_{\odot}$.
If it is instead on its 2nd or 3rd crossing with 13~M$_{\odot}$, the B$-$V colour index would
have either decreased from B$-$V$\simeq 1.5$ mag to 0.7 mag or it would have been roughly constant
(actually very slowly increading by $\sim 0.1$ mag in the last 5000 yr), respectively, 
in the last four millennia (indicated with the three lines in Fig.\,3). 

Wezen was noticed by Arabic Bedouines. Ibn Qutayba (AD 828--889) 
reported the following about one of their constellations: 
\begin{quotation}
The Maidenhead (al-$^{c}$udhra), the Maidenhead of al-Jawz\=a'a ($^{c}$udhrat al-jawz\=a'), are five white stars below 
the Shi$^{c}$r\=a-Who-Crossed-Over (ash-shi$^{c}$r\=a al-$^{c}$ab\=ur for Sirius) in the Milky Way (al-majarra). 
And they are called the Virgins (al-$^{c}$adh\=ar\=a)
\end{quotation}
as translated by D. Adams (priv. comm.) from the 1956 edition of Ibn Qutayba's work --
similarly in Kunitzsch (1959, pp. 103 and 140) and Kunitzsch (1961, pp. 115--116) in German, 
see also Adams (2018, pp. 125 and 254).\footnote{The constellation name al-$^{c}$adh\=ar\=a now pertains 
to its member star $\epsilon$ CMa=Adhara, while al-$^{c}$udhra was given to $\eta$ CMa=Aludra.}

Al-\d{S}\=uf\textit{\={\i}} (ca.\,AD 1000) identified four of the stars with $\delta, \epsilon, \eta$, and o$^{2}$ CMa;
the fifth may be $\sigma$ CMa (Adams 2018, p. 254), which is, however, 
not white (B$-$V=1.74 mag), but most certainly too faint for naked-eye colour detection (V=3.45 mag).
While $\epsilon, \eta$, and o$^{2}$ CMa are indeed blue-white and bright enough for naked-eye detection of colour, 
the central star of this group, $\delta$ CMa (Wezen, V=1.84 mag), now has B$-$V=0.70 mag, i.e. yellow after Table 1;
we could confirm a certain yellowish-orange tint by naked-eye inspection, in particular compared to the other stars of this group.
The primary name used for this star grouping (al-$^{c}$udhra') indicates the maidenhood or hymen of a woman, 
i.e. her state of virginity, or conversely her loss of virginity, with the root verb meaning `to cut', 
but also applied to male circumcision, for the same reason of bleeding (D. Adams, priv. comm.).
Therefore, it has been speculated that the central star Wezen may have been seen with some bloody orange-red colour (Adams 2018, pp. 125 and 254).
However, the statements by Ibn Qutayba and Al-\d{S}\=uf\textit{\={\i}} on those `five white stars' are reliable --
the two names of this star group, the Maidenhead and the Virgins, express this in a traditional manner
(white pertains to virginity). 

If Wezen was indeed white 
in the 9th century AD, in particular having a smaller B$-$V colour index than today,
then, of the three possible evolutions plotted in Fig. 3,
the scenario with a decreasing colour on the 2nd crossing with 13~M$_{\odot}$ can be clearly excluded --
and an almost constant colour appears unlikely (it would have been already yellow by the time of the Bedouines);
only the track with increasing B$-$V index from white to $\sim 0.7$ mag
in the past $\sim 3000$ yr is viable, so that Wezen is on the 1st crossing with 14 or 15~M$_{\odot}$.
Since we have just this single source, the evidence for an historically confirmed colour change is somewhat weakened. 

\subsection{Other stars in the Hertzsprung gap}

Other relatively massive stars located inside or towards the end of the Hertzsprung gap 
in Figs.\,1 and 2 are discussed in the following.
Of these, a reliable pre-telescopic report on colour is found only for Wezen (previous subsection).
For the others, some of which are quite far south, 
we discuss only the first crossing of the Hertzsprung gap;
see Fig.\,3 for their colour evolution (data in the appendix table).

For each such star, we determine its current age (and mass) from the location in the CMD (Figs. 1 \& 2)
and the best fitting MIST track. The latter also includes the colour evolution until its current age and colour,
which is plotted in Fig. 3. (The absolute ages are certainly somewhat uncertain, but do not matter much here,
as we are concerned with the relative age before the present time.) 

\begin{center}
\begin{figure*}
\includegraphics[angle=270,width=1\textwidth]{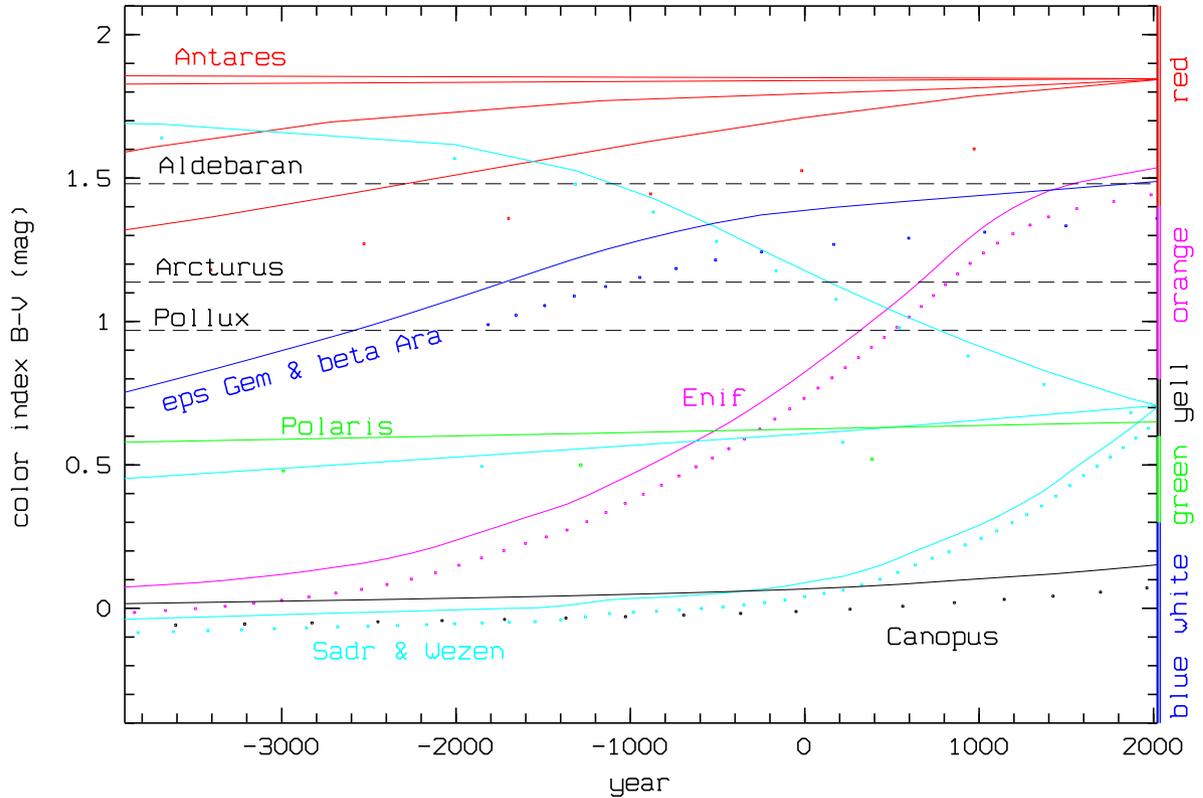}
\caption{{\bf Expected change over time of the colour index} of Antares 
(top, in red, from our best fits corresponding to 13 to 16~M$_{\odot}$; Fig. 4)
and for a few other bright stars as labeled (see Sects. 3.4 \& 3.5 for details).
Ages and masses of these stars were determined from Fig.\,1 using the MIST series, with
linear interpolation between the MIST time steps.
We plot the intrinsic colour indices (B$-$V)$_{0}$ (dotted lines) from the MIST tracks (Choi et al. 2016)
and also the observed colour indices B$-$V (full lines) as would be observed. The latter are calculated from the former by
adding interstellar extinction appropriate for the respective star.
The dotted lines also serve to illustrate the time resolution of the MIST models.
The full lines end at 2022 with the current observed colour index.
Enif ($\epsilon$ Peg, V=2.4 mag) may have changed from white over yellow and orange to red
within the last few millennia, but we found no colour reports from pre-telescopic times to support this.
Since Wezen ($\delta$ CMa) was really reported as white by the Bedouines in the 9th century, 
the evolution 
with decreasing or roughly constant colour index would be impossible (see Sect. 3.4).
For Polaris, we show the colour evolution for the primary at $\sim$5M$_{\odot}$.
See Sect. 3.4 for for details.
Aldebaran, Arcturus, and Pollux are plotted to show that these {\it other} Almagest stars described
there as `hypokirros' (i.e. somewhat orange-to-red, Sect. 3.2.2) indeed remained constant in colour according to
their location in the CMD (Fig. 1) and their MIST tracks.}
\end{figure*}
\end{center}

{\bf Enif} ($\epsilon$ Peg, HIP 107315, V=2.4 mag, B$-$V=1.53 mag, K2\,Ib-II) 
is plotted in Figs.\,1 \& 2 at (B$-$V)$_{0}\simeq 1.5$ mag and M$_{\rm V}\simeq -5$ mag (for 11 to 12~M$_{\odot}$)
as an unresolved star; its two companion candidates in the WDS are 6.2 and 10.3 mag fainter than the primary,
so that their effect is negligible. Enif is expected to have changed its colour from B$-$V$\simeq 0.9$ to 1.5 mag (now red) in 
the last two millennia (Fig. 3) -- and to have dimmed by $\sim 1.0$ mag (Fig. 2); no colour report was found. 

{\bf Ahadi:} As an unresolved star, Ahadi ($\pi$ Pup, HIP 35264, V=2.7 mag, B$-$V=1.53 mag, K4\,III) 
is plotted at a similar location in the CMD/HRD as Enif, but it has WDS companions only 3.6 and 5 mag fainter 
than the primary (WDS), which affect the primary's mass and age estimates. 

{\bf $\epsilon$ Gem} and {\bf $\beta$ Ara:}
The stars $\epsilon$ Gem (HIP 32246, V=3.0 mag, B$-$V=1.4 mag, G8\,Ib, its WDS companion is 6.5 mag fainter) 
and $\beta$ Ara (HIP 85258, V=2.8 mag, B$-$V=1.4 mag, K0\,IIb, single star)
are both plotted at (B$-$V)$_{0}\simeq 0.4$ mag and M$_{\rm V}\simeq -4$ mag ($\sim 9$~M$_{\odot}$). 
They are predicted to have reddened more slowly; no historical reports were found (both stars are relatively faint now
for colour detection, but should have been $\sim 0.5$ mag brighter two millennia ago). 

{\bf Avior} ($\epsilon$ Car, HIP 41037, V=1.9 mag) is placed at (B$-$V)$_{0}\simeq 1.1$ mag 
and M$_{\rm V}\simeq -5$ mag as an unresolved object, and although it appears to be in the gap,
it is actually a binary (WDS) made up of a red giant (K3\,III) past the gap and a B2\,V star before the gap. 

{\bf Sadr} ($\gamma$ Cyg, HIP 100453, V=2.23 mag, B$-$V=0.69 mag, F8\,Ib, two WDS companions are 8 to 9 mag fainter)
is plotted next to Wezen in Figs.\,1 \& 2 at (B$-$V)$_{0}\simeq 0.6$ mag and M$_{\rm V}\simeq-7$ mag (13 to 15~M$_{\odot}$),
so that it is also expected to have evolved from whitish to yellow by 
increasing its B$-$V index by $\sim 0.7$ mag
in the past $\sim 3000$ yr, if on the 1st crossing with 14 or 15~M$_{\odot}$.
If it were instead on its 2nd or 3rd crossing with 13~M$_{\odot}$, the B$-$V colour index would
have either decreased from B$-$V$\simeq 1.5$ mag to 0.7 mag, or it would have been roughly constant, respectively, 
in the last four millennia (indicated with the three lines in Fig.\,3).

{\bf Capella} ($\alpha$ Aur, HIP 24608, G3\,III, V=0.08 mag, 
B$-$V=0.80 mag, i.e. at the border between yellow and orange, Table 1), whose colour is clearly seen by the naked eye, 
falls in the Hertzsprung gap, but is actually a binary. 
Once the components are resolved, Capella B (2.5~M$_{\odot}$) is still located in the gap, 
while Capella A (2.6~M$_{\odot}$) has just passed the gap. 
Capella B is predicted to have cooled at a rate of $\sim 1130$ K over the past Myr,
or just $\sim 1^{\circ}$ per millennium (Torres et al. 2015). 

{\bf Polaris} ($\alpha$ UMi, HIP 11767, F8\,Ib, 
B$-$V=0.62 mag as an unresolved multiple) 
is considered a triple star with masses of 5.4, 1.39, and 1.26~M$_{\odot}$ (Evans et al. 2008, Fadeyev 2015), 
so that the primary yields most of the visible light. 
The combined light of the three components does not make it appear much higher in the CMD, and would only imply a slightly higher mass of
6 to 7~M$_{\odot}$ instead of 5.4~M$_{\odot}$. Polaris is a Cepheid in the instability strip. 
We would expect a small colour change over the last few millennia, but we did not find
any pre-telescopic reports confirming such an evolution. 
It was suggested that its brightness may have increased slightly
over the past two millennia (Engle, Guinan, \& Koch 2004);
however, the 3rd magnitude estimate given in Ptolemy's Almagest corresponds to a modern V$=3.17 \pm 0.67$ mag 
(Schaefer 2013), which is within $1\sigma$ of Hevelius' value (visual $2.4 \pm 0.2$ mag) and only $2\sigma$ deviant from 
today (V=2.0 mag). 

We note, finally, that
{\bf Canopus} ($\alpha$ Car, HIP 30438, A9\,II, V=$-0.63$ mag), {\bf iota Car} (HIP 45556, A7\,I, V=2.25 mag), 
and {\bf upsilon Car} (HIP 48002, O9+B7, V=3 mag) are just entering the gap, and beginning to shift to the red.

\section{Final results for Betelgeuse and Antares}

In this section, we check whether the historically recorded colours of Betelgeuse and Antares
are consistent with their location in the CMD, i.e. with their evolutionary state --
and we estimate their masses and ages.

\subsection{Betelgeuse and Antares -- mass and age} 

Betelgeuse and Antares A are similar in intrinsic colour, but due to different extinctions,
the observed B$-$V colours are different by up to 0.1 mag.
And since the spectral type of Betelgeuse is in the {\it range} M1--2, while Antares A has M1.5,
the uncertainty in spectral type, and hence in intrinsic colour index, may be larger for Betelgeuse than for Antares A.
Given that Betelgeuse and Antares A also show similar absolute magnitudes (Figs.\,1 \& 2),
they may also have similar masses and ages.
However, while Antares was always reported `red' for at least three millennia --
e.g. the Greek Ant-ares means `like Mars' (Ares) in colour -- Betelgeuse was reported with a different colour two millennia ago.
If they are currently on different crossings of the Hertzsprung gap (e.g., Betelgeuse on the 1st crossing
and Antares on the 2nd or 3rd crossing), then they can have significantly different ages and/or masses.

Next, we determine the masses of Betelgeuse and Antares considering the historical constraints on colour.

We select those mass tracks from the CMD (Figs. 1 \& 2) which are possibly consistent with Betelgeuse and Antares.
Both Betelgeuse and Antares A lie near the 13--16 M$_{\odot}$ MIST tracks (Figs. 1 \& 2): 
within 1$\sigma$, only masses of 13 M$_{\odot}$ (for the 2nd and 3rd crossing) 
and 15 M$_{\odot}$ are possible; within 1.5$\sigma$, 
also 14 and 16 M$_{\odot}$ match (17 and 18 M$_{\odot}$ would also be possible for Betelgeuse for a proposed 
larger distance of 197 pc; Fig. 2).
For those masses (13~M$_{\odot}$ first, second, and third crossing of the gap, as
well as 14, 15, and 16~M$_{\odot}$), we use the MIST tracks to determine their
current age and then to plot colour index versus time (Figs. 4 \& 5).\footnote{The absolute age
is certainly somewhat uncertain, but does not matter much here: we are concerned with
the relative age (before the present) and colour evolution.} 
Both Betelgeuse and Antares should now (in 2022) have a colour index
as observed (Table 2). Comparing the colour curves with the historical data from
Sect. 3 (Table 3), we find that for Betelgeuse, only the 14~M$_{\odot}$ mass track
is consistent, while for Antares, several solutions are possible (see Table 4).

\begin{center}
\begin{figure*}
\includegraphics[angle=270,width=1\textwidth]{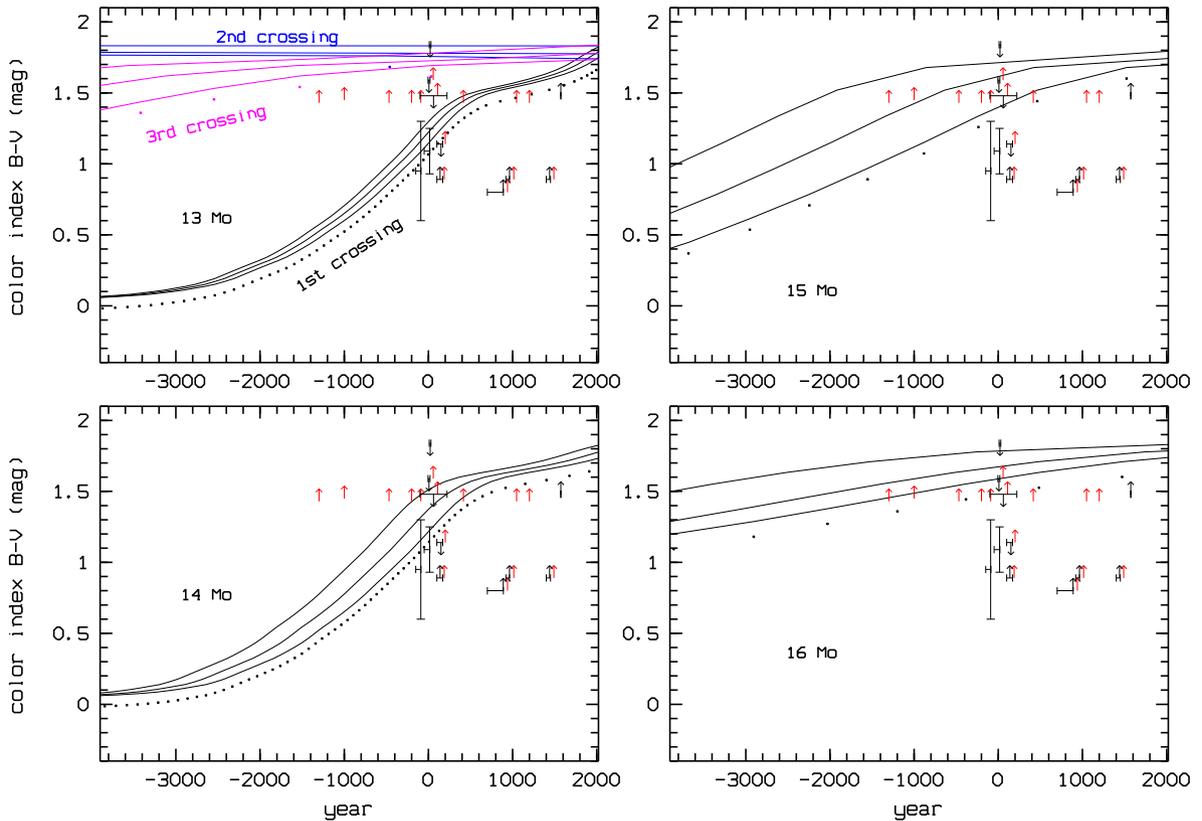}
\caption{{\bf Predicted colour index change over time} according to the MIST models (Choi et al. 2016)
for a star like Betelgeuse or Antares, for masses of 13 to 16~M$_{\odot}$:
All lines end in 2022 with the current B$-$V colour index,
B$-$V=1.73 to 1.83 mag, shown as three lines for values of 1.73, 1.78, and 1.83 mag
for Betelgeuse (the values are very similar for Antares).
In the upper left we show the three crossings of the Hertzsprung gap for a star of 13~M$_{\odot}$: 
2nd crossing at 16.1 Myr (top, blue lines, 
slightly decreasing B$-$V colour index),
3rd crossing at 17.3 Myr (middle, pink, 
slightly increasing B$-$V colour index),
and 1st crossing at 15.7 Myr (bottom, red, 
significantly increasing B$-$V colour index over the last few millennia).
In the lower left we show the colour evolution over the last few millennia for 14~M$_{\odot}$,
in the upper right for 15~M$_{\odot}$, and in the lower right for 16~M$_{\odot}$, where there is
always only one crossing according to these models.
Intrinsic colour indices (B$-$V)$_{0}$ are shown with dotted lines,
and the evolution of the B$-$V index as would be observed is represented with full lines. 
We calculate the latter from the former by
adding interstellar reddening appropriate for the respective star.
The dotted lines also serve to indicate the time resolution of the MIST models (shown only for a few cases, for clarity).
The black data points and limits are those reported for Betelgeuse (Table 3)
and the red lower limits are those for 
Antares -- some shifted by 50 yr for clarity.
While for Betelgeuse the first crossing with 13~M$_{\odot}$ would be consistent with the historical data,
its absolute magnitude (see Fig.\,1) 
is not consistent with this crossing.
Therefore, among all possible tracks for Betelgeuse within $1.5~\sigma$ of its current location in the CMD 
(in both absolute magnitude and colour),
only the track for 14~M$_{\odot}$ is also consistent with the rapid colour evolution in the last two millennia.
For Antares, the tracks for 13 (2nd or 3rd crossing), 15, or 16~M$_{\odot}$ are consistent 
with both the CMD location (absolute magnitude and colour index) and 
its always red colour,
but neither the 1st crossing with 13~M$_{\odot}$ nor the 14~M$_{\odot}$ track satisfy those conditions.}
\end{figure*}
\end{center}

\begin{center}
\begin{figure*}
\includegraphics[angle=270,width=1\textwidth]{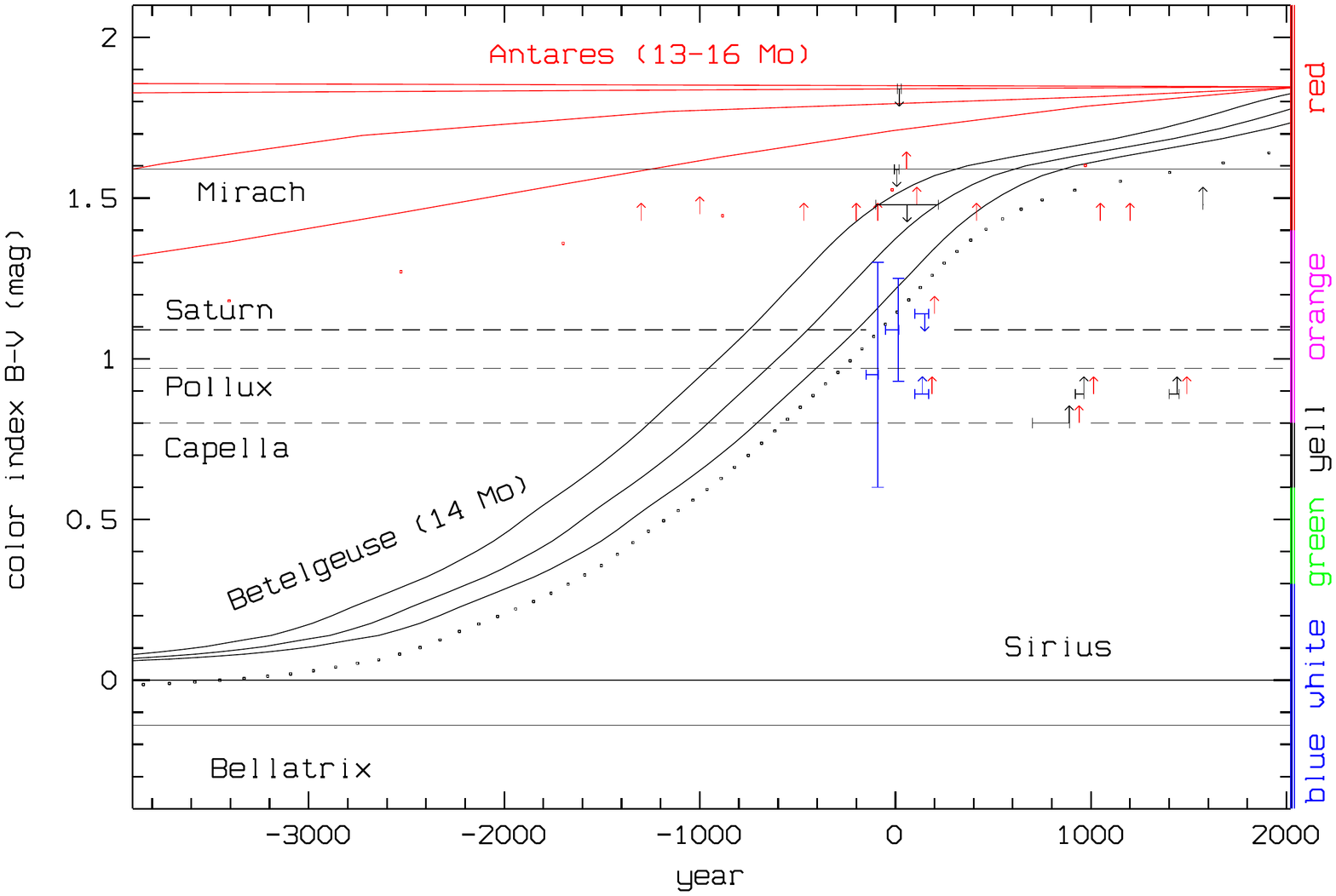}
\caption{{\bf Colour evolution of Betelgeuse and Antares:}
We plot the intrinsic colour indices (B$-$V)$_{0}$ from the MIST tracks 
(dotted lines for Antares and Betelgeuse indicating the MIST time resolution)
and the colour indices B$-$V as would be observed (full or dashed lines),
calculated by adding A$_{\rm V}$/3.1 to account for extinction -- and also Saturn. 
The full lines end at 2022 with the current colour index.
For Betelgeuse, three black lines indicate its colour uncertainty.
Antares hardly changed in colour,
while Betelgeuse evolved rapidly in the last few millennia,
consistent with the historical observations (blue and black data and limits for Betelgeuse
from Table 3, red lower limits for Antares from Sect. 3.3); 
for Betelgeuse, the blue data and limits from the Simas, Hyginus, Almagest, and {\it Tetrabiblos} (left to right)
constrain the colour index to be significantly lower two millennia ago than now.
Sirius, Bellatrix, and Mirach, the other stars mentioned by the Simas (Sect. 3.2.7) 
are added to show that these were indeed constant in colour.}
\end{figure*}
\end{center}

For Betelgeuse, the best fit for both the CMD (Fig. 1) and the historical colour evolution (Fig. 5)
corresponds to the 14 M$_{\odot}$ track according to these models. Then, the lowest of the three solid B$-$V curves 
in Fig. 5 is most consistent with the historical data.
	
If Antares A were 14 M$_{\odot}$ or were crossing the gap at 13 M$_{\odot}$ for the first time, it would have evolved in the last few millennia 
from yellow to red (Fig. 5), which is inconsistent with the historical reports. For 15 and 16 M$_{\odot}$ as well as the 2nd and 3rd crossing 
of the gap with 13 M$_{\odot}$, Antares A was almost constant in colour (red) over the last few millennia (Figs. 4 \& 5).
It was crossing the Hertzsprung gap slowly, or was on its way up the Red Giant Branch in the H-R diagram --
all consistent with historical transmissions, 
and coeval with Antares B (Fig. 1, Table 4). Hence, historical records of an always red Antares 
together with theoretical models can constrain the evolutionary status of Antares A to be on the 13 M$_{\odot}$ blue loop or on the 
15 or 16 M$_{\odot}$ track (Fig. 2).
From their location on the CMD (Fig. 1) and the MIST tracks, we can infer the remaining
time until the supernova to be $\sim 1.5$ Myr for Betelgeuse and $\sim 1.0$ to 1.4 Myr for Antares.

Our mass estimates for Betelgeuse ($\sim 14$~M$_{\odot}$) and Antares A ($\sim$13, 15, or 16~M$_{\odot}$),
based on consistency with both the CMD (Fig.\,1) and the historical colour evolution as observed (Table 3),
are in agreement with recent independent determinations by other methods: 
For Betelgeuse, $11.6^{+5.0}_{-3.9}$~M$_{\odot}$ was obtained from stellar limb darkening observations (Neilson, Lester, \& Haubois 2011),
$\sim 16.5$ to 19~M$_{\odot}$ from a combined evolutionary, asteroseismic, and hydrodynamic investigation (Joyce et al. 2020),
and 17 to 25~M$_{\odot}$ by Dolan et al. (2016), although they used the larger distance from Harper, Brown, \& Guinan (2008).
They also wrote (their section 3):
`in a rotating model with an initial rotation with $v_{ini}/v_{crit}=0.4$, the current observed properties
are best fit with a progenitor mass of around 15~M$_{\odot}$' (Dolan et al. 2016).
In the dust clump hypothesis studied for the Great Dimming (Montarg\`es et al. 2021),
Betelgeuse was modeled `as a PHOENIX photosphere for a 15~M$_{\odot}$ red supergiant with temperature 3700~K
(3800~K for March 2020) ... according to parameters derived from the spectrophotometric measurements'
(see appendix in Montarg\`es et al. 2021),
so that the mass they gave is compatible with our result.
For Antares A, a mass estimate of $15 \pm 5$~M$_{\odot}$ was obtained from high-resolution interferometric
imaging of the CO first overtones (Ohnaka et al. 2013).

Furthermore, the CNO abundances of Betelgeuse indicate that it has already experienced the first dredge-up
(Lambert 1984), resulting in enhanced N and depleted C and O. This places it on the ascending supergiant branch,
after the base of the horizontal branch, i.e. on its way up:
The point with the faintest brightness on the MIST track for 14~M$_{\odot}$ at the end of the Hertzsprung gap
is at an age of 14.0880 to 14.0882 Myr with M$_{\rm V}=-5.15$ mag and (B$-$V)$_{0}=$1.67 to 1.69 mag.
The current intrinsic colour index of Betelgeuse is (B$-$V)$_{0}$=1.65 to 1.75 mag (Table 2);
the nominal value (1.70 mag) is reached on the MIST track just $\sim 100$ to 400 yr after that base.
This base in brightness is seen in Fig. 1 in the inlay as a broad minimum (for 14~M$_{\odot}$).
From the base onward, it takes several millennia for a star to increase in brightness by just 0.1 mag.
But in the two millennia before the base, a star on that track would dim by 0.7 mag (MIST track, Fig. 1).
In such a case, Betelgeuse had not only a lower colour index at around the BC/AD turn,
but was also brighter than now. 
This is not necessarily in conflict with Ptolemy assigning a `faint 1st mag' for Betelgeuse in the Almagest.
The reasons are not only its strong variability 
or because the letters $\epsilon \lambda^{\varsigma}$ (for the Greek word {\it elachistoteros}
or possibly {\it elass\=on}, both meaning {\it smaller/fainter})
could be a late addition,
but also because the Almagest magnitudes scatter by about $\pm 1$ mag around today's values (Hearnshaw 1999; Schaefer 2013), 
even more for the brightest bin (1st mag) including, e.g., Sirius (V=$-1.44$ mag) and Denebola (2.13 mag).
Furthermore, manuscript D, one of the four used in the edition in Heiberg (1903, p. 133),
gives $\mu^{\epsilon}$ for Greek {\it meiz\=on} for {\it larger/brighter} --
this manuscript was preferred in the English translation by Toomer (1984)
(but Sirius is given as plain 1st mag and `the brightest' on sky).\footnote{While 
the CMD location of Betelgeuse is not consistent with the first crossing
on the 13~M$_{\odot}$ MIST track (Fig. 1), the colour evolution according to that track
would be consistent with the historical colour observations (Fig. 4). On the 13~M$_{\odot}$ track
(with different time resolution in MIST), the brigthness minimum (base) was reached $\sim 400$ yr ago 
with (B$-$V)$_{0}=1.58$ mag, and Betelgeuse would have increased in both colour index
and brightness by $\sim 0.13$ mag, too small to be confirmed by historical observations
given the variability. In the last two millennia, it would have dimmed by 0.8 mag.} 
  
\begin{table*}
\begin{tabular}{llllll} \hline
\multicolumn{6}{l}{{\bf Table 4: Consistency of historical reports and theoretical models (a):}} \\ \hline
Mass                        & \multicolumn{2}{c}{Betelgeuse:} &  Antares: & colour al- & Antares A \& B \\ 
                            & from CMD? & got redder?         & from CMD? & ways red? & coeval? (b) \\ \hline
13 M$_{\odot}$ 1st crossing & no           & $\checkmark$    & no           & no               & $\checkmark$ (A: $\sim 15.7$ Myr) \\
13 M$_{\odot}$ 2nd crossing & $\checkmark$ & no              & $\checkmark$ & $\checkmark$     & $\checkmark$ (A: $\sim 16.3$ Myr) \\
13 M$_{\odot}$ 3rd crossing & $\checkmark$ & no              & $\checkmark$ & $\checkmark$     & $\checkmark$ (A: $\sim 17.3$ Myr) \\
14 M$_{\odot}$ & $\checkmark$ (14 Myr)   & $\checkmark$    & $\checkmark$ & no               & $\checkmark$ (A: $\sim 14.1$ Myr) \\
15 M$_{\odot}$ & $\checkmark$ & no              & $\checkmark$ & $\checkmark$     & $\checkmark$ (A: $\sim 12.8$ Myr) \\
16 M$_{\odot}$ & $\checkmark$ & no & $\checkmark$ & $\checkmark$ & $\checkmark$ (A: $\sim 11.8$ Myr) \\ 
17 M$_{\odot}$ & ($\checkmark$ 197 pc) & no & no & $\checkmark$ & no (A: $\sim 11.0$ Myr) \\
18 M$_{\odot}$ & ($\checkmark$ 197 pc) & no & no & $\checkmark$ & no (A: $\sim 10.2$ Myr) \\ \hline
\end{tabular}

Notes: 
(a) We indicate whether the given MIST track is consistent with the CMD location in both absolute magnitude and colour index, and
whether the historically observed colour evolution ($\alpha$ Ori getting red, $\alpha$ Sco staying red)
is consistent with the MIST track given the current CMD location.  \\
(b) We give the inferred age for Antares A, which should in all cases be coeval 
with its companion. Antares B is near the $\sim$6 to 8~M$_{\odot}$ tracks in Fig.\,1, 
where the evolution is so slow 
that the age is not well constrained, and a range of 12 to 29 Myr is possible.
\end{table*} 

\begin{center}
\begin{figure*}
\includegraphics[angle=270,width=1\textwidth]{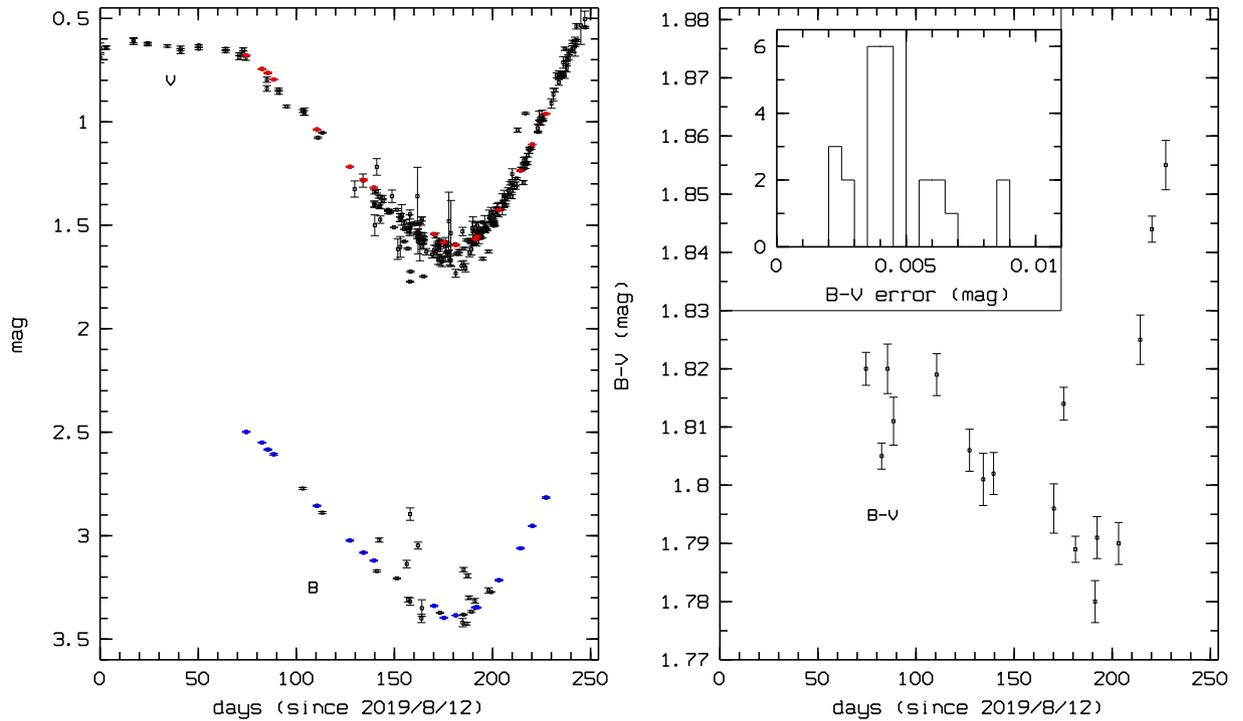}
\caption{{\bf Evolution of the magnitude and colour of Betelgeuse during the Great Dimming of 2019--2020:}
We plot V- and B-band magnitudes (left) and colour index B$-$V (right) based on AAVSO data 
collected between 2019 Aug 12 and 2020 Apr 15 (www.aavso.org),
i.e., until conjunction with the Sun. There was no further brightening after this.
For B$-$V, we use data for V and B closest in time, always within one night.
Given that the colour index error bars show a bimodal distribution (inlay),
we plot here only those with error bars below $\pm 0.005$ mag
(these selected data are shown in colour in the light curves on the left).
These data show a small decrease in B$-$V
from $1.82 \pm 0.01$ to $1.78 \pm 0.01$ mag from day $\sim 100$ until
the minimum in mid Feb 2020.
Therefore, the index did not decrease much during the dimming,
so that the colour detectable by the naked eye was still red even in Feb 2020.
Since other studies have shown a temperature drop during the Great Dimming 
(Guinan, Wasatonic, \& Calderwood 2019; Guinan et al. 2020; Guinan \& Massey 2020;
Levesque et al. 2020; Dupree et al. 2020),
associated with dust formation (Montarg\`es et al. 2021), 
the colour drop in B$-$V by $0.04 \pm 0.02$ mag could be due to a modified
reddening law (parameter R) for the additional dust.
After the minimum, Betelgeuse brightened to a similar B-band magnitude as just before the dimming
(see also AAVSO), but got slightly brighter in V than before (by 0.1 mag), 
leading to an increase in B$-$V to $1.855 \pm 0.004$ mag.}
\end{figure*}
\end{center}

In the alternative to a merger scenario for Betelgeuse (see Sect. 4.2) given by Wheeler et al. (2017),
the star would also be located within $\sim 1000$ yr after the lowest luminosity point 
before the rise up to the red supergiant branch (Sect. 4.2), consistent with our findings.
Our best fit for Betelgeuse on the 14~M$_{\odot}$ track would correspond to a distance of $\sim 125$ pc, 
consistent within $1.4\sigma$ with the revised Hipparcos distance (van Leeuwen 2007), and
fully consistent within $1\sigma$ with a mass of $\sim 14.2$ to 15.5~M$_{\odot}$ (interpolating between MIST tracks).

N.B.: Even if the reporting of Betelgeuse's colour in antiquity had been affected by 
the star being in the midst of a dimming episode, 
the colour index of Betelgeuse during the Great Dimming in February 2020
changed only slightly from B$-$V=$1.82 \pm 0.01$ to $1.78 \pm 0.01$ mag (Fig. 6). 
Historically, there is no indication that the observations studied here were taken during another dimming
episode; the dimming phase itself (as in 2019/20)
is short and strong enough to be noticed with the unaided eye -- and Betelgeuse remains bright enough for naked-eye
colour detection.

Finally, we compare the bolometric luminosity $\log$(L/L$_{\odot}$)=$4.7 \pm 0.1$ at the revised Hipparcos distance 
of $151 \pm 19$ pc, van Leeuwen 2007) and effective temperature (T$_{\rm eff}=3600 \pm 200$ K, Joyce et al. 2020) 
of Betelgeuse
against a different set of evolutionary models from the Geneva series (Ekstr\"om et al. 2012), see Fig. 7. 
The luminosity is based on an absolute magnitude of M$_{\rm V}=-5.59 \pm 0.28$ mag at 151 pc (Table 2),
and a bolometric correction of BC=$-1.43$ mag for an M1.5\,I star (Levesque et al. 2005).
At the larger distance of 197 pc, it would be $\log$(L/L$_{\odot})=4.90 \pm 0.11$.

The data only marginally agree with recent colour evolution for T$_{\rm eff} \simeq 3800$ K, 
but the rate of evolution from T$_{\rm eff}$ $\sim$4400--3800 K (B$-$V$\simeq$1.3--1.6 mag) in $\sim 10^{4}$ yr 
(Ekstr\"om et al. 2012) is too slow. Furthermore, stars 
with initial or actual masses of 14.5--15.5 M$_{\odot}$ have colour indices of up to (B$-$V)$_{0}$=1.62 to 1.64 mag 
(Ekstr\"om et al. 2012), i.e. too blue. 
In considering Betelgeuse's rotational velocity together with broad luminosity and temperature ranges, 
it was suggested to be located within $\sim$1000 yr after the turn from the horizontal to the ascending branch (Wheeler et al. 2017). 
Unusual CNO abundances in Betelgeuse indicate that it has experienced its first dredge-up (Lambert 1984). 
Both are fully consistent with our result placing the star less than a millennium past the base of the horizontal branch (Figs. 1 \& 8). 
MIST tracks for $\sim$14 M$_{\odot}$ supergiants, both for absolute magnitude versus colour (Fig. 1) and luminosity versus temperature (Fig. 8), 
can match Betelgeuse's recent cooling (reddening) better than other models, possibly because of a different treatment of rotational 
mixing or other physical ingredients (Choi et al. 2016). 

Following three-dimensional modeling of the bow shock ahead of the runaway star Betelgeuse with constant wind,
Mohamed, Mackey, \& Langer (2012) proposed that the combination of the low mass inferred for the structure and its more spherical shape compared 
to expectations for steady-state bow shocks could be accounted for if the shock is very young, suggesting Betelgeuse moved onto to the
red supergiant branch within the last 30 kyr.
Modelling with variable wind might explain the additional bar as a shell-shell collision (Mackey et al. 2012),
in which case Betelgeuse would have been on the blue loop until recently.
However, there are also a number of assumptions and uncertainties involved in this argument:
the mass and distance of Betelgeuse were poorly constrained
(they used the larger VLA distance of $\sim 200$ pc, while the rapid historical colour evolution 
shown in our work points to the smaller Hipparcos distance $\sim 150$ pc), blue supergiant wind parameters
are unconstrained and had to be assumed, and possible interstellar medium (ISM) density variations and the ambient 
magnetic field were not considered.
Clumps on the bow shock simulations by Mohamed, Mackey, \& Langer (2012) could be smoothed by
considering the interstellar medium magnetic field (Meyer et al. 2021); 
see also van Marle, Decin, \& Meliani (2014) and Mackey et al. (2014).
This study further shows that the bar could be of interstellar nature and that no inner shell-shell collision 
region of blue loop origin is needed (Meyer et al. 2021).
Our constraint on the rapid recent colour change of Betelgeuse is consistent with both
Mohamed, Mackey, \& Langer (2012) that the bow shock is young, and with Meyer et al. (2021)
that the bow shock is smooth and that the bar is an ISM feature.
(It remains unknown, whether stars like Betelgeuse experience a blue loop, which is not seen in all
MIST mass tracks, but only in some, and is also not seen in the Ekstr\"om et al. (2012) models.)

\begin{center}
\begin{figure*}
\includegraphics[angle=270,width=1\textwidth]{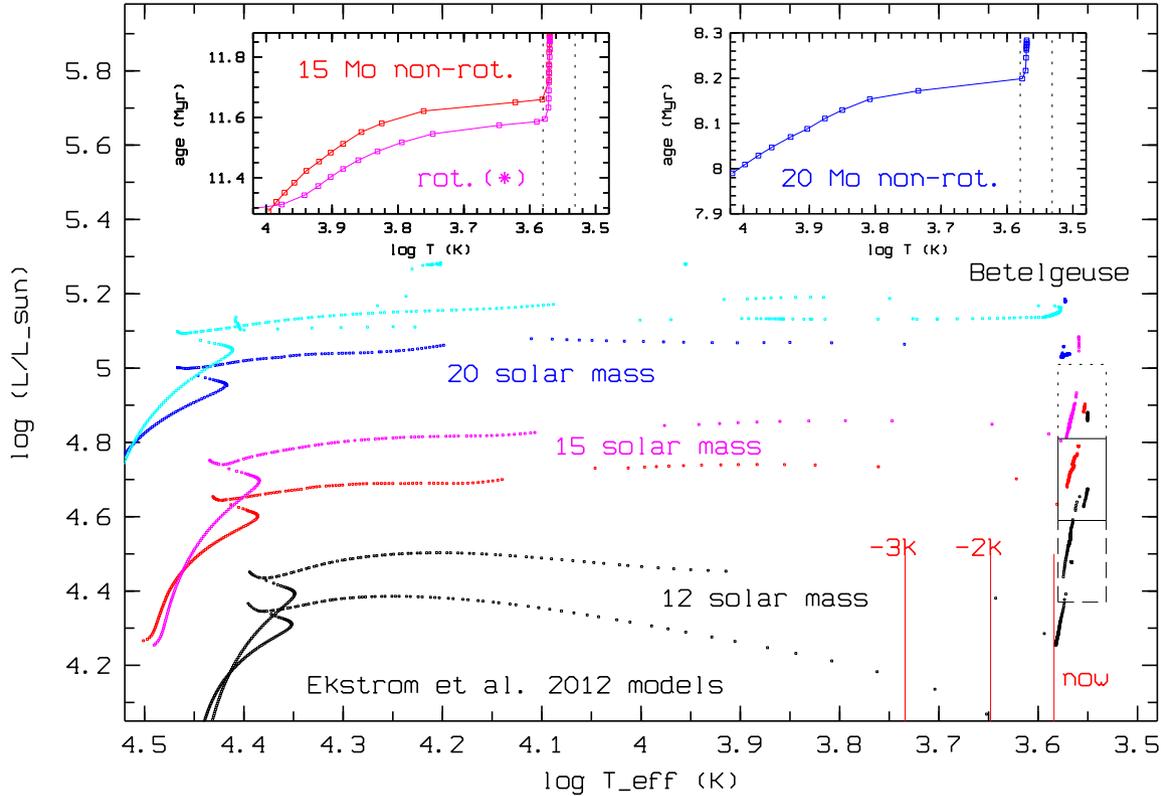}
\caption{{\bf Evolutionary tracks from the Geneva models} (Ekstr\"om et al. 2012)
in the HRD for the parameter range of Betelgeuse (Table 2). 
We show tracks for 12~M$_{\odot}$ (bottom, black),
15~M$_{\odot}$ (middle, red and pink), and
20~M$_{\odot}$ (top, cyan and blue);
rotating models are at the top, and non-rotating ones at the bottom.
The vertical red lines on the lower right show the temperature of Betelgeuse from the MIST series for
the present time, 2000 ($-2$k), and 3000 ($-3$k) years ago.
The black rectangle shows the $1\sigma$ data range possible for Betelgeuse at a distance of 151 pc
(short-dashed extension upwards for 197 pc, long-dashed extension downwards: $3\sigma$ error bar).
The models are inconsistent with significant recent temperature and colour change for Betelgeuse (i.e., evolution along the
abscissa) for the non-rotating 20~M$_{\odot}$ track as well as for the rotating and non-rotating 15~M$_{\odot}$ tracks. 
Furthermore, the time difference between the first point within the error box and the last few
just before the box is $\sim 10^{4}$ yr in these models, which is inconsistent with rapid colour
evolution over the last few millennia.
The upper inlays show the temperature evolution for those cases, but it is too slow,
given the historical reports, for the colour of the star to have been orange $\sim$2000 yr ago (i.e., B$-$V=0.8 to 1.4 mag, 
corresponding to temperatures of
$\sim$5350 to 4300 K for supergiants, according to Drilling \& Landolt 2000).
(*) In the upper left plot, the age for the rotating model (pink) is shifted downward for clarity by 2.32 Myr 
to fit on the same scale as the non-rotating model
(the true range is 13.6 to 14.2 Myr).}
\end{figure*}
\end{center}

\begin{center}
\begin{figure*}
\includegraphics[angle=270,width=1\textwidth]{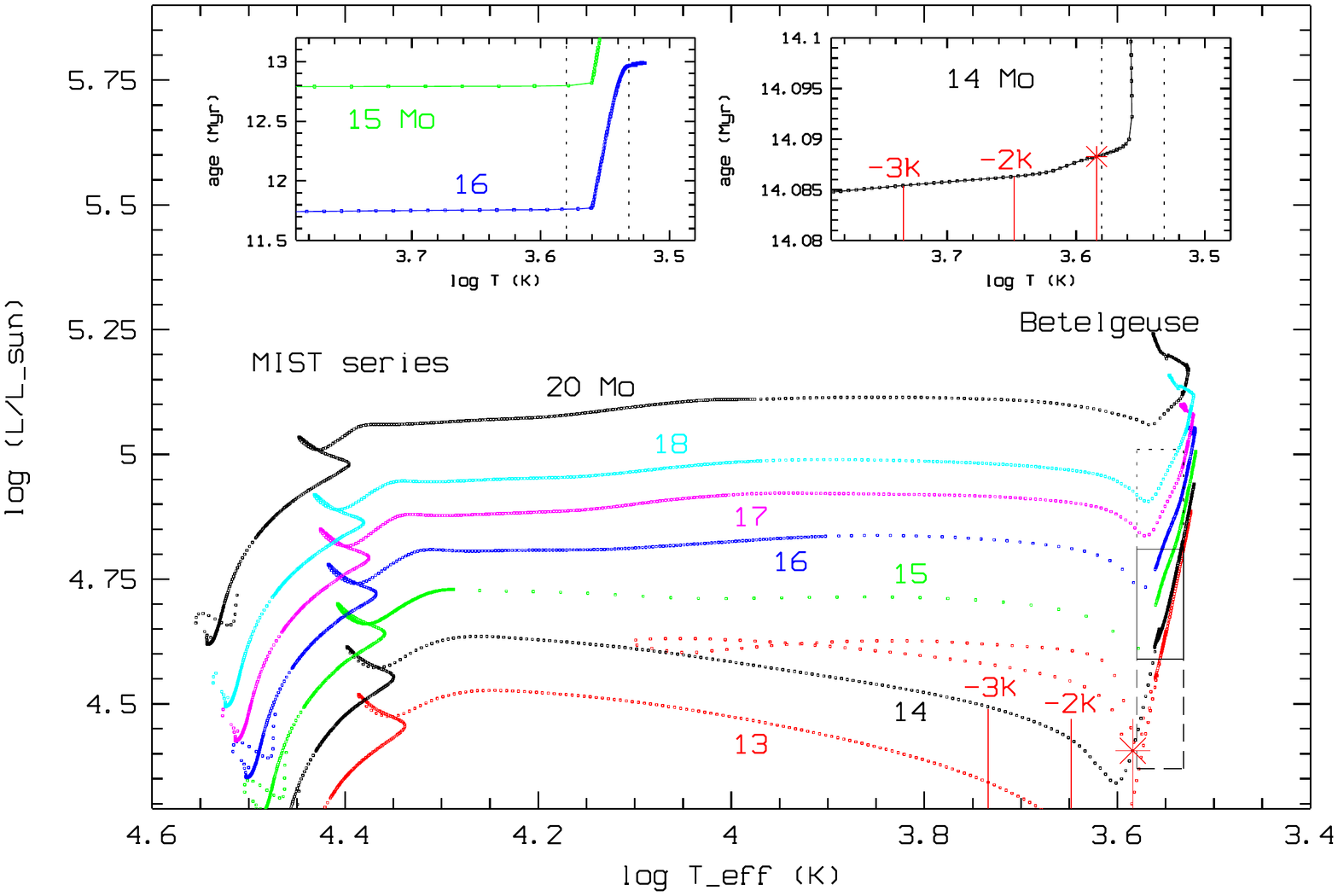}
\caption{{\bf MIST evolutionary tracks (Choi et al. 2016) in the HRD} 
(luminosity vs.\,effective temperature) from 13 to 20~M$_{\odot}$ (labeled): 
The vertical red lines on the lower right on the 14~M$_{\odot}$ track show the 
temperature of Betelgeuse at the present time, 2000 ($-2$k), and 3000 ($-3$k) years ago.
The black rectangle shows the $1\sigma$ data range possible for Betelgeuse at a distance of 151 pc
(short-dashed extension upwards for 197 pc, long-dashed extension downwards: $3\sigma$ error bar).
The upper inlays show the temperature evolution for 14 (right) as well as 15 and 16~M$_{\odot}$ (left).
Only on the 14~M$_{\odot}$ track is the evolution as fast as observed.
The current location of Betelgeuse (lower right red star symbol) is just a few centuries
past the base of the track, now on the ascending red supergiant branch
(consistent with its CNO abundances);
Betelgeuse is here at an effective temperature of $\sim 3800$~K.
At lower temperatures, the tracks would not allow significant recent temperature evolution,
which is inferred from 
historical transmissions by Hyginus, Ptolemy, and the Simas.}
\end{figure*}
\end{center}

\subsection{Betelgeuse merger models}

In the above we have explicitly considered Betelgeuse as a single star. 
However, this may not have always been the case. The unusually high rotational velocity measured for such 
an evolved star as Betelgeuse 
(observed projected velocity $v \sin i = 5.45 \pm 0.25$ km/s, Dupree et al. 1987; Kervella et al. 2018;
equatorial velocity $v_{\rm rot} \simeq 15$ km/s) 
has led to the suggestion that it may have been spun up by accreting a binary companion 
(Wheeler et al. 2017; Chatzopoulos et al. 2020; Sullivan, Nance, \& Wheeler 2020).
Wheeler et al. (2017) wrote that the models fitting the observed radius $R$, 
temperature T$_{\rm eff}$, luminosity $L$, and rotation $v_{\rm rot}$
`represent the evolutionary stage after the models have crossed the Hertzsprung gap and are at
very nearly the point of minimum luminosity before the rise up to the RSB' (red supergiant branch)
and that the `only models that formally fit the observed data on $L$, $R$, T$_{\rm eff}$, and $v_{\rm rot}$ are required to sit at
a very special short-lived point in the evolution ... The interval in which the models are predicted
to have conditions similar to Betelgeuse and v$_{\rm rot}$ between 1 and 50 km s$^{-1}$ is $\sim 1000$ yr' (their section 3.2).
While it may be unlikely for an ensemble to be in such a special short-lived state, we deal here only with one
object (Betelgeuse), where a-priori any special 1000-yr phase is not less likely than any other.
Furthermore, there are alternative explanations for the high rotational velocity:
`there may be more viscosity than computed in the default prescriptions in MESA that would allow
a greater transport of angular momentum from the rotating core to the envelope',
or that the velocity `measurements were affected, perhaps even dominated, by the large-scale
convective motions of the envelope for which Betelgeuse is famous' (Wheeler et al. 2017, their section 3.3).

Detailed modeling of the merger scenario (Sullivan, Nance, \& Wheeler 2020)
indicates that the end product would be remarkably insensitive 
to the mass of the lower-mass companion and to the epoch of accretion, and that after the event, it may not 
look much different than a normal supergiant. Therefore, our finding that the colour of Betelgeuse has 
changed significantly over the last few millennia, as we have shown in Sect. 3.2, does not conflict 
with this merger scenario, if it occurred prior to the earliest historical colour reports (ca. BC 100), 
and would justify our use of single-star evolutionary models to describe the evolution of Betelgeuse since then.

Alternatively, the smaller B$-$V colour index
of Betelgeuse two millennia ago could be seen as evidence that it was indeed a binary prior 
to the time of the historical records. In this interpretation, the colour change could be a direct consequence of the merger. 
While the original masses of the binary components are not known, many combinations can be found of a massive primary 
and a somewhat less massive secondary that lead to a lower B$-$V colour index for 
the combined light, consistent with that recorded some 2000 years ago.

A merger during historical times is unlikely, in particular within the last 2--3 millennia,
because it would almost certainly have been noticed and recorded as an exceptional change in colour 
(and certainly a significant brightening) 
on a relatively short time-scale.  
We note that none of the many `guest star' records from 
China (Xu, Jiang, \& Pankenier 2000) over the last two millennia (some interpreted today as supernovae, novae, etc.,
but see also R. Neuh\"auser, D.L. Neuh\"auser, \& Chapman 2021) seems to report any sudden brightness or colour changes in any of 
the fixed stars visible to the unaided eye
(except due to atmospheric scintillation).\footnote{For one historical `new star', it has indeed been discussed 
whether it became visible due to a merger: `Nova' Vulpeculae 1670 (CK Vul) was observed with a peak 
brightness of 2.5 to 3 mag for 10 days in AD 1671 in Europe by Hevelius and Cassini, and is now considered to be the result
of a merger of a white dwarf and a brown dwarf (Kami\'nski et al. 2015) rather than a nova.}

If one assumed that our present Betelgeuse with its relatively large rotational velocity resulted 
from a merger with a lower-mass star, our constraint on a colour change 
within the last two millennia must be fulfilled.
There is yet another alternative: given the high space motion of $\sim 30$ to 36 km/s
(with the Hipparcos proper motion at a distance range of 150 to 200 pc 
and a radial velocity of $21.91 \pm 0.51$ km/s, Famaey et al. 2005), Betelgeuse is a runaway star.
Since its motion does not point back to any star forming region (Bally 2008), both fast motion
and fast rotation could be explained by a two-step ejection process: first a dynamical ejection as a binary from
some star forming region, followed by a supernova explosion of the primary, so that Betelgeuse as the former companion
continued to move with orbital speed and direction as at the moment of the supernova. During the late
evolution of the supernova progenitor, mass and angular momentum could have been transfered to Betelgeuse
spinning it up and thus explaining the currently large rotational velocity.
See also Chatzopoulos et al. (2020) for a discussion of this scenario, which may admittedly require some
fine-tuning, but is not definitely ruled out. A merger would also need fine-tuning to avoid becoming a
blue straggler, to arrive at the current parameters of Betelgeuse, and to fulfill our new colour change constraint.
A dynamically ejected close massive binary can in principle end either in a supernova (with break-up or forming a 
high-mass X-ray binary) or in a merger; in case of the latter, though, the current space motion should
point back to a star forming region (which is not the case). The probability of a merger of a massive
binary is $22^{+26}_{-8}\%$, but $86^{+11}_{-9}\%$ for break-up during the first supernova (Renzo et al. 2019).   
The latter scenario can in principle be probed by searching for the neutron star likely formed in that 
supernova (see R. Neuh\"auser, Gie\ss ler, \& Hambaryan 2020 for the example of $\zeta$ Oph and PSR1706) 
or by detecting supernova debris in Betelgeuse's atmosphere
(e.g. $^{14}$N is enriched, although this could also be due to dredge-up; see Lambert 1984).

\section{Summary and conclusion}

In this work, we have considered the secular colour evolution of bright stars over millennia
by two independent approaches: We have placed all stars down to V=3.3 mag on
the CMD to select those inside or towards the end of the Hertzsprung gap.
In addition, we have consulted pre-telescopic colour reports from Europe,
the Mediterranean, West Asia (Near East) and East Asia, as well as First Nations around the world,
to find stars, that have evolved in colour in the last few millennia.

We have obtained the following main results:
\begin{itemize}
\item With our novel historically-critical approach, we have been able to quantify textual statements on star colour from antiquity
in terms of the B$-$V colour index (Table 3). 
\item Colour {\it differences} in B$-$V to within $\pm$\,0.1 to 0.2 mag 
were noticed and reported in pre-telescopic times (see Sect. 3.2.4 with footnote 5 on Brahe).
\item Betelgeuse's colour index B$-$V was 
at least 5.1$\sigma$ different (non-red) 
two millennia ago compared to now (red),
as derived 
from two independent sources, Hyginus from Rome and the Simas from China (Sect. 3.2).
That Betelgeuse had a different colour two millennia ago
is supported by further upper and lower limits from other ancient Mediterranean scholars.
E.g., Ptolemy's upper (B$-$V$> 0.80$ mag, Almagest) and lower limits ($<1.14$ mag, {\it Tetrabiblos})  
together yield a colour index of $0.97 \pm 0.17$ mag, which is itself 4.6$\sigma$ different from today's Betelgeuse.
\item We have also shown that Antares has been reported as having a red colour for more than three millennia (Sect. 3.3). 
\item Betelgeuse's evolutionary track must accommodate the rapid colour evolution over the last two millennia,
so that it is now located just after the base of the horizontal track (Sect. 4.1).
\item The colour evolution constrains the mass of Betelgeuse to 14~M$_{\odot}$;
the (by and large) constant colour of Antares shows that it is either on the 15 or 16~M$_{\odot}$ track,
or that it is on the 13~M$_{\odot}$ blue loop, see Sect. 4.1, according to the MESA MIST models.
\item We also present the B$-$V colour index evolution during the Betelgeuse Great Dimming in Feb 2020.
It dropped slightly
from B$-$V=$1.82 \pm 0.01$ to $1.78 \pm 0.01$ mag, possibly due to a variation in the reddening parameter R (Fig. 6).
\item In addition to Betelgeuse, Wezen ($\delta$ CMa) apparently has changed in colour since the 9th century
(Ibn Qutayba for the Bedouines: 
`white' -- but now yellow with B$-$V=0.7 mag), so that we can 
constrain its mass and evolutionary phase (Sect. 3.4, Fig. 3).
\item We also expect that Enif ($\epsilon$ Peg) has changed its colour significantly in the last two millennia from
yellow over orange to red (Fig. 3), but we have found no historical records to support this (Sect. 3.5).
\item For the other stars with pre-telescopic colour reports mentioned (e.g. by Ptolemy or the Simas), 
the derived historical B$-$V colour index is consistent with their current value --
and a constant colour is also expected from their locations in the CMD given the MIST tracks (Figs. 3 \& 5).
\item For Betelgeuse and Enif, we also noticed that they should have dimmed by almost 1 mag in the
last two millennia (Figs. 1 \& 2), according to their MIST tracks, but we have not yet found any confirming records (Sect. 4.1).
\end{itemize}

Colour {\it evolution} through the Hertzsprung gap is usually too slow compared to the human lifetime to have been noticed,
but when studying pre-telescopic records on star colours, secular colour evolution certainly has to be taken into account --
and also even when identifying stars in pre-modern records partly by the mentioned colour. 
While colour perception is somewhat subjective and some observers might consider Betelgeuse and Antares as orange 
(strong absorption in the red), historical reports present Betelgeuse's colour as clearly different from Antares 
and significantly less red than today. 

Our main conclusion is that all historical transmissions since antiquity along with the MIST track for $\sim$14~M$_{\odot}$ 
consistently show a rapid colour evolution of Betelgeuse. 
Quantitative details on the colour evolution of Betelgeuse (as expected from the position on its evolutionary track)
depend sensitively on the input data (parallax, brightness, colour index, extinction).
In the future, the mass and other parameters can be further constrained:
Joint use of all observable properties 
of Antares and Betelgeuse (present brightness, colour, temperature, chemical composition, rotation, extinction, distance, 
pulsation properties, etc.) and their uncertainties, again with calibrated historical colour (and brightness) reports, should help 
to pinpoint their masses with even higher precision. This could provide further insight into the physics of stellar interiors and 
late evolution of supergiants (and the time left until they go supernova).
The historical colour evolution is a new tight constraint on either the single-star evolutionary models
or the Betelgeuse merger models.

\bigskip

{\bf Acknowledgements.}
We acknowledge the American Association for Variable Star Observers for their observations during the 
recent Great Dimming of Betelgeuse used in Fig. 6. We used the SIMBAD, VizieR, Hipparcos, and Tycho catalogs. 
The thank Andrea Dupree (Smithsonian Astrophysical Observatory) and Valeri Hambaryan (U Jena) for helpful comments.
We acknowledge Danielle Adams (Lowell) and Pouyan Rezvani (PAL Munich) for help with Arabic texts,
and Martha Noyes (Hawaii) for consultation on Polynesian records.
RN would also like to acknowledge H.U. Keller (Stuttgart) for pointing us on the literature on Polaris.
We would like to thank Salvador Bar\'a (U Santiago de Compostela), Larry Thibos (Indiana U), and Michael Geymeier (then U Jena)
for discussion about naked-eye star colour detection.
We acknowledge Dominique Meyer (U Potsdam) and Shazrene Mohamed (U Miami Coral Gable) for discussion about the bow shock of Betelgeuse.
We would like to thank the referee, Jacco van Loon, for 
very careful reading and good recommendations.
MC's contribution for this article has received funding from the European Union's Horizon 2020 research and 
innovation program under the Marie Sklodowska-Curie Grant agreement No. 844152: `SN1604. The Ophiuchus Supernova: 
Post-Aristotelian Stargazing in the European Context (1604--1654)'. \\
Authors' contributions: RN searched the pre-telescopic literature for star colours, conducted the main research, 
prepared the figures, and wrote the manuscript with DLN and GT. All co-authors read and commented on the whole 
manuscript. Furthermore, GT studied and discussed the MIST tracks, MM compiled the data for the bright stars and 
calculated intrinsic colour indices and absolute magnitudes, DLN analyzed the historical records, JC searched for 
and translated Chinese texts, DL translated Latin and Greek texts, and MC contributed to the literature 
search for historical star colour records. 

\bigskip

{\bf Data availability.}
All data are given either in this paper or in the references.

{}

\newpage

\newpage

\section{Appendix: Input data for all 236 bright stars}

Here, we list all input data for the 236 bright stars down to V=3.3 mag,
for which the colour may in principle be detectable by the naked eye.

\begin{onecolumn}

\begin{longtable}{lrrrcrcrl}
\multicolumn{9}{l}{{\bf Table 5:} Data for all bright stars at least as bright as V=3.3 mag (sorted by V)} \\ \hline
Star   & V       & \multicolumn{3}{c}{colour index [mag]} & Parallax  & A$_{\rm V}$ & abs. mag.     & Notes \\
Hip    & [mag]   & B$-$V & (B$-$V)$_{0}$ & error         & [mas]     & [mag]       & M$_{\rm V}$ [mag] &  \\ \hline

\endfirsthead
Star   & V       & \multicolumn{3}{c}{colour index [mag]} & Parallax  & A$_{\rm V}$ & abs. mag.     & Notes \\
Hip    & [mag]   & B$-$V & (B$-$V)$_{0}$ & error         & [mas]     & [mag]       & M$_{\rm V}$ [mag] &  \\ \hline
\endhead
\hline
\endfoot
\endlastfoot
           32349 &$ -1.44 $&$  0.009 $&$  0.003 $&$\pm 0.007 $&$ 379.2 \pm   1.6 $&  0.02 &$  1.43 \pm  0.01 $& Hip \\
           30438 &$ -0.63 $&$  0.173 $&$  0.096 $&$\pm 0.024 $&$  10.6 \pm   0.6 $&  0.24 &$ -5.75 \pm  0.12 $&   \\
           71683 &$ -0.01 $&$  0.710 $&$  0.707 $&$\pm 0.040 $&$ 754.8 \pm   4.1 $&  0.01 &$  4.37 \pm  0.01 $& Hip \\
           91262 &$  0.03 $&$ -0.001 $&$ -0.011 $&$\pm 0.005 $&$ 130.2 \pm   0.4 $&  0.03 &$  0.57 \pm  0.01 $& Hip \\
           24608 &$  0.08 $&$  0.795 $&$  0.779 $&$\pm 0.019 $&$  76.2 \pm   0.5 $&  0.05 &$ -0.56 \pm  0.01 $& Hip \\
           69673 &$  0.16 $&$  1.137 $&$  1.131 $&$\pm 0.011 $&$  88.8 \pm   0.5 $&  0.02 &$ -0.12 \pm  0.01 $&   \\
           24436 &$  0.28 $&$  0.023 $&$ -0.054 $&$\pm 0.014 $&$   3.8 \pm   0.3 $&  0.24 &$ -7.07 \pm  0.20 $&   \\
           37279 &$  0.40 $&$  0.432 $&$  0.426 $&$\pm 0.015 $&$ 284.6 \pm   1.3 $&  0.02 &$  2.65 \pm  0.01 $& Hip \\
            7588 &$  0.54 $&$ -0.058 $&$ -0.100 $&$\pm 0.012 $&$  23.4 \pm   0.6 $&  0.13 &$ -2.74 \pm  0.05 $&   \\
           27989 &$  0.57 $&$  1.78  $&$  1.70  $&$\pm 0.05  $&$   6.6 \pm   0.8 $&  0.24 &$ -5.59 \pm  0.28 $& Tab. 3 \\
           68702 &$  0.64 $&$ -0.153 $&$ -0.250 $&$\pm 0.017 $&$   8.3 \pm   0.5 $&  0.30 &$ -5.06 \pm  0.13 $&   \\
           97649 &$  0.93 $&$  0.271 $&$  0.265 $&$\pm 0.015 $&$ 194.9 \pm   0.6 $&  0.02 &$  2.36 \pm  0.01 $&   \\
           21421 &$  0.99 $&$  1.480 $&$  1.457 $&$\pm 0.010 $&$  48.9 \pm   0.8 $&  0.07 &$ -0.63 \pm  0.03 $&   \\
           65474 &$  1.06 $&$ -0.136 $&$ -0.155 $&$\pm 0.014 $&$  13.1 \pm   0.7 $&  0.06 &$ -3.42 \pm  0.12 $&   \\ 
           80763 &$  1.07 $&$  1.844 $&$  1.660 $&$\pm 0.014 $&$   5.9 \pm   1.0 $&  0.57 &$ -5.65 \pm  0.37 $& Tab. 3 \\
           37826 &$  1.22 $&$  0.969 $&$  0.959 $&$\pm 0.005 $&$  96.5 \pm   0.3 $&  0.03 &$  1.11 \pm  0.01 $&   \\
          113368 &$  1.23 $&$  0.142 $&$  0.132 $&$\pm 0.011 $&$ 129.8 \pm   0.5 $&  0.03 &$  1.77 \pm  0.01 $&   \\
           60718A&$  1.28 $&$ -0.169 $&$ -0.285 $&$\pm 0.018 $&$  10.1 \pm   0.5 $&  0.36 &$ -4.05 \pm  0.11 $&   \\
           62434 &$  1.31 $&$ -0.158 $&$ -0.271 $&$\pm 0.009 $&$  11.7 \pm   1.0 $&  0.35 &$ -3.70 \pm  0.18 $&   \\
          102098 &$  1.33 $&$  0.112 $&$ -0.340 $&$\pm 0.010 $&$   2.3 \pm   0.3 $&  1.40 &$ -8.25 \pm  0.30 $& (a) \\
           71681 &$  1.35 $&$  0.900 $&$  0.897 $&$\pm 0.020 $&$ 796.9 \pm  25.9 $&  0.01 &$  5.85 \pm  0.07 $& Hip \\
           49669 &$  1.41 $&$ -0.036 $&$ -0.055 $&$\pm 0.011 $&$  41.1 \pm   0.4 $&  0.06 &$ -0.58 \pm  0.02 $&   \\
           33579 &$  1.53 $&$ -0.132 $&$ -0.229 $&$\pm 0.007 $&$   8.1 \pm   0.1 $&  0.30 &$ -4.24 \pm  0.04 $&   \\
           60718B&$  1.58 $&$ -0.159 $&$ -0.275 $&$\pm 0.018 $&$  10.1 \pm   0.5 $&  0.36 &$ -3.75 \pm  0.11 $&   \\
           85927 &$  1.63 $&$ -0.142 $&$ -0.297 $&$\pm 0.011 $&$   5.7 \pm   0.8 $&  0.48 &$ -5.07 \pm  0.29 $&   \\
           61084 &$  1.65 $&$  1.516 $&$  1.464 $&$\pm 0.005 $&$  36.8 \pm   0.2 $&  0.16 &$ -0.68 \pm  0.01 $&   \\
           25336 &$  1.66 $&$ -0.142 $&$ -0.210 $&$\pm 0.013 $&$  12.9 \pm   0.5 $&  0.21 &$ -2.99 \pm  0.09 $&   \\
           45238 &$  1.67 $&$  0.060 $&$ -0.008 $&$\pm 0.007 $&$  28.8 \pm   0.1 $&  0.21 &$ -1.24 \pm  0.01 $&   \\
           25428 &$  1.68 $&$ -0.059 $&$ -0.117 $&$\pm 0.011 $&$  24.4 \pm   0.3 $&  0.18 &$ -1.57 \pm  0.03 $&   \\
           26311 &$  1.72 $&$ -0.123 $&$ -0.213 $&$\pm 0.007 $&$   1.7 \pm   0.5 $&  0.28 &$ -7.47 \pm  0.59 $& (a) \\
           62956 &$  1.76 $&$  0.041 $&$  0.031 $&$\pm 0.007 $&$  39.5 \pm   0.2 $&  0.03 &$ -0.29 \pm  0.01 $&   \\
          109268 &$  1.77 $&$ -0.073 $&$ -0.105 $&$\pm 0.005 $&$  32.3 \pm   0.2 $&  0.10 &$ -0.78 \pm  0.01 $&   \\
           15863 &$  1.81 $&$  0.512 $&$  0.399 $&$\pm 0.004 $&$   6.4 \pm   0.2 $&  0.35 &$ -4.50 \pm  0.06 $&   \\
           90185 &$  1.81 $&$  0.018 $&$ -0.053 $&$\pm 0.009 $&$  22.8 \pm   0.2 $&  0.22 &$ -1.62 \pm  0.02 $&   \\
           39953 &$  1.82 $&$ -0.146 $&$ -0.236 $&$\pm 0.007 $&$   2.9 \pm   0.3 $&  0.28 &$ -6.13 \pm  0.22 $&   \\
           54061 &$  1.82 $&$  1.064 $&$  1.054 $&$\pm 0.003 $&$  26.5 \pm   0.5 $&  0.03 &$ -1.09 \pm  0.04 $&   \\
           34444 &$  1.84 $&$  0.701 $&$  0.653 $&$\pm 0.003 $&$   2.0 \pm   0.4 $&  0.15 &$ -6.77 \pm  0.41 $& (a) \\
           86228 &$  1.86 $&$  0.442 $&$  0.339 $&$\pm 0.007 $&$  10.9 \pm   1.5 $&  0.32 &$ -3.28 \pm  0.30 $&   \\
           67301 &$  1.86 $&$ -0.102 $&$ -0.112 $&$\pm 0.007 $&$  31.4 \pm   0.2 $&  0.03 &$ -0.69 \pm  0.02 $&   \\
           26727 &$  1.90 $&$ -0.104 $&$ -0.217 $&$\pm 0.006 $&$   4.4 \pm   0.6 $&  0.35 &$ -5.22 \pm  0.31 $&   \\
           28360 &$  1.90 $&$  0.072 $&$  0.046 $&$\pm 0.006 $&$  40.2 \pm   0.2 $&  0.08 &$ -0.16 \pm  0.01 $&   \\
           82273 &$  1.91 $&$  1.447 $&$  1.382 $&$\pm 0.002 $&$   8.4 \pm   0.2 $&  0.20 &$ -3.68 \pm  0.04 $&   \\
          100751 &$  1.92 $&$ -0.120 $&$ -0.175 $&$\pm 0.008 $&$  18.2 \pm   0.5 $&  0.17 &$ -1.94 \pm  0.06 $&   \\
           31681 &$  1.93 $&$  0.001 $&$ -0.054 $&$\pm 0.005 $&$  29.8 \pm   2.2 $&  0.17 &$ -0.87 \pm  0.16 $& Hip \\
           42913 &$  1.94 $&$  0.086 $&$  0.044 $&$\pm 0.003 $&$  40.5 \pm   0.4 $&  0.13 &$ -0.15 \pm  0.02 $&   \\
           41037 &$  1.95 $&$  1.195 $&$  1.098 $&$\pm 0.003 $&$   5.4 \pm   0.4 $&  0.30 &$ -4.69 \pm  0.17 $&   \\
           30324 &$  1.96 $&$ -0.164 $&$ -0.274 $&$\pm 0.007 $&$   6.6 \pm   0.2 $&  0.34 &$ -4.28 \pm  0.07 $&   \\
           36850A&$  1.98 $&$  0.030 $&$  0.014 $&$\pm 0.010 $&$  64.1 \pm   3.8 $&  0.05 &$  0.96 \pm  0.13 $& (b) \\
           46390 &$  1.99 $&$  1.453 $&$  1.385 $&$\pm 0.004 $&$  18.1 \pm   0.2 $&  0.21 &$ -1.93 \pm  0.02 $&   \\
           11767 &$  2.00 $&$  0.620 $&$  0.585 $&$\pm 0.003 $&$   7.5 \pm   0.1 $&  0.11 &$ -3.72 \pm  0.03 $&   \\
            9884 &$  2.02 $&$  1.160 $&$  1.137 $&$\pm 0.004 $&$  49.6 \pm   0.3 $&  0.07 &$  0.43 \pm  0.01 $&   \\
            3419 &$  2.05 $&$  1.026 $&$  1.003 $&$\pm 0.004 $&$  33.9 \pm   0.2 $&  0.07 &$ -0.37 \pm  0.01 $&   \\
           27366 &$  2.06 $&$ -0.116 $&$ -0.190 $&$\pm 0.006 $&$   5.0 \pm   0.2 $&  0.23 &$ -4.66 \pm  0.09 $&   \\
             677 &$  2.06 $&$ -0.043 $&$ -0.072 $&$\pm 0.004 $&$  33.6 \pm   0.4 $&  0.09 &$ -0.40 \pm  0.02 $&   \\
           72607 &$  2.06 $&$  1.480 $&$  1.454 $&$\pm 0.003 $&$  24.9 \pm   0.1 $&  0.08 &$ -1.04 \pm  0.01 $&   \\
           92855 &$  2.07 $&$ -0.138 $&$ -0.235 $&$\pm 0.006 $&$  14.3 \pm   0.3 $&  0.30 &$ -2.45 \pm  0.04 $&   \\
            5447 &$  2.08 $&$  1.586 $&$  1.554 $&$\pm 0.006 $&$  16.5 \pm   0.6 $&  0.10 &$ -1.93 \pm  0.07 $&   \\
           68933 &$  2.08 $&$  1.019 $&$  0.990 $&$\pm 0.006 $&$  55.5 \pm   0.2 $&  0.09 &$  0.71 \pm  0.01 $&   \\
           86032 &$  2.09 $&$  0.190 $&$  0.171 $&$\pm 0.004 $&$  67.1 \pm   1.1 $&  0.06 &$  1.16 \pm  0.03 $&   \\
           14576 &$  2.11 $&$ -0.006 $&$ -0.041 $&$\pm 0.004 $&$  36.3 \pm   1.4 $&  0.11 &$ -0.20 \pm  0.08 $&   \\
          112122 &$  2.12 $&$  1.535 $&$  1.496 $&$\pm 0.011 $&$  18.4 \pm   0.4 $&  0.12 &$ -1.67 \pm  0.05 $&   \\
           57632 &$  2.13 $&$  0.140 $&$  0.134 $&$\pm 0.005 $&$  90.9 \pm   0.5 $&  0.02 &$  1.90 \pm  0.01 $&   \\
            9640 &$  2.17 $&$  1.400 $&$  1.345 $&$\pm 0.006 $&$   8.3 \pm   1.0 $&  0.17 &$ -3.40 \pm  0.27 $&   \\
            4427 &$  2.17 $&$ -0.051 $&$ -0.145 $&$\pm 0.003 $&$   5.9 \pm   0.1 $&  0.29 &$ -4.24 \pm  0.04 $&   \\
           44816 &$  2.21 $&$  1.685 $&$  1.614 $&$\pm 0.006 $&$   6.0 \pm   0.1 $&  0.22 &$ -4.12 \pm  0.04 $&   \\
           39429 &$  2.22 $&$ -0.203 $&$ -0.316 $&$\pm 0.004 $&$   3.0 \pm   0.1 $&  0.35 &$ -5.74 \pm  0.07 $&   \\
           65378 &$  2.22 $&$  0.053 $&$  0.043 $&$\pm 0.003 $&$  38.0 \pm   1.7 $&  0.03 &$  0.09 \pm  0.10 $&   \\
           76267 &$  2.22 $&$  0.028 $&$  0.009 $&$\pm 0.003 $&$  43.5 \pm   0.3 $&  0.06 &$  0.35 \pm  0.01 $&   \\
           25930 &$  2.23 $&$ -0.166 $&$ -0.269 $&$\pm 0.005 $&$   4.7 \pm   0.6 $&  0.32 &$ -4.72 \pm  0.27 $&   \\
           87833 &$  2.23 $&$  1.535 $&$  1.483 $&$\pm 0.004 $&$  21.1 \pm   0.1 $&  0.16 &$ -1.30 \pm  0.01 $&   \\
           50583 &$  2.23 $&$  1.203 $&$  1.190 $&$\pm 0.064 $&$  25.1 \pm   0.5 $&  0.04 &$ -0.81 \pm  0.05 $&   \\
          100453 &$  2.23 $&$  0.690 $&$  0.574 $&$\pm 0.003 $&$   1.8 \pm   0.3 $&  0.36 &$ -6.88 \pm  0.33 $& (d) \\
            3179 &$  2.25 $&$  1.175 $&$  1.130 $&$\pm 0.002 $&$  14.3 \pm   0.2 $&  0.14 &$ -2.11 \pm  0.02 $&   \\
           45556 &$  2.25 $&$  0.202 $&$  0.067 $&$\pm 0.005 $&$   4.3 \pm   0.1 $&  0.42 &$ -5.02 \pm  0.05 $&   \\
           66657 &$  2.28 $&$ -0.173 $&$ -0.283 $&$\pm 0.003 $&$   7.6 \pm   0.5 $&  0.34 &$ -3.65 \pm  0.14 $&   \\
             746 &$  2.28 $&$  0.376 $&$  0.357 $&$\pm 0.004 $&$  59.6 \pm   0.4 $&  0.06 &$  1.10 \pm  0.01 $&   \\
           82396 &$  2.29 $&$  1.144 $&$  1.115 $&$\pm 0.032 $&$  51.2 \pm   0.2 $&  0.09 &$  0.75 \pm  0.01 $& Hip \\
           71860 &$  2.29 $&$ -0.156 $&$ -0.243 $&$\pm 0.002 $&$   7.0 \pm   0.2 $&  0.27 &$ -3.75 \pm  0.05 $&   \\
           78401 &$  2.30 $&$ -0.090 $&$ -0.232 $&$\pm 0.004 $&$   6.6 \pm   0.9 $&  0.44 &$ -4.03 \pm  0.29 $&   \\
           71352 &$  2.34 $&$ -0.160 $&$ -0.244 $&$\pm 0.004 $&$  10.7 \pm   0.2 $&  0.26 &$ -2.78 \pm  0.04 $&   \\
           53910 &$  2.35 $&$  0.029 $&$  0.019 $&$\pm 0.002 $&$  40.9 \pm   0.2 $&  0.03 &$  0.38 \pm  0.01 $&   \\
          107315 &$  2.39 $&$  1.526 $&$  1.429 $&$\pm 0.004 $&$   4.7 \pm   0.2 $&  0.30 &$ -4.54 \pm  0.08 $&   \\
           86670 &$  2.39 $&$ -0.173 $&$ -0.308 $&$\pm 0.006 $&$   6.8 \pm   0.2 $&  0.42 &$ -3.88 \pm  0.05 $&   \\
            2081 &$  2.40 $&$  1.090 $&$  1.067 $&$\pm 0.003 $&$  38.5 \pm   0.7 $&  0.07 &$  0.26 \pm  0.04 $&   \\
           61932A&$  2.42 $&$ -0.122 $&$ -0.196 $&$\pm 0.011 $&$  25.1 \pm   0.3 $&  0.23 &$ -0.81 \pm  0.02 $& (c) \\
           58001 &$  2.43 $&$  0.041 $&$  0.031 $&$\pm 0.002 $&$  39.2 \pm   0.4 $&  0.03 &$  0.37 \pm  0.02 $&   \\
           84012 &$  2.43 $&$  0.098 $&$  0.059 $&$\pm 0.004 $&$  36.9 \pm   0.8 $&  0.12 &$  0.15 \pm  0.05 $&   \\
           35904 &$  2.46 $&$ -0.048 $&$ -0.142 $&$\pm 0.004 $&$   1.6 \pm   0.4 $&  0.29 &$ -6.76 \pm  0.53 $& (a) \\
          113881 &$  2.47 $&$  1.628 $&$  1.586 $&$\pm 0.010 $&$  16.6 \pm   0.2 $&  0.13 &$ -1.55 \pm  0.02 $&   \\
          105199 &$  2.47 $&$  0.249 $&$  0.236 $&$\pm 0.003 $&$  66.5 \pm   0.1 $&  0.04 &$  1.54 \pm  0.00 $&   \\
           45941 &$  2.48 $&$ -0.143 $&$ -0.204 $&$\pm 0.004 $&$   5.7 \pm   0.3 $&  0.19 &$ -3.93 \pm  0.11 $&   \\
          113963 &$  2.49 $&$ -0.008 $&$ -0.043 $&$\pm 0.003 $&$  24.5 \pm   0.2 $&  0.11 &$ -0.68 \pm  0.02 $&   \\
          102488 &$  2.49 $&$  1.036 $&$  1.010 $&$\pm 0.002 $&$  44.9 \pm   0.1 $&  0.08 &$  0.67 \pm  0.01 $&   \\
           72105 &$  2.50 $&$  1.129 $&$  1.110 $&$\pm 0.005 $&$  16.1 \pm   0.7 $&  0.06 &$ -1.53 \pm  0.09 $&   \\
           68002 &$  2.53 $&$ -0.178 $&$ -0.262 $&$\pm 0.005 $&$   8.5 \pm   0.1 $&  0.26 &$ -3.07 \pm  0.03 $&   \\
           14135 &$  2.55 $&$  1.627 $&$  1.598 $&$\pm 0.005 $&$  13.1 \pm   0.4 $&  0.09 &$ -1.96 \pm  0.07 $&   \\
           54872 &$  2.56 $&$  0.152 $&$  0.142 $&$\pm 0.003 $&$  55.8 \pm   0.3 $&  0.03 &$  1.26 \pm  0.01 $&   \\
           59196 &$  2.57 $&$ -0.103 $&$ -0.203 $&$\pm 0.005 $&$   7.9 \pm   0.5 $&  0.31 &$ -3.26 \pm  0.13 $&   \\
           81377 &$  2.58 $&$  0.015 $&$ -0.104 $&$\pm 0.004 $&$   8.9 \pm   0.2 $&  0.37 &$ -3.04 \pm  0.05 $&   \\
           59803 &$  2.59 $&$ -0.073 $&$ -0.112 $&$\pm 0.004 $&$  21.2 \pm   0.2 $&  0.12 &$ -0.90 \pm  0.02 $&   \\
           25985 &$  2.59 $&$  0.238 $&$  0.219 $&$\pm 0.002 $&$   1.5 \pm   0.1 $&  0.06 &$ -6.63 \pm  0.21 $& (d) \\
           93506 &$  2.61 $&$  0.099 $&$  0.060 $&$\pm 0.003 $&$  37.0 \pm   0.9 $&  0.12 &$  0.33 \pm  0.05 $&   \\
           74785 &$  2.61 $&$ -0.074 $&$ -0.129 $&$\pm 0.004 $&$  17.6 \pm   0.2 $&  0.17 &$ -1.33 \pm  0.02 $&   \\
           78820 &$  2.62 $&$ -0.072 $&$ -0.188 $&$\pm 0.005 $&$   8.1 \pm   0.8 $&  0.36 &$ -3.21 \pm  0.21 $&   \\
           77070 &$  2.63 $&$  1.183 $&$  1.160 $&$\pm 0.003 $&$  44.1 \pm   0.2 $&  0.07 &$  0.78 \pm  0.01 $&   \\
           28380 &$  2.65 $&$ -0.060 $&$ -0.115 $&$\pm 0.009 $&$  19.7 \pm   0.2 $&  0.17 &$ -1.05 \pm  0.02 $&   \\
           61359 &$  2.66 $&$  0.895 $&$  0.847 $&$\pm 0.003 $&$  22.4 \pm   0.2 $&  0.15 &$ -0.74 \pm  0.02 $&   \\
            8903 &$  2.66 $&$  0.162 $&$  0.139 $&$\pm 0.003 $&$  55.6 \pm   0.6 $&  0.07 &$  1.32 \pm  0.02 $&   \\
           26634 &$  2.66 $&$ -0.091 $&$ -0.172 $&$\pm 0.003 $&$  12.5 \pm   0.4 $&  0.25 &$ -2.11 \pm  0.06 $&   \\
           85696 &$  2.68 $&$ -0.180 $&$ -0.335 $&$\pm 0.004 $&$   5.7 \pm   0.2 $&  0.48 &$ -4.04 \pm  0.07 $&   \\
           23015 &$  2.68 $&$  1.520 $&$  1.378 $&$\pm 0.004 $&$   6.6 \pm   0.4 $&  0.44 &$ -3.66 \pm  0.12 $&   \\
           73273 &$  2.68 $&$ -0.185 $&$ -0.262 $&$\pm 0.002 $&$   8.5 \pm   0.2 $&  0.24 &$ -2.91 \pm  0.05 $&   \\
            6686 &$  2.68 $&$  0.157 $&$  0.122 $&$\pm 0.002 $&$  32.8 \pm   0.1 $&  0.11 &$  0.15 \pm  0.01 $&   \\
           67927 &$  2.68 $&$  0.605 $&$  0.599 $&$\pm 0.002 $&$  87.8 \pm   1.2 $&  0.02 &$  2.38 \pm  0.03 $&   \\
           61585 &$  2.69 $&$ -0.178 $&$ -0.284 $&$\pm 0.002 $&$  10.3 \pm   0.1 $&  0.33 &$ -2.57 \pm  0.02 $&   \\
           89931 &$  2.70 $&$  1.390 $&$  1.248 $&$\pm 0.004 $&$   9.4 \pm   0.2 $&  0.44 &$ -2.88 \pm  0.04 $&   \\
           35264 &$  2.71 $&$  1.592 $&$  1.502 $&$\pm 0.004 $&$   4.0 \pm   0.3 $&  0.28 &$ -4.54 \pm  0.18 $&   \\
           97278 &$  2.71 $&$  1.522 $&$  1.425 $&$\pm 0.002 $&$   8.3 \pm   0.2 $&  0.30 &$ -3.01 \pm  0.04 $&   \\
           52727 &$  2.72 $&$  0.925 $&$  0.854 $&$\pm 0.005 $&$  27.8 \pm   0.4 $&  0.22 &$ -0.28 \pm  0.03 $&   \\
           79593 &$  2.73 $&$  1.581 $&$  1.520 $&$\pm 0.005 $&$  19.1 \pm   0.2 $&  0.19 &$ -1.06 \pm  0.02 $&   \\
           80331 &$  2.73 $&$  0.914 $&$  0.898 $&$\pm 0.002 $&$  35.4 \pm   0.1 $&  0.05 &$  0.43 \pm  0.01 $&   \\
           61941 &$  2.74 $&$  0.368 $&$  0.358 $&$\pm 0.017 $&$  85.6 \pm   0.6 $&  0.03 &$  2.37 \pm  0.02 $& Hip \\
           52419 &$  2.74 $&$ -0.200 $&$ -0.294 $&$\pm 0.002 $&$   7.2 \pm   0.2 $&  0.29 &$ -3.28 \pm  0.06 $&   \\
           72622 &$  2.75 $&$  0.169 $&$  0.140 $&$\pm 0.003 $&$  43.0 \pm   0.2 $&  0.09 &$  0.83 \pm  0.01 $&   \\
           65109 &$  2.76 $&$  0.064 $&$  0.035 $&$\pm 0.004 $&$  55.5 \pm   0.2 $&  0.09 &$  1.39 \pm  0.01 $&   \\
           86742 &$  2.77 $&$  1.186 $&$  1.151 $&$\pm 0.002 $&$  39.9 \pm   0.2 $&  0.11 &$  0.66 \pm  0.01 $&   \\
           80816 &$  2.78 $&$  0.941 $&$  0.899 $&$\pm 0.002 $&$  23.4 \pm   0.6 $&  0.13 &$ -0.50 \pm  0.05 $&   \\
           26241 &$  2.78 $&$ -0.211 $&$ -0.298 $&$\pm 0.004 $&$   1.4 \pm   0.2 $&  0.27 &$ -6.76 \pm  0.34 $& (a) \\
           76297 &$  2.78 $&$ -0.185 $&$ -0.275 $&$\pm 0.004 $&$   7.8 \pm   0.5 $&  0.28 &$ -3.05 \pm  0.14 $&   \\
           59747 &$  2.79 $&$ -0.194 $&$ -0.310 $&$\pm 0.002 $&$   9.5 \pm   0.2 $&  0.36 &$ -2.69 \pm  0.03 $&   \\
           23875 &$  2.79 $&$  0.151 $&$  0.109 $&$\pm 0.003 $&$  36.5 \pm   0.4 $&  0.13 &$  0.47 \pm  0.02 $&   \\
           85670 &$  2.80 $&$  0.958 $&$  0.890 $&$\pm 0.002 $&$   8.6 \pm   0.1 $&  0.21 &$ -2.74 \pm  0.03 $&   \\
           85258 &$  2.82 $&$  1.478 $&$  1.349 $&$\pm 0.003 $&$   5.1 \pm   0.6 $&  0.40 &$ -4.06 \pm  0.28 $&   \\
            2021 &$  2.82 $&$  0.618 $&$  0.605 $&$\pm 0.008 $&$ 134.1 \pm   0.1 $&  0.04 &$  3.42 \pm  0.00 $& Hip \\
           81266 &$  2.83 $&$ -0.207 $&$ -0.365 $&$\pm 0.002 $&$   6.9 \pm   0.5 $&  0.49 &$ -3.47 \pm  0.17 $&   \\
           90496 &$  2.83 $&$  1.056 $&$  1.021 $&$\pm 0.004 $&$  41.7 \pm   0.2 $&  0.11 &$  0.82 \pm  0.01 $&   \\
           39757 &$  2.83 $&$  0.458 $&$  0.426 $&$\pm 0.003 $&$  51.3 \pm   0.2 $&  0.10 &$  1.28 \pm  0.01 $&   \\
           77952 &$  2.84 $&$  0.313 $&$  0.290 $&$\pm 0.003 $&$  80.8 \pm   0.2 $&  0.07 &$  2.31 \pm  0.00 $&   \\
           25606 &$  2.84 $&$  0.821 $&$  0.747 $&$\pm 0.002 $&$  20.3 \pm   0.2 $&  0.23 &$ -0.85 \pm  0.02 $&   \\
           63608 &$  2.84 $&$  0.934 $&$  0.924 $&$\pm 0.003 $&$  29.8 \pm   0.1 $&  0.03 &$  0.18 \pm  0.01 $&   \\
            1067 &$  2.84 $&$ -0.191 $&$ -0.239 $&$\pm 0.002 $&$   8.3 \pm   0.5 $&  0.15 &$ -2.71 \pm  0.14 $&   \\
          107556 &$  2.85 $&$  0.180 $&$  0.164 $&$\pm 0.150 $&$  84.3 \pm   0.2 $&  0.05 &$  2.43 \pm  0.00 $& Hip \\
           81693 &$  2.85 $&$  0.651 $&$  0.641 $&$\pm 0.003 $&$  93.3 \pm   0.5 $&  0.03 &$  2.67 \pm  0.01 $&   \\
           85792 &$  2.85 $&$ -0.113 $&$ -0.207 $&$\pm 0.004 $&$  12.2 \pm   0.9 $&  0.29 &$ -2.01 \pm  0.15 $&   \\
            9236 &$  2.86 $&$  0.308 $&$  0.282 $&$\pm 0.003 $&$  45.4 \pm   0.4 $&  0.08 &$  1.07 \pm  0.02 $&   \\
          110130 &$  2.86 $&$  1.393 $&$  1.351 $&$\pm 0.003 $&$  16.3 \pm   0.6 $&  0.13 &$ -1.21 \pm  0.08 $&   \\
           61932B&$  2.87 $&$  0.042 $&$ -0.032 $&$\pm 0.018 $&$  25.1 \pm   0.3 $&  0.23 &$ -0.37 \pm  0.02 $& (c) \\
           17702 &$  2.88 $&$ -0.071 $&$ -0.155 $&$\pm 0.004 $&$   8.1 \pm   0.4 $&  0.26 &$ -2.84 \pm  0.11 $&   \\
           74946 &$  2.88 $&$  0.026 $&$ -0.080 $&$\pm 0.002 $&$  17.7 \pm   0.1 $&  0.33 &$ -1.21 \pm  0.01 $&   \\
           18246 &$  2.88 $&$  0.082 $&$ -0.118 $&$\pm 0.002 $&$   4.3 \pm   0.2 $&  0.62 &$ -4.55 \pm  0.10 $&   \\
           36850B&$  2.88 $&$  0.040 $&$  0.024 $&$\pm 0.010 $&$  64.1 \pm   3.8 $&  0.05 &$  1.86 \pm  0.13 $& (b) \\
          106278 &$  2.89 $&$  0.837 $&$  0.753 $&$\pm 0.002 $&$   6.1 \pm   0.2 $&  0.26 &$ -3.45 \pm  0.08 $&   \\
           63125 &$  2.89 $&$ -0.058 $&$ -0.064 $&$\pm 0.004 $&$  28.4 \pm   0.9 $&  0.02 &$  0.14 \pm  0.07 $&   \\
           36188 &$  2.89 $&$ -0.075 $&$ -0.152 $&$\pm 0.004 $&$  20.2 \pm   0.2 $&  0.24 &$ -0.83 \pm  0.02 $&   \\
           78265 &$  2.89 $&$ -0.160 $&$ -0.331 $&$\pm 0.002 $&$   5.6 \pm   0.6 $&  0.53 &$ -3.91 \pm  0.25 $&   \\
           94141 &$  2.90 $&$  0.377 $&$  0.222 $&$\pm 0.003 $&$   6.4 \pm   0.4 $&  0.48 &$ -3.55 \pm  0.15 $&   \\
           30343 &$  2.90 $&$  1.592 $&$  1.521 $&$\pm 0.007 $&$  14.1 \pm   0.7 $&  0.22 &$ -1.58 \pm  0.11 $&   \\
           97165 &$  2.91 $&$ -0.022 $&$ -0.087 $&$\pm 0.003 $&$  19.8 \pm   0.5 $&  0.20 &$ -0.81 \pm  0.05 $&   \\
           80112 &$  2.91 $&$  0.099 $&$ -0.124 $&$\pm 0.004 $&$   4.7 \pm   0.6 $&  0.69 &$ -4.43 \pm  0.28 $&   \\
           18532 &$  2.91 $&$ -0.174 $&$ -0.322 $&$\pm 0.003 $&$   5.1 \pm   0.2 $&  0.46 &$ -4.01 \pm  0.10 $&   \\
           32768 &$  2.94 $&$  1.206 $&$  1.109 $&$\pm 0.003 $&$  17.9 \pm   0.4 $&  0.30 &$ -1.09 \pm  0.05 $&   \\
          112158 &$  2.94 $&$  0.851 $&$  0.803 $&$\pm 0.002 $&$  15.2 \pm   0.7 $&  0.15 &$ -1.30 \pm  0.10 $&   \\
          109074 &$  2.94 $&$  0.964 $&$  0.880 $&$\pm 0.004 $&$   6.2 \pm   0.2 $&  0.26 &$ -3.35 \pm  0.07 $&   \\
           14328 &$  2.94 $&$  0.699 $&$  0.647 $&$\pm 0.002 $&$  13.4 \pm   0.5 $&  0.16 &$ -1.58 \pm  0.08 $&   \\
           18543 &$  2.96 $&$  1.586 $&$  1.544 $&$\pm 0.005 $&$  16.0 \pm   0.6 $&  0.13 &$ -1.14 \pm  0.08 $&   \\
           47908 &$  2.97 $&$  0.808 $&$  0.766 $&$\pm 0.003 $&$  13.2 \pm   0.2 $&  0.13 &$ -1.55 \pm  0.02 $&   \\
           60965 &$  2.97 $&$ -0.015 $&$ -0.044 $&$\pm 0.004 $&$  37.6 \pm   0.2 $&  0.09 &$  0.75 \pm  0.01 $&   \\
           88635 &$  2.98 $&$  0.981 $&$  0.936 $&$\pm 0.047 $&$  33.7 \pm   0.2 $&  0.14 &$  0.48 \pm  0.01 $& Hip \\
           93747 &$  2.99 $&$  0.027 $&$ -0.008 $&$\pm 0.003 $&$  39.3 \pm   0.2 $&  0.11 &$  0.85 \pm  0.01 $&   \\
           82514 &$  3.00 $&$ -0.175 $&$ -0.320 $&$\pm 0.010 $&$   6.5 \pm   0.9 $&  0.45 &$ -3.38 \pm  0.30 $&   \\
           64962 &$  3.00 $&$  0.914 $&$  0.866 $&$\pm 0.003 $&$  24.4 \pm   0.2 $&  0.15 &$ -0.22 \pm  0.01 $&   \\
           26451 &$  3.00 $&$ -0.150 $&$ -0.247 $&$\pm 0.009 $&$   7.3 \pm   0.8 $&  0.30 &$ -2.97 \pm  0.24 $&   \\
           54539 &$  3.00 $&$  1.152 $&$  1.152 $&$\pm 0.003 $&$  22.6 \pm   0.1 $&  0.00 &$ -0.23 \pm  0.01 $&   \\
           48002 &$  3.01 $&$  0.267 $&$  0.180 $&$\pm 0.004 $&$   2.3 \pm   0.3 $&  0.27 &$ -5.48 \pm  0.27 $& (d) \\
          108085 &$  3.01 $&$ -0.087 $&$ -0.126 $&$\pm 0.003 $&$  15.4 \pm   0.7 $&  0.12 &$ -1.17 \pm  0.09 $&   \\
           59316 &$  3.01 $&$  1.338 $&$  1.290 $&$\pm 0.003 $&$  10.3 \pm   0.2 $&  0.15 &$ -2.08 \pm  0.03 $&   \\
           87073 &$  3.01 $&$  0.509 $&$  0.296 $&$\pm 0.004 $&$   1.7 \pm   0.2 $&  0.66 &$ -6.51 \pm  0.19 $& (d) \\
           32246 &$  3.01 $&$  1.405 $&$  1.318 $&$\pm 0.003 $&$   3.9 \pm   0.2 $&  0.27 &$ -4.33 \pm  0.10 $&   \\
           10064 &$  3.02 $&$  0.155 $&$  0.113 $&$\pm 0.004 $&$  25.7 \pm   0.3 $&  0.13 &$ -0.06 \pm  0.03 $&   \\
           17358 &$  3.02 $&$ -0.106 $&$ -0.219 $&$\pm 0.002 $&$   6.3 \pm   0.5 $&  0.35 &$ -3.33 \pm  0.16 $&   \\
           30122 &$  3.02 $&$ -0.163 $&$ -0.253 $&$\pm 0.002 $&$   9.0 \pm   0.1 $&  0.28 &$ -2.49 \pm  0.03 $&   \\
           23416 &$  3.03 $&$  0.529 $&$  0.190 $&$\pm 0.007 $&$   1.5 \pm   1.3 $&  1.05 &$ -7.10 \pm  1.83 $& (d) \\
           75097 &$  3.03 $&$  0.058 $&$ -0.003 $&$\pm 0.004 $&$   6.7 \pm   0.1 $&  0.19 &$ -3.03 \pm  0.04 $&   \\
           33977 &$  3.04 $&$ -0.089 $&$ -0.144 $&$\pm 0.004 $&$   1.2 \pm   0.4 $&  0.17 &$ -6.77 \pm  0.74 $& (a) \\
           71075 &$  3.04 $&$  0.209 $&$  0.196 $&$\pm 0.003 $&$  37.6 \pm   0.1 $&  0.04 &$  0.87 \pm  0.01 $&   \\
           50801 &$  3.05 $&$  1.574 $&$  1.555 $&$\pm 0.006 $&$  14.2 \pm   0.5 $&  0.06 &$ -1.25 \pm  0.08 $&   \\
           95947 &$  3.08 $&$  1.086 $&$  1.041 $&$\pm 0.002 $&$   7.5 \pm   0.3 $&  0.14 &$ -2.68 \pm  0.10 $&   \\
           94376 &$  3.08 $&$  1.003 $&$  0.977 $&$\pm 0.002 $&$  33.5 \pm   0.1 $&  0.08 &$  0.62 \pm  0.01 $&   \\
           62322 &$  3.08 $&$ -0.166 $&$ -0.272 $&$\pm 0.014 $&$   9.6 \pm   0.4 $&  0.33 &$ -2.35 \pm  0.09 $&   \\
          100345 &$  3.09 $&$  0.765 $&$  0.684 $&$\pm 0.003 $&$  10.0 \pm   1.1 $&  0.25 &$ -2.16 \pm  0.24 $&   \\
           27628 &$  3.11 $&$  1.170 $&$  1.128 $&$\pm 0.003 $&$  37.4 \pm   0.1 $&  0.13 &$  0.84 \pm  0.01 $&   \\
           83081 &$  3.11 $&$  1.611 $&$  1.514 $&$\pm 0.005 $&$   6.7 \pm   0.2 $&  0.30 &$ -3.06 \pm  0.06 $&   \\
          101772 &$  3.11 $&$  1.000 $&$  0.961 $&$\pm 0.002 $&$  33.2 \pm   0.2 $&  0.12 &$  0.59 \pm  0.01 $&   \\
           52943 &$  3.11 $&$  1.250 $&$  1.205 $&$\pm 0.003 $&$  22.7 \pm   0.2 $&  0.14 &$ -0.25 \pm  0.02 $&   \\
           43813 &$  3.12 $&$  1.001 $&$  0.940 $&$\pm 0.004 $&$  19.5 \pm   0.2 $&  0.19 &$ -0.62 \pm  0.02 $&   \\
           56561 &$  3.12 $&$ -0.039 $&$ -0.145 $&$\pm 0.002 $&$   7.8 \pm   0.3 $&  0.33 &$ -2.76 \pm  0.10 $&   \\
           73334 &$  3.13 $&$ -0.179 $&$ -0.253 $&$\pm 0.003 $&$   8.5 \pm   0.5 $&  0.23 &$ -2.45 \pm  0.14 $&   \\
           45860 &$  3.13 $&$  1.555 $&$  1.526 $&$\pm 0.005 $&$  16.1 \pm   0.2 $&  0.09 &$ -0.93 \pm  0.02 $&   \\
           89642 &$  3.13 $&$  1.528 $&$  1.457 $&$\pm 0.005 $&$  22.4 \pm   0.2 $&  0.22 &$ -0.34 \pm  0.02 $&   \\
           84379 &$  3.13 $&$  0.091 $&$  0.065 $&$\pm 0.003 $&$  43.4 \pm   0.2 $&  0.08 &$  1.24 \pm  0.01 $&   \\
           44127 &$  3.14 $&$  0.214 $&$  0.204 $&$\pm 0.003 $&$  68.9 \pm   0.2 $&  0.03 &$  2.30 \pm  0.01 $&   \\
           84380 &$  3.14 $&$  1.433 $&$  1.362 $&$\pm 0.002 $&$   8.7 \pm   0.1 $&  0.22 &$ -2.39 \pm  0.03 $&   \\
           46701 &$  3.16 $&$  1.541 $&$  1.425 $&$\pm 0.003 $&$  13.7 \pm   0.1 $&  0.36 &$ -1.52 \pm  0.02 $&   \\
           23767 &$  3.17 $&$ -0.152 $&$ -0.226 $&$\pm 0.004 $&$  13.4 \pm   0.2 $&  0.23 &$ -1.42 \pm  0.03 $&   \\
           92041 &$  3.17 $&$ -0.092 $&$ -0.195 $&$\pm 0.004 $&$  13.6 \pm   0.2 $&  0.32 &$ -1.48 \pm  0.03 $&   \\
           23685 &$  3.18 $&$  1.470 $&$  1.402 $&$\pm 0.003 $&$  15.3 \pm   0.2 $&  0.21 &$ -1.11 \pm  0.03 $&   \\
           46853 &$  3.18 $&$  0.484 $&$  0.474 $&$\pm 0.004 $&$  74.2 \pm   0.1 $&  0.03 &$  2.50 \pm  0.00 $&   \\
           83895 &$  3.18 $&$ -0.107 $&$ -0.152 $&$\pm 0.002 $&$   9.9 \pm   0.4 $&  0.14 &$ -1.98 \pm  0.08 $&   \\
           71908 &$  3.18 $&$  0.269 $&$  0.237 $&$\pm 0.003 $&$  60.4 \pm   0.1 $&  0.10 &$  1.98 \pm  0.01 $&   \\
           31685 &$  3.18 $&$ -0.098 $&$ -0.175 $&$\pm 0.003 $&$   8.8 \pm   0.3 $&  0.24 &$ -2.34 \pm  0.06 $&   \\
           83000 &$  3.19 $&$  1.158 $&$  1.123 $&$\pm 0.003 $&$  35.7 \pm   0.2 $&  0.11 &$  0.84 \pm  0.01 $&   \\
           22449 &$  3.19 $&$  0.474 $&$  0.464 $&$\pm 0.003 $&$ 123.9 \pm   0.2 $&  0.03 &$  3.63 \pm  0.00 $&   \\
           87261 &$  3.19 $&$  1.191 $&$  1.126 $&$\pm 0.003 $&$  25.9 \pm   0.2 $&  0.20 &$  0.06 \pm  0.01 $&   \\
          104732 &$  3.20 $&$  0.990 $&$  0.935 $&$\pm 0.002 $&$  22.8 \pm   0.4 $&  0.17 &$ -0.18 \pm  0.03 $&   \\
           13847 &$  3.22 $&$  0.156 $&$  0.114 $&$\pm 0.003 $&$  20.2 \pm   0.6 $&  0.13 &$ -0.38 \pm  0.06 $&   \\
          116727 &$  3.22 $&$  1.036 $&$  1.020 $&$\pm 0.003 $&$  70.9 \pm   0.4 $&  0.05 &$  2.42 \pm  0.01 $&   \\
           75141 &$  3.22 $&$ -0.201 $&$ -0.369 $&$\pm 0.004 $&$   3.7 \pm   0.5 $&  0.52 &$ -4.46 \pm  0.32 $&   \\
          106032 &$  3.23 $&$ -0.207 $&$ -0.278 $&$\pm 0.004 $&$   4.8 \pm   0.3 $&  0.22 &$ -3.60 \pm  0.14 $&   \\
           79882 &$  3.24 $&$  0.969 $&$  0.927 $&$\pm 0.005 $&$  30.6 \pm   0.2 $&  0.13 &$  0.54 \pm  0.01 $&   \\
           99473 &$  3.25 $&$ -0.047 $&$ -0.134 $&$\pm 0.002 $&$  11.4 \pm   0.2 $&  0.27 &$ -1.74 \pm  0.05 $&   \\
           89962 &$  3.25 $&$  0.943 $&$  0.917 $&$\pm 0.005 $&$  53.9 \pm   0.2 $&  0.08 &$  1.83 \pm  0.01 $&   \\
           32607 &$  3.25 $&$  0.229 $&$  0.184 $&$\pm 0.003 $&$  33.8 \pm   1.8 $&  0.14 &$  0.75 \pm  0.11 $&   \\
           93194 &$  3.25 $&$ -0.036 $&$ -0.094 $&$\pm 0.002 $&$   5.3 \pm   0.3 $&  0.18 &$ -3.33 \pm  0.11 $&   \\
           36377 &$  3.26 $&$  1.509 $&$  1.403 $&$\pm 0.004 $&$  16.8 \pm   0.5 $&  0.33 &$ -0.94 \pm  0.06 $&   \\
           68895 &$  3.26 $&$  1.116 $&$  1.071 $&$\pm 0.003 $&$  32.3 \pm   0.2 $&  0.14 &$  0.67 \pm  0.01 $&   \\
           21281 &$  3.26 $&$ -0.079 $&$ -0.137 $&$\pm 0.004 $&$  19.3 \pm   0.3 $&  0.18 &$ -0.49 \pm  0.03 $&   \\
           17678 &$  3.26 $&$  1.593 $&$  1.528 $&$\pm 0.004 $&$  15.2 \pm   0.1 $&  0.20 &$ -1.03 \pm  0.02 $&   \\
           84970 &$  3.26 $&$ -0.187 $&$ -0.348 $&$\pm 0.002 $&$   7.5 \pm   0.2 $&  0.50 &$ -2.87 \pm  0.05 $&   \\
            3092 &$  3.27 $&$  1.282 $&$  1.250 $&$\pm 0.003 $&$  30.9 \pm   0.2 $&  0.10 &$  0.62 \pm  0.01 $&   \\
          113136 &$  3.27 $&$  0.075 $&$  0.033 $&$\pm 0.002 $&$  20.3 \pm   1.8 $&  0.13 &$ -0.32 \pm  0.19 $&   \\
           73714 &$  3.28 $&$  1.621 $&$  1.544 $&$\pm 0.013 $&$  11.3 \pm   0.3 $&  0.24 &$ -1.69 \pm  0.05 $&   \\
           24305 &$  3.29 $&$ -0.095 $&$ -0.163 $&$\pm 0.002 $&$  17.5 \pm   0.6 $&  0.21 &$ -0.70 \pm  0.07 $&   \\
           75458 & $ 3.30 $&$  1.176 $&$  1.163 $&$\pm 0.003 $&$  32.2 \pm   0.1 $&  0.04 &$  0.80 \pm  0.01 $&   \\ 
           50099 & $ 3.30 $&$ -0.070 $&$ -0.180 $&$\pm 0.003 $&$   9.5 \pm   0.1 $&  0.34 &$ -2.14 \pm  0.02 $&   \\
           59774 & $ 3.30 $&$  0.092 $&$  0.082 $&$\pm 0.003 $&$  40.5 \pm   0.2 $&  0.03 &$  1.31 \pm  0.01 $&   \\
           29655 & $ 3.30 $&$  1.560 $&$  1.486 $&$\pm 0.012 $&$   8.5 \pm   1.2 $&  0.23 &$ -2.29 \pm  0.31 $& \\ \hline
\end{longtable}

\vspace{-.5cm}

\end{onecolumn}

\noindent Notes and references: `Hip' means that V and B$-$V and their errors are from the Hipparcos rather than the Tycho-2 catalogs;
references are either Tycho-2 or as otherwise given. 
Extinctions are from Goncharov \& Mosenkov (2017), unless otherwise given. 
Parallaxes are from the new Hipparcos catalog (van Leeuwen 2007).
Others: (a) Bobylev, Goncharov, \& Bajkova (2006), (b) Barrado (1998), (c) Kharchenko 2009, (d) Melnik \& Dambis (2020).
Intrinsic colours (B$-$V)$_{0}$ and absolute magnitudes M$_{\rm V}$ are calculated by us.

\end{document}